\documentclass{aa}  
\usepackage{graphicx}
\usepackage{txfonts}
\begin{document}
   \title{Cold gas in the Perseus cluster core: \\ 
Excitation of molecular gas in filaments}
   \author{Salom\'e, P.  \inst{1}\fnmsep\thanks{Based on observations carried out 
       at the IRAM 30m telescope, with the help of IRAM members, Pico Veleta, Spain.}
     \and Combes, F.  \inst{2}
     \and Revaz Y. \inst{2}
     \and Edge A.C. \inst{3}
     \and Hatch N.A. \inst{5}
     \and Fabian A.C.  \inst{4}
     \and Johnstone R.M. \inst{4}
   }
   \offprints{salome@iram.fr}
   \institute{Institut of Radio Astronomy (IRAM), Domaine
              Universitaire, 300, rue de la piscine, F-38400 St Martin
              d'H\`eres, France\\ \email{salome@iram.fr} \and
              Observatoire de Paris,LERMA, 61 Av. de l'Observatoire,
              F-75014, Paris, France 
              \and Department of Physics, University of Durham, South
              Road, Durham DH1 3LEi, UK
              \and Institute of
              Astronomy, Madingley Road, Cambridge CB3 OHA, UK \and
              Leiden Observatory, NL-2300 RA Leiden, The Netherlands}
   \date{Received 4 February 2008 / Accepted 14 March 2008}
   \abstract {We have recently detected CO lines in the well-known
    filaments around NGC 1275, the galaxy at the centre of the Perseus
    cluster of galaxies.  These previous observations, with the HERA
    multi-beam array at the IRAM 30m telescope enabled us to make a large
    map of the CO(2--1) line and to see hints of molecular gas far
    away from the cluster centre. To confirm the presence of CO
    emission lines in the outer filaments and to study the
    CO(2--1)/CO(1--0) line ratio, we observed seven regions of
    interest again with the 30m telescope in both CO(1--0) and
    CO(2--1). The regions we observed were: the eastern filament, the
    horseshoe, the northern filament and a southern extension, all
    selected from H$\alpha$ emission line mapping. Molecular gas is
    detected in all the observed regions. This result confirms the
    large extent of the cold molecular gas filaments.  We discuss the
    CO(2--1)/CO(1--0) ratios in the filaments. The eastern filament
    has optically thick gas, whereas further away, the line ratio
    increases close to values expected for a warmer optically thin
    medium.  We also show CO(1--0) and CO(2--1) lines in 9 regions
    closer to the centre.  The kinematics of the CO is studied here in
    more detail and confirms that it follows the motions of the warm
    H$_2$ gas found in the near-infrared. Finally, we searched for
    dense gas tracers around 3C84 and claim here the first detection
    of HCN(3-2).

    \keywords{Galaxies: cD, cooling flows, intergalactic medium,
   Galaxies: individual: NGC 1275}
     }
    \maketitle
%
\section{Introduction}
%
Over the past decade X-ray observations have shown a lack of cool X-ray
emitting gas in cooling flow clusters (e.g. Peterson et al 2003),
and many feedback models have been proposed to explain the required
energy injection into the intracluster medium (e.g. Binney \& Tabor
1995; Omma \& Binney 2004). At the same time, large amounts of
molecular gas have been found in many Brightest Cluster Galaxies
(BCG) (Edge 2001; Salom\'e \& Combes 2003), providing evidence
that some of the ICM may to cool to very low temperatures. Detailed
studies of this cold molecular gas reservoir are a complementary way to
probe feedback processes.

The giant cD galaxy NGC 1275 is the central galaxy of the Perseus cluster
(Abell 426) and lies at a redshift of 0.01756. At this distance, 1$''$ is 350
pc (H$_0$ = 71 km/s/Mpc, $\Omega_M$=0.27, $\Omega_\lambda$=0.73).  
This object is famous for the huge filamentary structure detected in
the optical (Hu et al., 1983, Conselice et al. 2001).  These bright
H$\alpha$ emitting filaments are known to be directly/indirectly
associated with cooling cores (Crawford et al., 1999; Edwards et al.,
2007).
We have shown previously (Salom\'e et al., 2006; hereafter S06) that the
molecular gas is also detected in CO(2--1) emission in the centre of
NGC 1275 with morphology and dynamics identical to that of the
H$\alpha$ emission. 

We present here the results of follow-up observations aimed at confirming
the presence of cold gas associated with the optical filaments further
out from the galaxy. We looked for CO(1--0) and CO(2--1) emission
lines in 7 regions selected in the optical filaments. We also observed
the central region at both 1.3 and 3mm with high sensitivity
receivers. Section 2 describes the observations. Section 3 presents
the results and discusses the spectra obtained in the different
regions of interest. We then compare the CO results (morphology,
kinematics, line ratios) with data at other wavelengths in section
4. Section 5 summarizes our conclusions.
%
\section{Observations}
%
The observations were made with the IRAM 30m telescope on Pico Veleta,
Spain during two different runs in 2006. We used the wobbler switching
mode with two 3mm and two 1.3mm receivers operating simultaneously.
The beam throw was 4 arcmin. Receivers were tuned to the CO(1--0) and
CO(2--1) emission lines, redshifted to the velocity of NGC~1275.  Frequent
pointings were done on 3C84, the radio source at the centre of NGC~1275. 
At 3mm, we used two 512\,$\times\,$1\,MHz filter-banks. This
gives a total band of $\sim$ 1300km/s for the CO(1--0) line setup.  In
addition, we used the two 250\,$\times\,$\,4 MHz resolution
filter-banks for the 1.3mm receivers, providing a 1 GHz band width, for
the 1.3mm receivers which also corresponds to a 1300km/s
bandwidth. The beams of the 30m telescope, at 3mm (113.282 GHz) and
1.3mm (226.559 GHz), are respectively 22$^{\prime\prime}$ and
11$^{\prime\prime}$.

Regions close to the centre were very quickly detected (in 10-20 min),
while the more distant regions required more time ($\sim$ 2 hours).
The signals are expressed here in main beam brightness temperatures.
The main-beam efficiency of the 30m telescope is the ratio of the antenna
temperature to the main beam temperature: ${\rm T_{\rm A}^*/T_{\rm mb} =
B_{eff}/F_{eff}}$ with the ratio of the beam efficiency to the forward 
efficiency being: ${\rm B_{eff}/F_{eff}}$= 0.75/0.95 at 3mm and 0.52/0.91 at
1.3mm (cf IRAM-30m site http://www.iram.es/).

The data were calibrated with the MIRA software and reduced with the
CLASS90 package. Spiky channels and bad scans were dropped and linear
baselines were subtracted for each spectrum.  After averaging all the
spectra for each line at each position, the data were Hanning smoothed to a 42
km/s resolution. 

The data are summarized in Tables \ref{table-filaments} and
\ref{table2-center}. The CO(2--1) lines of the filaments use the
present observations and the data obtained with HERA (S06): the
CO(2--1) spectra were convolved with the beam pattern at 3mm to
compare the CO(1--0) and CO(2--1) temperatures.
%
\begin{figure*} 
\centering
\includegraphics[width=13cm, angle=-90]{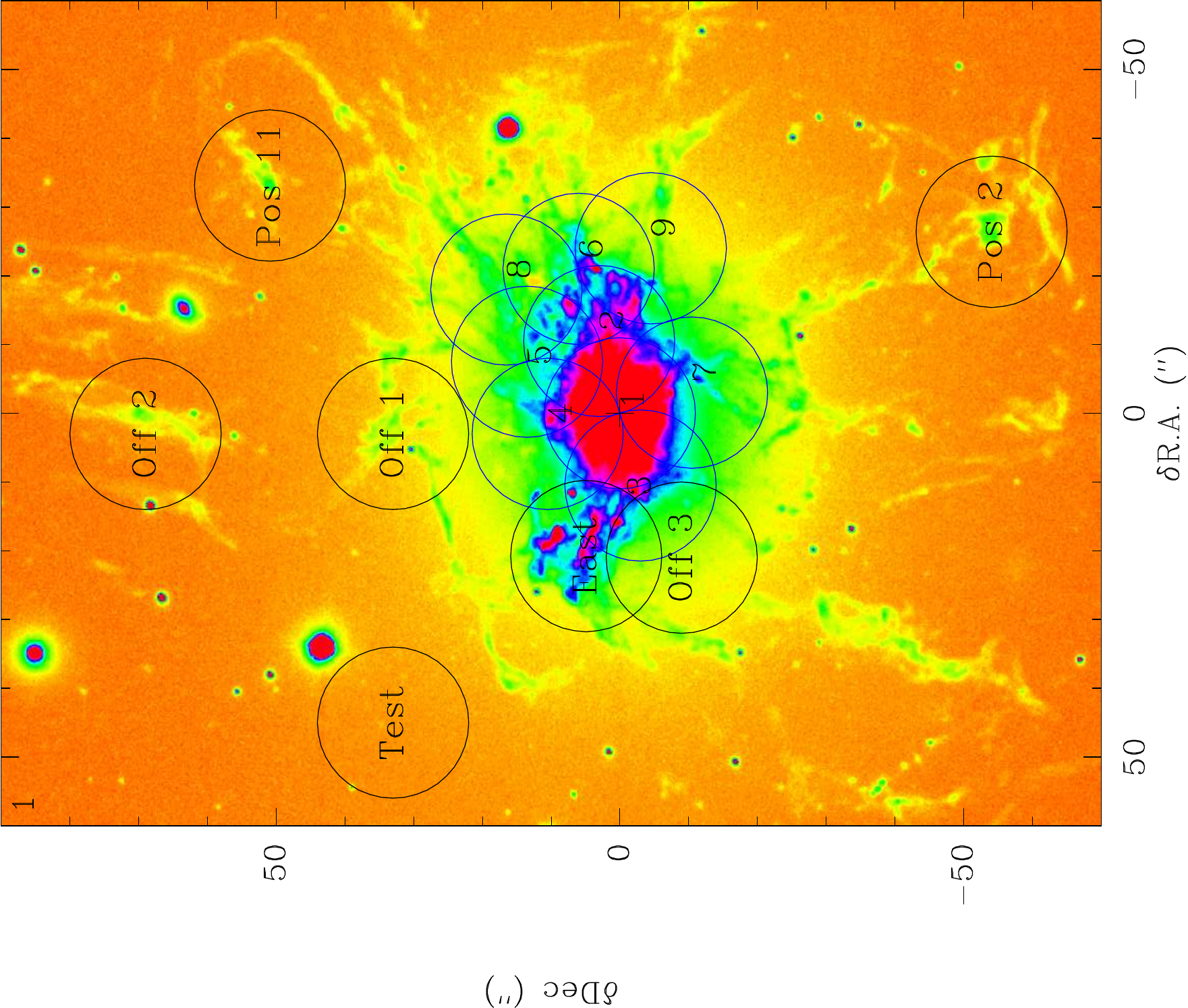} \\
\caption{H$\alpha$ image of the filament system in NGC1275 (Conselice et al.,
2001). Overlaid are the regions re-observed with the 30m
telescope. The circles represent the beam size at 3mm.}
\label{pointings} 
\end{figure*} 
%
%
\section{Results}
%
%
\begin{table*} 
\caption{Results of the observations.} 
\begin{center} 
\begin{tabular}{ccccccccc} 
\hline 
\hline 
Position & Offsets & Line & T$_{{\rm mb}}$ & Velocity & Width & I$_{{\rm CO}}$ & M$_{{\rm gas}}$ & T$_{21}$/T$_{10}$ \\ 
 & [$\prime\prime$ $\times$ $\prime\prime$] &  & [mK] & [km/s] & [km/s] & [K.km/s] & [10$^8$M$_\odot$] & \\ 
\hline 
Off1& [3, 33] & CO(1-0) & 2 $\pm$ 1.0 &33.3 $\pm$ 37.1 &237.6 $\pm$ 59.8 &0.5 $\pm$ 0.1 & 1.1 & \\
Off1& [3, 33] & CO(2-1) & 5.3 $\pm$ 1.8 &-55.2 $\pm$ 26 &275 $\pm$ 47.7 &1.5 $\pm$ 0.3 & & \\
Off1& [3, 33] & CO(2-1) & 4.7 $\pm$ 0.8 &-18.4 $\pm$ 12.9 &247.1 $\pm$ 26.7 &1.2 $\pm$ 0.1 & & 2.4\\
\hline 
Off2& [3, 69] & CO(1-0) & 2.5 $\pm$ 0.6 &49 $\pm$ 13.4 &128.5 $\pm$ 31.1 &0.3 $\pm$ 0.1 & 0.8 & \\
Off2& [3, 69] & CO(2-1) & 7.8 $\pm$ 2.3  &-21.8 $\pm$ 20.7 &210.6 $\pm$ 43 &1.8 $\pm$ 0.3 & & \\
Off2& [3, 69] & CO(2-1) & 4 $\pm$ 1.2  &-29.3 $\pm$ 25 &187.3 $\pm$ 73.9 &0.8 $\pm$ 0.2 & & 1.6 \\
\hline 
Off3& [21, -9] & CO(1-0) & 15.2 $\pm$ 2.9 &-91.2 $\pm$ 8.4 &93 $\pm$ 16.9 &1.5 $\pm$ 0.3 & 3.5 & \\
Off3& [21, -9] & CO(2-1) & 9.4 $\pm$ 3.1 &-88.1 $\pm$ 24.7 &276.6 $\pm$ 48.6 &2.77 $\pm$ 0.5 & & \\
Off3& [21, -9] & CO(2-1) & 8.2 $\pm$ 0.9 &-62.8 $\pm$ 6.5 &153.8 $\pm$ 17.7 &1.34 $\pm$ 0.1 & & 0.5\\
\hline 
Pos11& [-45, 51] & CO(1-0) & 1.5 $\pm$ 0.7 &65 $\pm$ 32.2 &272.6 $\pm$ 119.9 &0.44 $\pm$ 0.13 & 1 & \\
Pos11& [-45, 51] & CO(2-1) & 3 $\pm$ 1.5 &55 $\pm$ 21.5 &96.2 $\pm$ 45.6 &0.31 $\pm$ 0.13 & & 2$^*$ \\
Pos11& [-45, 51] & CO(2-1) & $\le$3$\times$ 1.3 & - & - & - & & \\
\hline 
Pos2& [-45, -51] & CO(1-0) & 1.9 $\pm$ 0.9 &-47.8 $\pm$ 25.7 &116.9 $\pm$ 48.1 &0.2 $\pm$ 0.1 & 0.5 & \\
Pos2& [-45, -51] & CO(2-1) & 3.8 $\pm$ 1.6 &-65.4 $\pm$ 27.9 &205 $\pm$ 53.2 &0.8 $\pm$ 0.2 & & \\
Pos2& [-45, -51] & CO(2-1) & 2.3 $\pm$ 1.0 &-64.4 $\pm$ 23.2 &140.6 $\pm$ 51.1 &0.3 $\pm$ 0.1 & & 1.2\\
\hline 
East& [27, 3] & CO(1-0) & 23.4 $\pm$ 4.5 &-109.1 $\pm$ 9.1 &106.2 $\pm$ 20.1 &2.6 $\pm$ 0.5 & 6.2 & \\
East& [27, 3] & CO(2-1) & 36.2 $\pm$ 4.5 &-84.3 $\pm$ 6.4 &124.3 $\pm$ 15.9 &4.7 $\pm$ 0.5 & & \\
East& [27, 3] & CO(2-1) & 13.4 $\pm$ 1.0 &-71.4 $\pm$ 3.7 &127.5 $\pm$ 9.9 &1.8 $\pm$ 0.1 & & 0.6\\
\hline 
Test& [45, 33] & CO10 &  $\le$3$\times$1.6  & -& -& - &  - \\
Test& [45, 33] & CO21 &  $\le$3$\times$5.9  & -& -& - &  \\
\hline 
\multicolumn{9}{l}{$^*$Estimated without the HERA data. Note that the CO(1--0) could be half the value in the 
table here, see spectra on Fig \ref{spectra-filaments}.}\\
\end{tabular} 
\end{center}
\label{table-filaments}
\end{table*}
%
We detected CO in all the regions observed inside the H$\alpha$
filaments. Offsets are relative to the 3C84 position : RA 03:19:48.15,
Dec 41:30:42.1 (J2000). In the central region (indicated as centre 1
to 9 in Fig \ref{pointings} and Table \ref{table2-center}), the 3mm
data suffered from baselines instabilities due to the strong continuum
of 3C84. The 3mm spectra shown on the left hand side of Fig
\ref{spectra-center2} are affected by those baseline ripples.  The
1.3mm data, not affected by baseline ripples give a more reliable idea
of the line shape. Nevertheless {\itshape CO lines are detected in all
regions except the region outside the H$\alpha$ filaments}. This is
the region named test, which has no H$\alpha$ emitting counterpart. Our earlier
HERA map (S06) showed a hint of a detection at this place, but this was
not confirmed by the new observations.

In order to compute the molecular gas content from the integrated CO intensity
I$_{\rm CO}$ (K.km/s), we used 
\begin{equation}
{\rm I}_{\rm CO}= \int T_{\rm mb}{\rm (CO)}\, dV \,,
\end{equation}  
where $T_{\rm mb}{\rm (CO)}$ is the main beam antenna brightness
temperature, obtained with the CO emission line and $dV$ is the line
width.  We then adopted the standard Milky Way conversion factor commonly
used for N(H$_2$) estimates:
\begin{equation}
{\rm N(H_2)} = 2.3 \, 10^{20} \, {\rm I}_{\rm CO}\, \,
{\rm(molecules/cm^2)} \,.
\end{equation}
From this equation the mass of molecular hydrogen, contained in
one beam is:
\begin{equation}
\rm M_{gas}\, (M_{\odot}) \, = 1.36\times 2.95 \, 10^{-19} \, {\rm I}_{\rm CO}\,
\theta^2 \, D^2\, \frac{N(H_{2})}{I_{CO}}\, ,
\end{equation}
I$_{\rm CO}$ is the integrated intensity in K\,km/s, $\theta$ is the
beamsize of the telescope in arcsec, and D is the distance of the
galaxy, taken to be 72.6 Mpc. We also included a factor 1.36 to take into
account the He contribution to the gas mass.
%
\begin{figure*}[h]
\centering 
\vspace{-1cm}
\begin{tabular}{ccc} 
\vspace{-1cm}
\includegraphics[width=4cm,angle=-90]{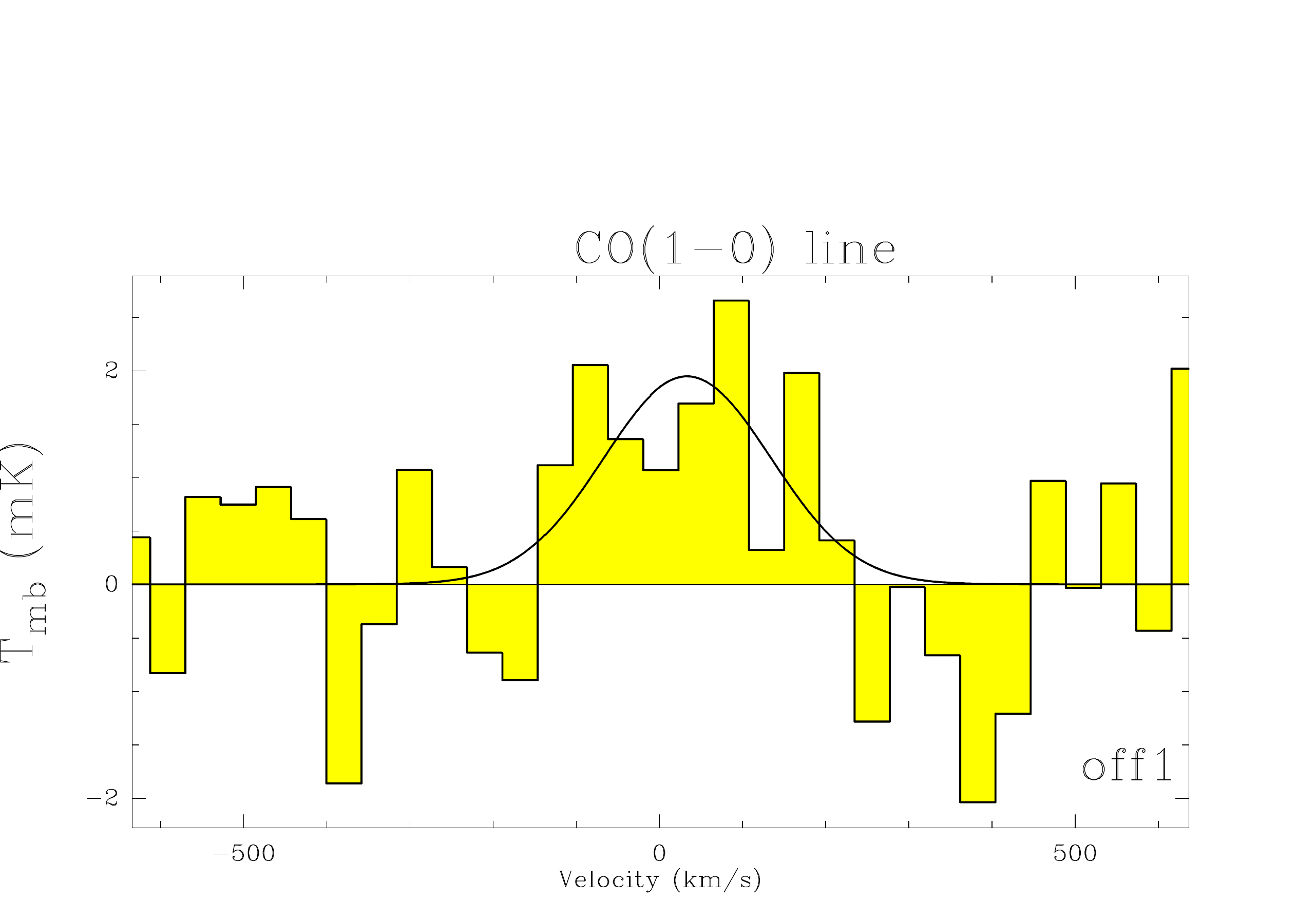} & 
\includegraphics[width=4cm,angle=-90]{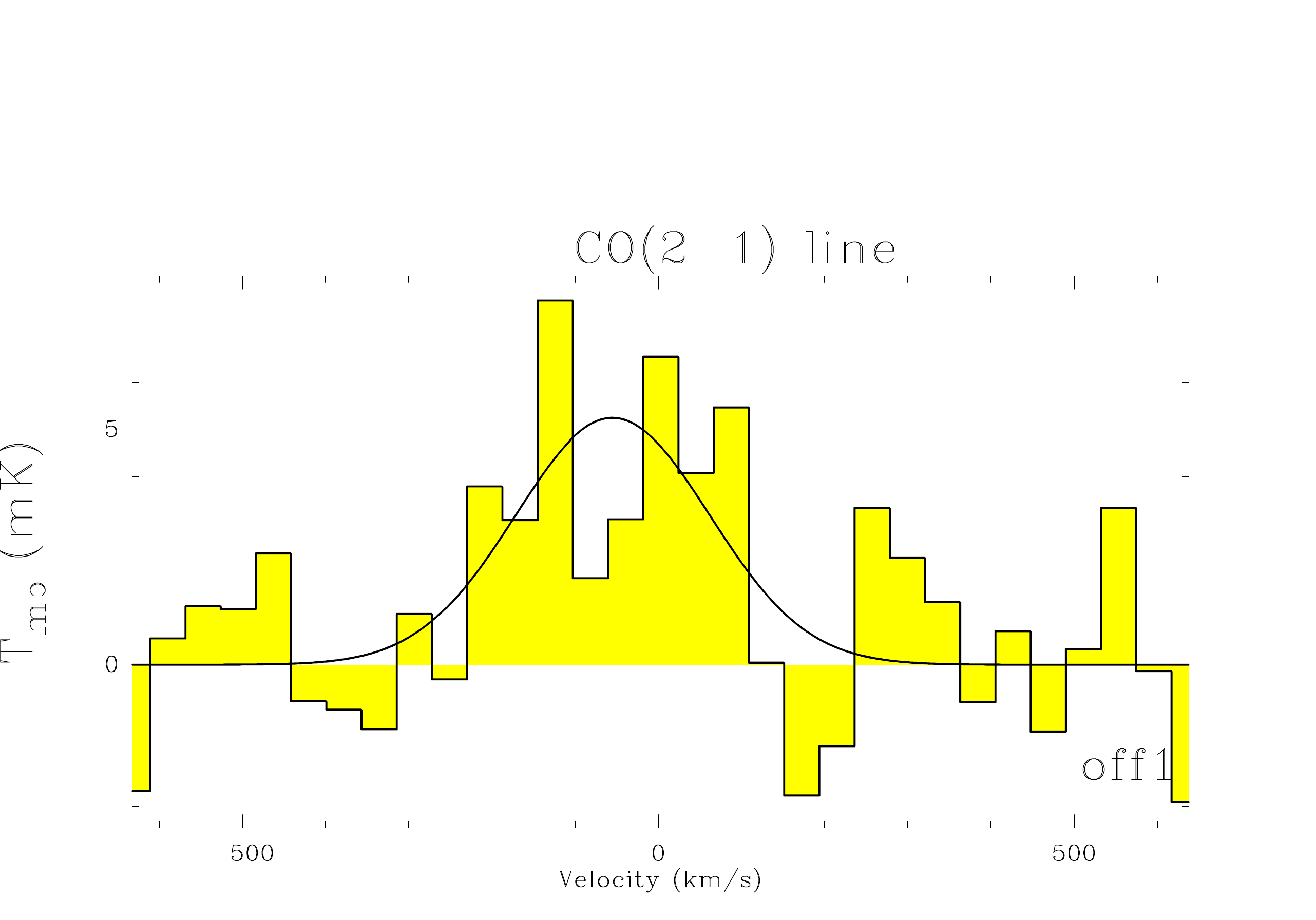} & 
\includegraphics[width=4cm,angle=-90]{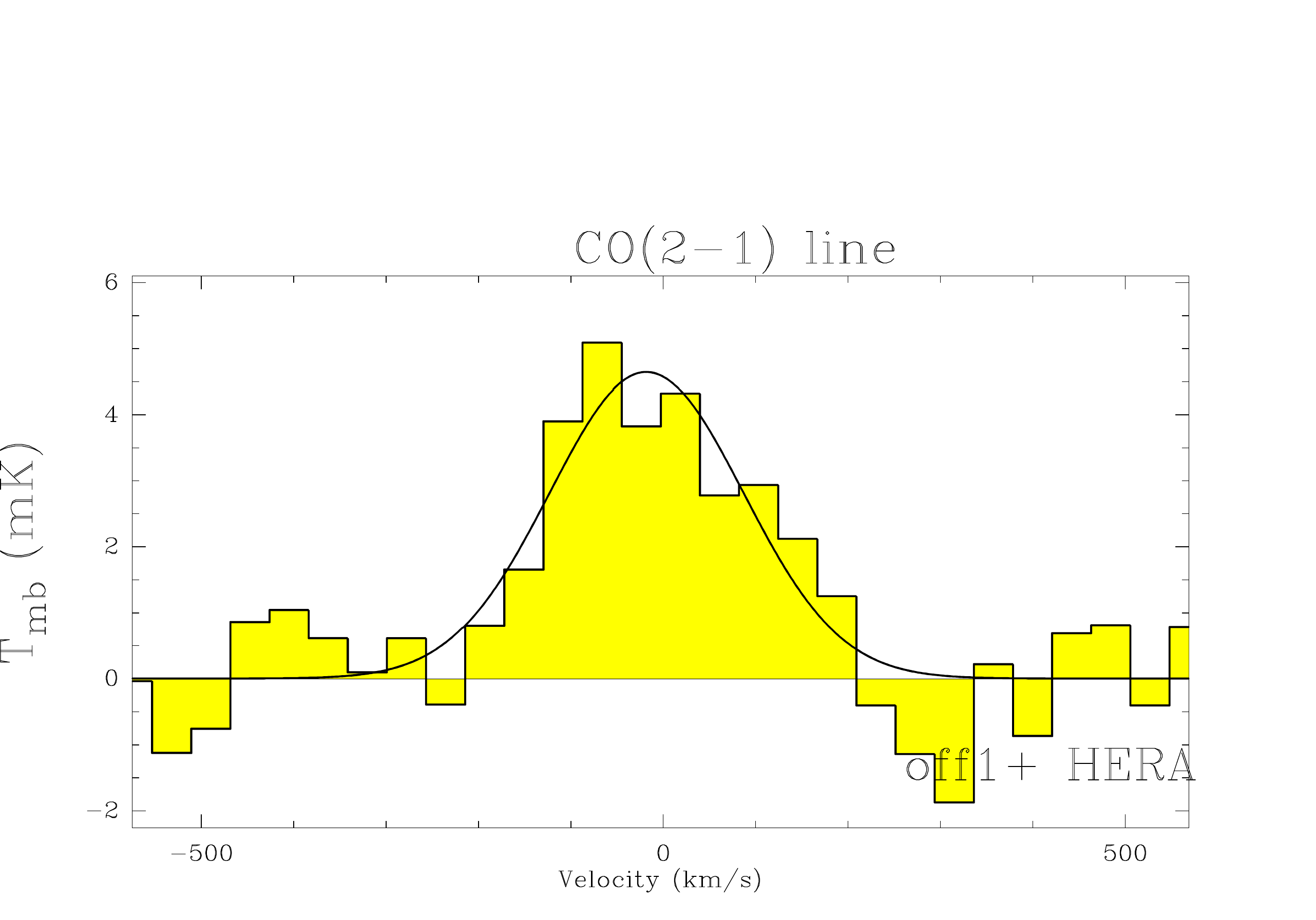} \\ 
\vspace{-1cm}
\includegraphics[width=4cm,angle=-90]{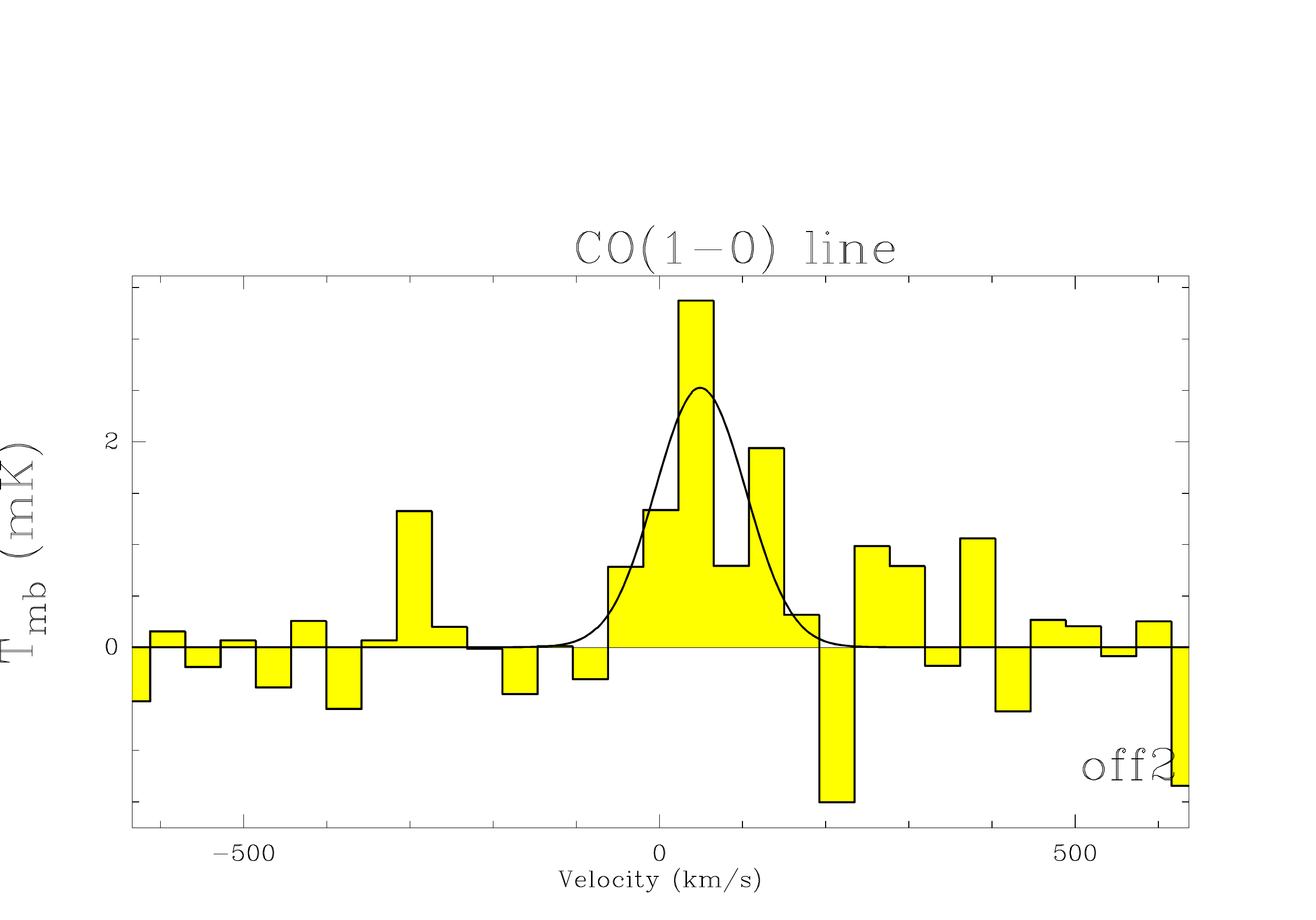} &  
\includegraphics[width=4cm,angle=-90]{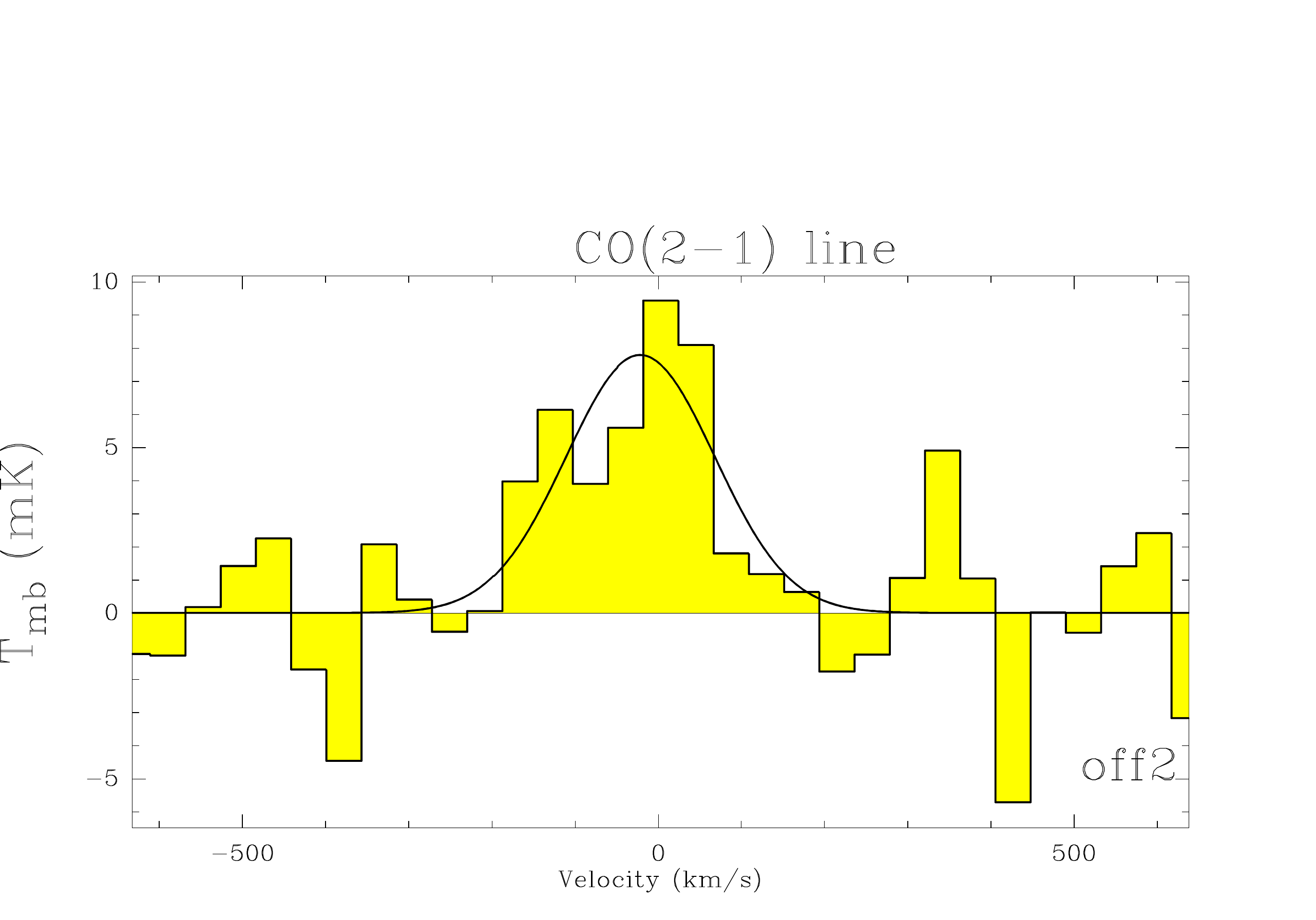} & 
\includegraphics[width=4cm,angle=-90]{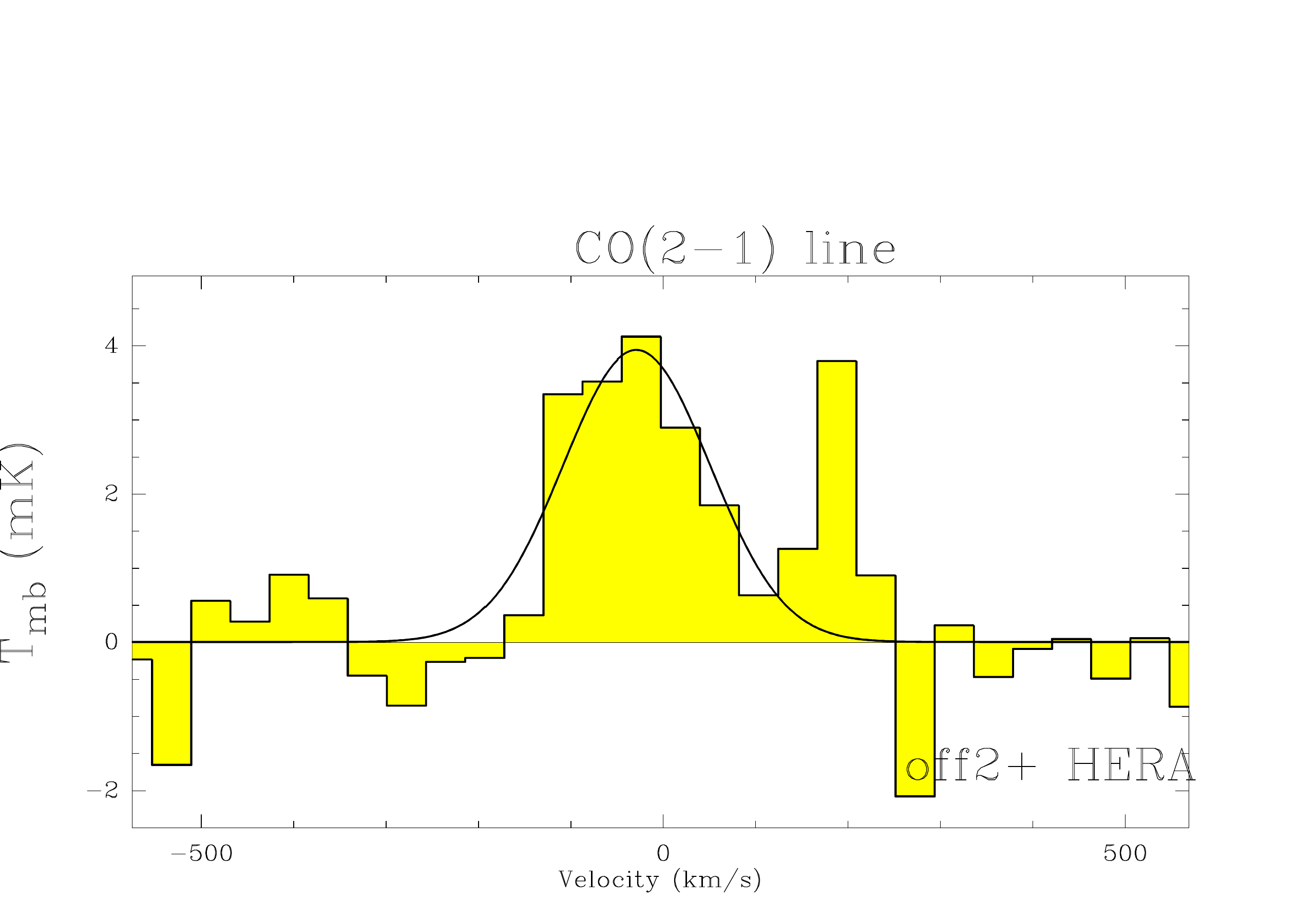} \\ 
\vspace{-1cm}
\includegraphics[width=4cm,angle=-90]{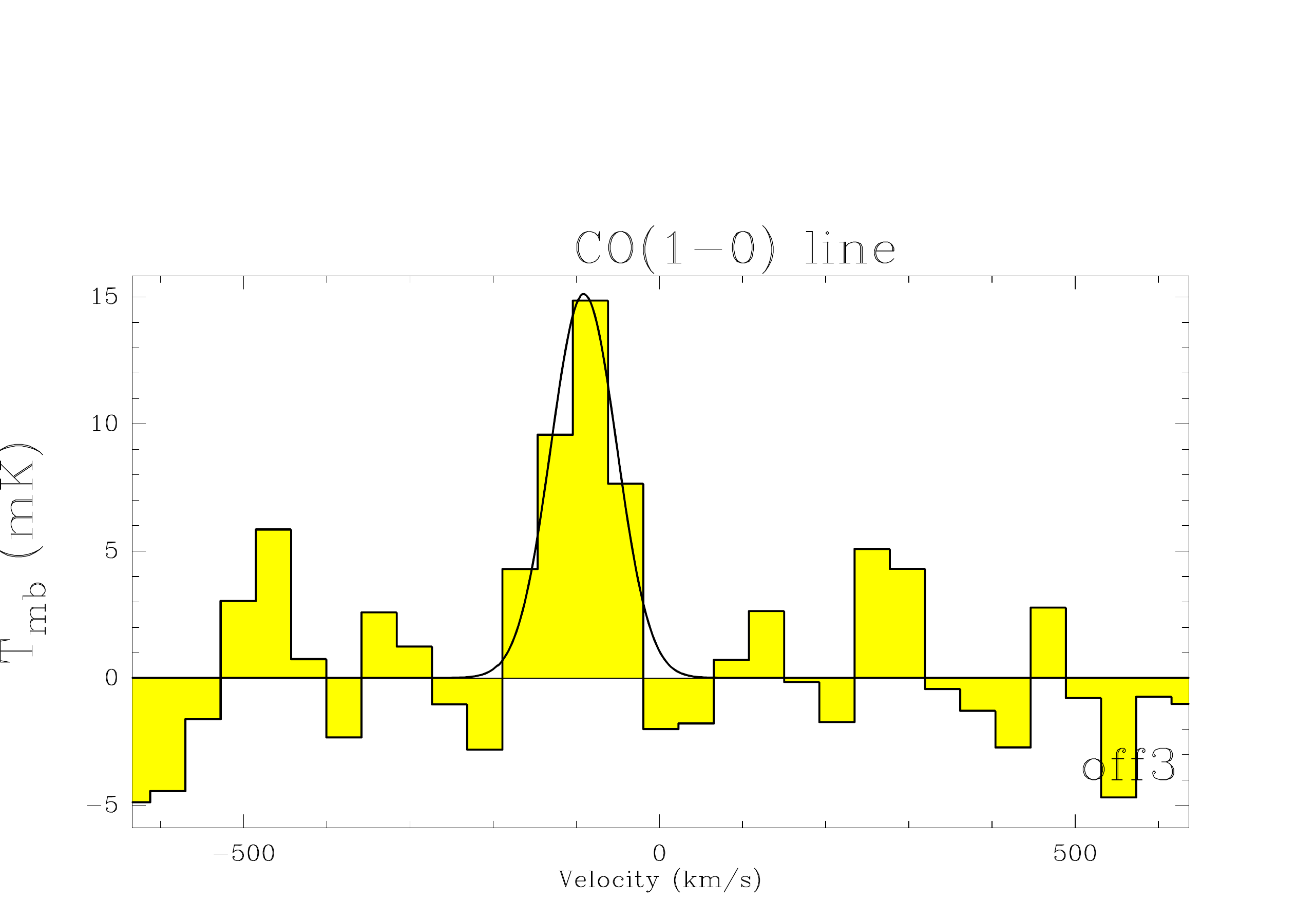} &  
\includegraphics[width=4cm,angle=-90]{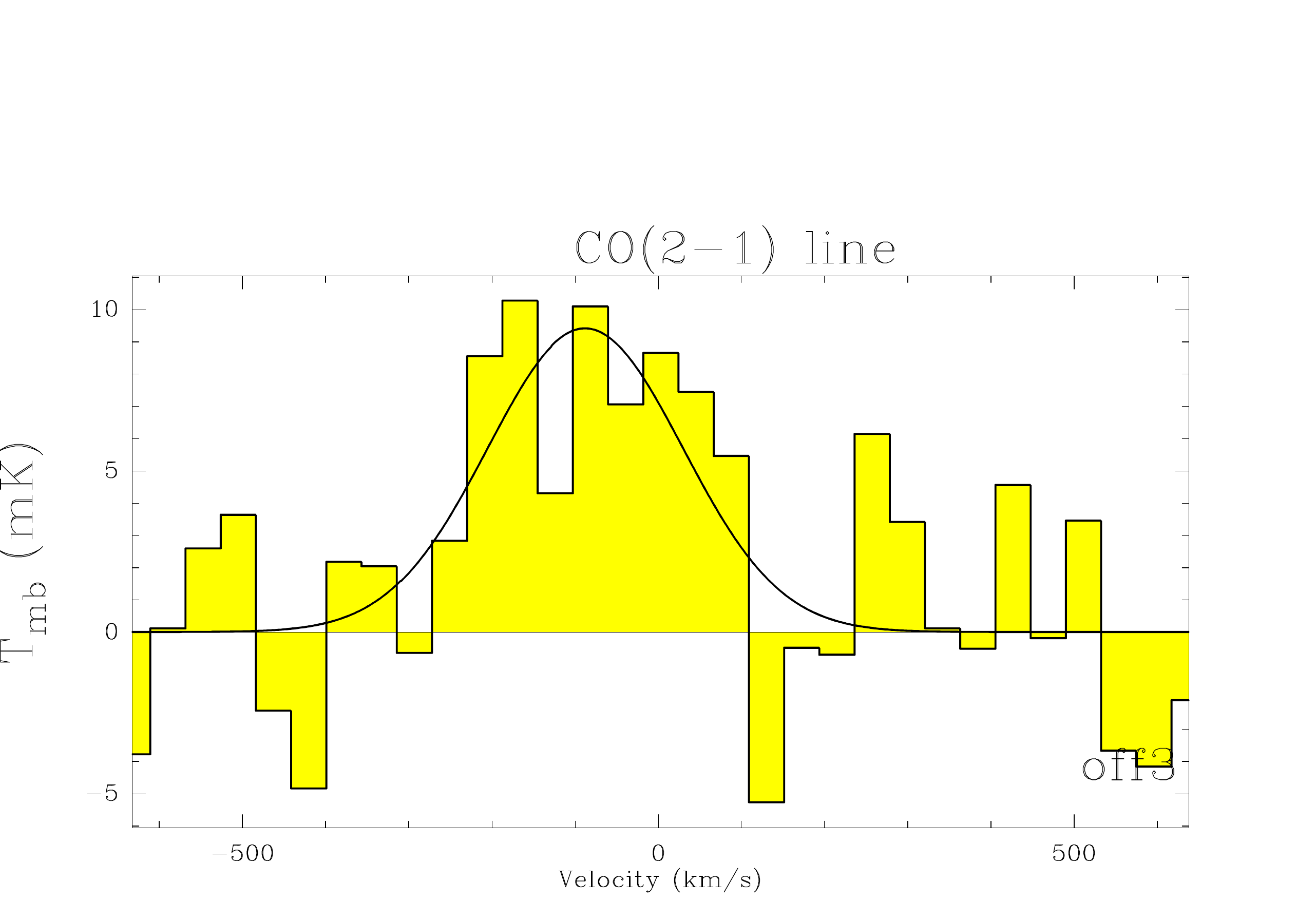} & 
\includegraphics[width=4cm,angle=-90]{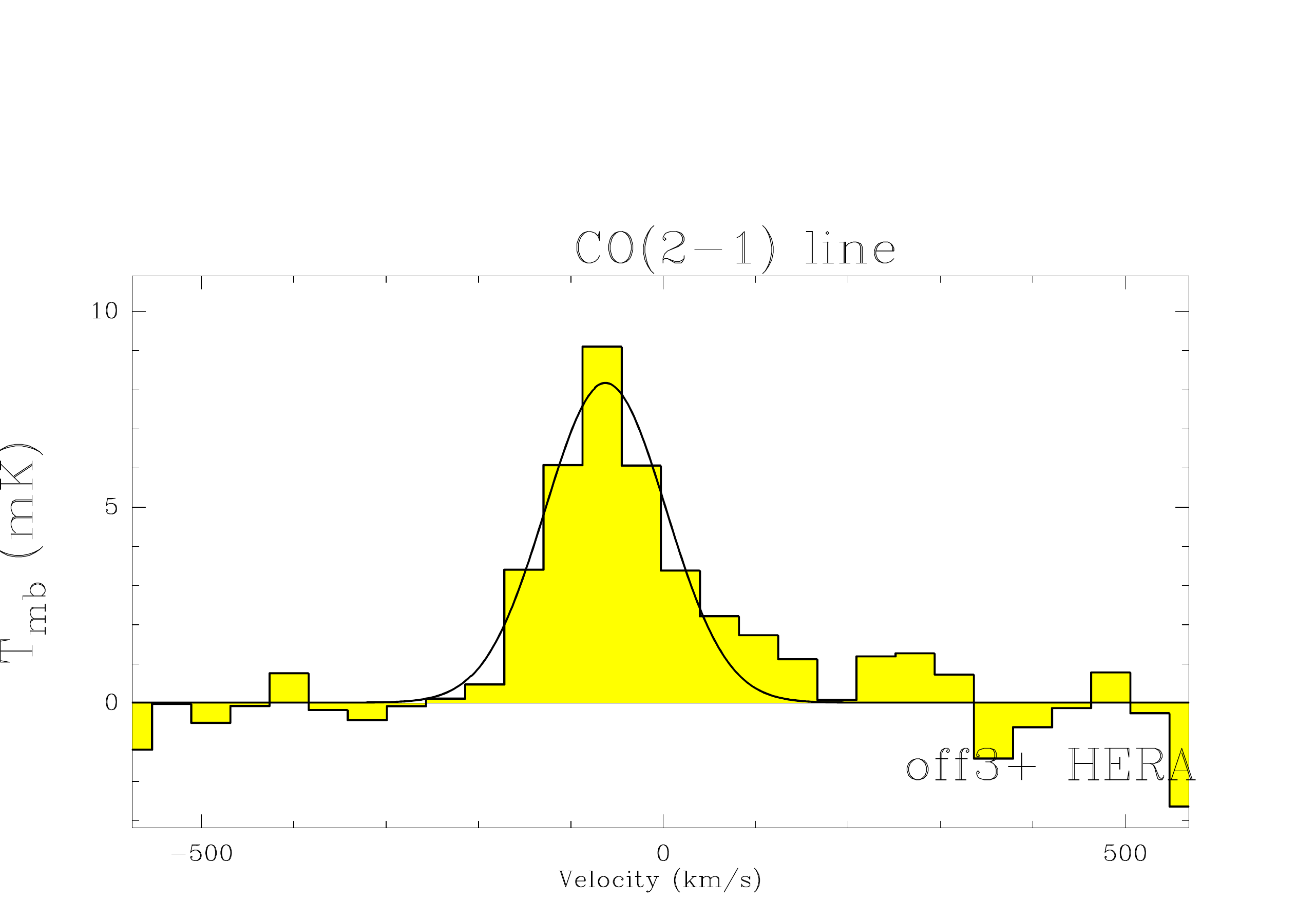} \\ 
\vspace{-1cm}
\includegraphics[width=4cm,angle=-90]{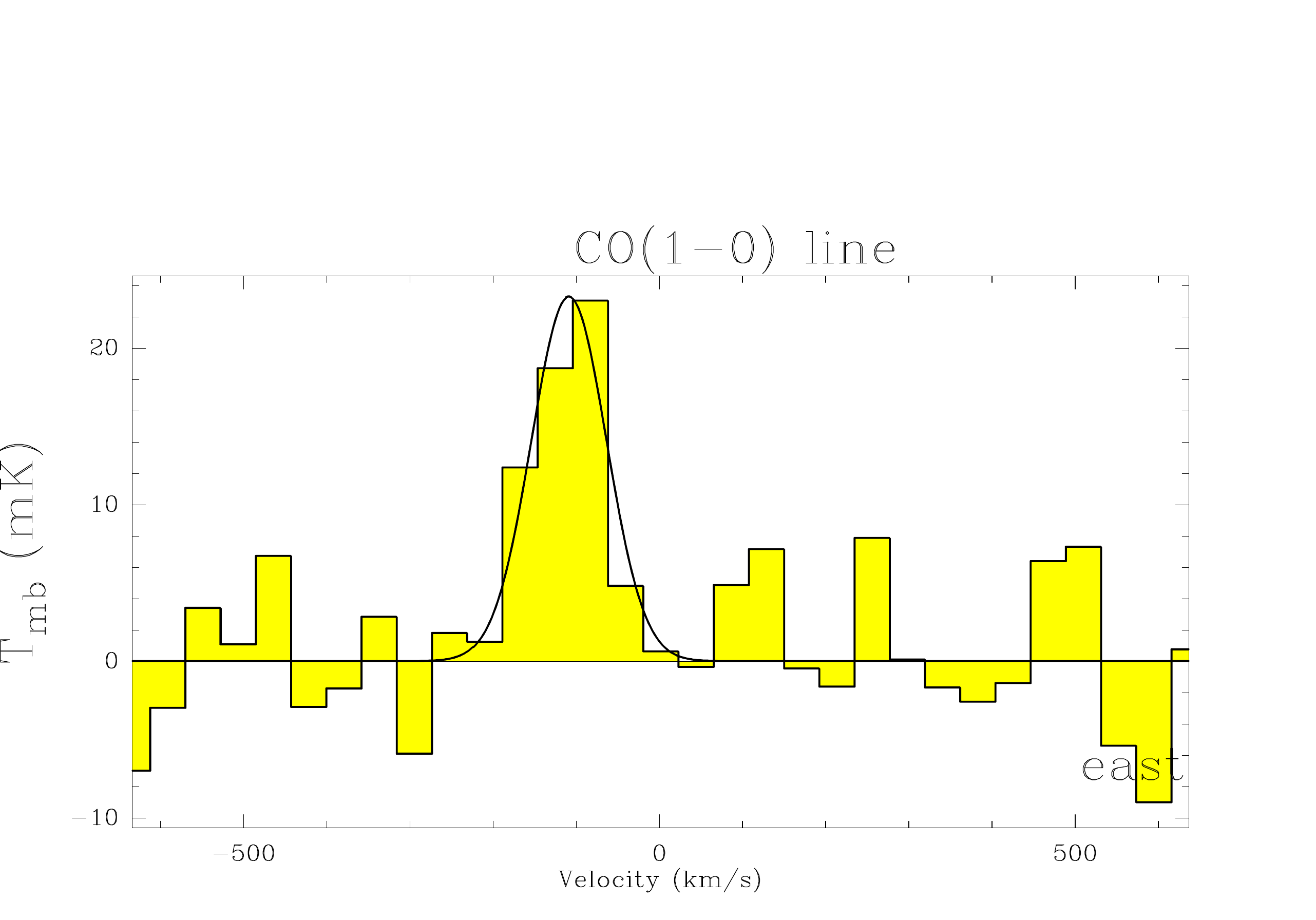} &  
\includegraphics[width=4cm,angle=-90]{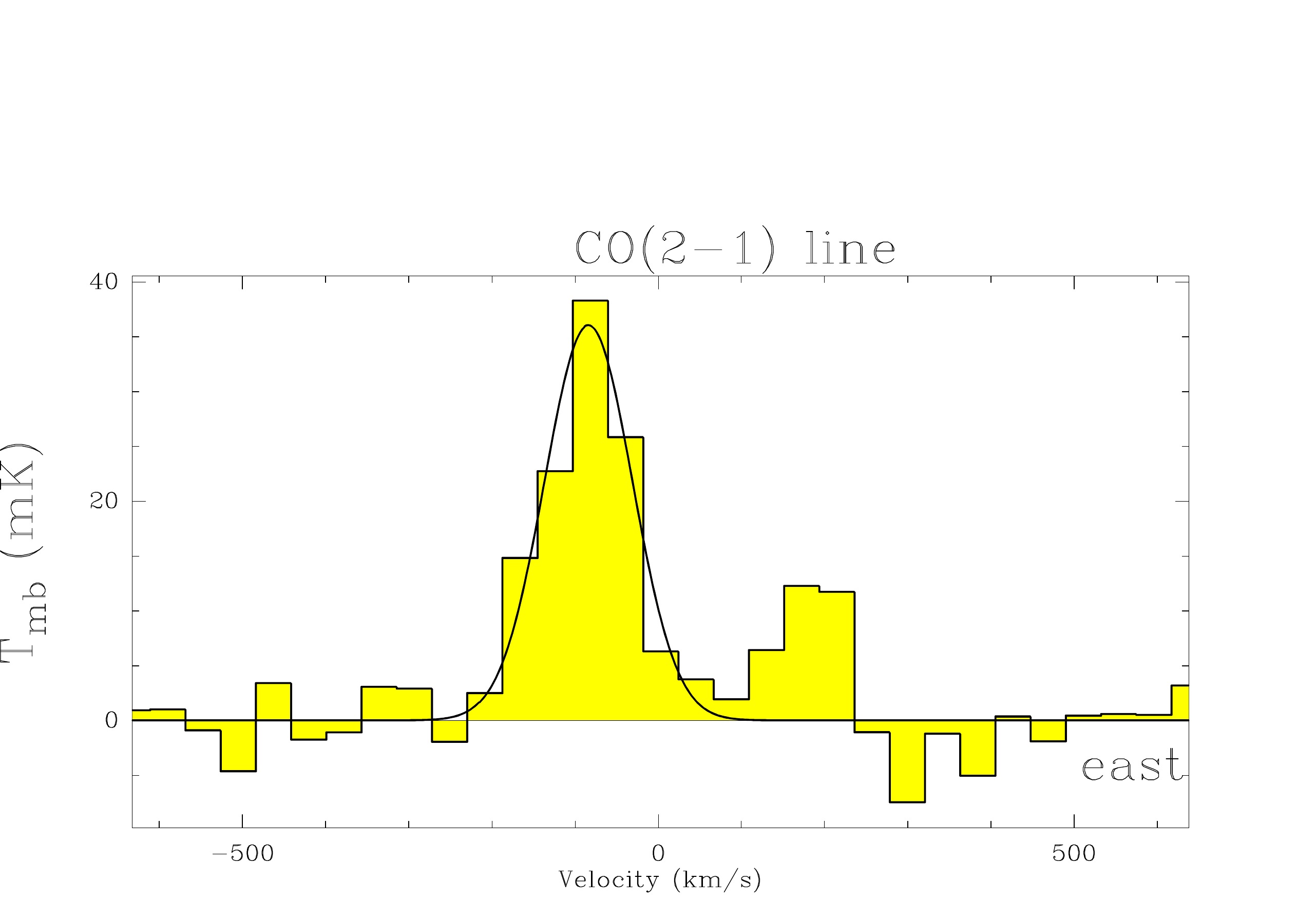} & 
\includegraphics[width=4cm,angle=-90]{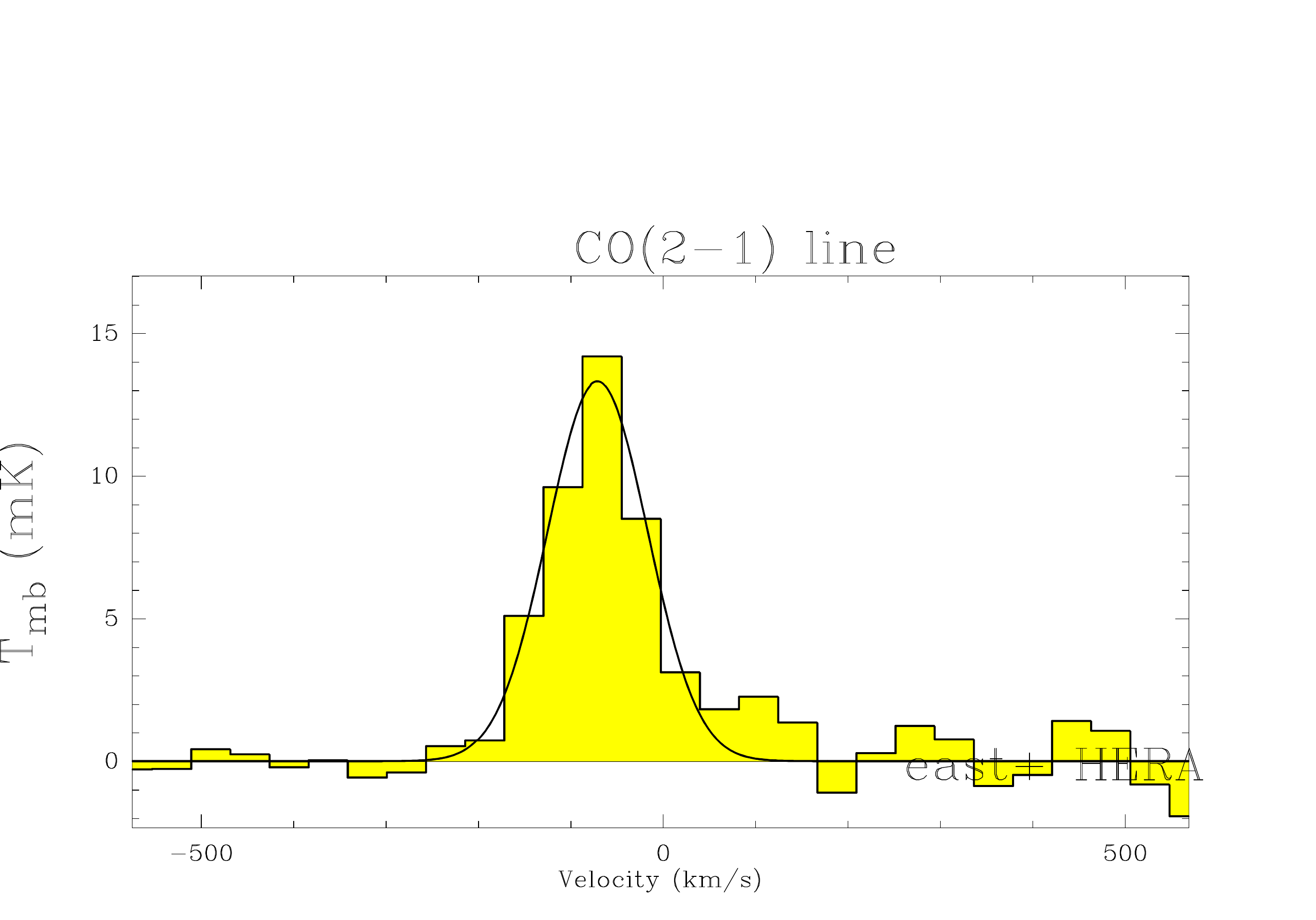} \\ 
\vspace{-1cm}
\includegraphics[width=4cm,angle=-90]{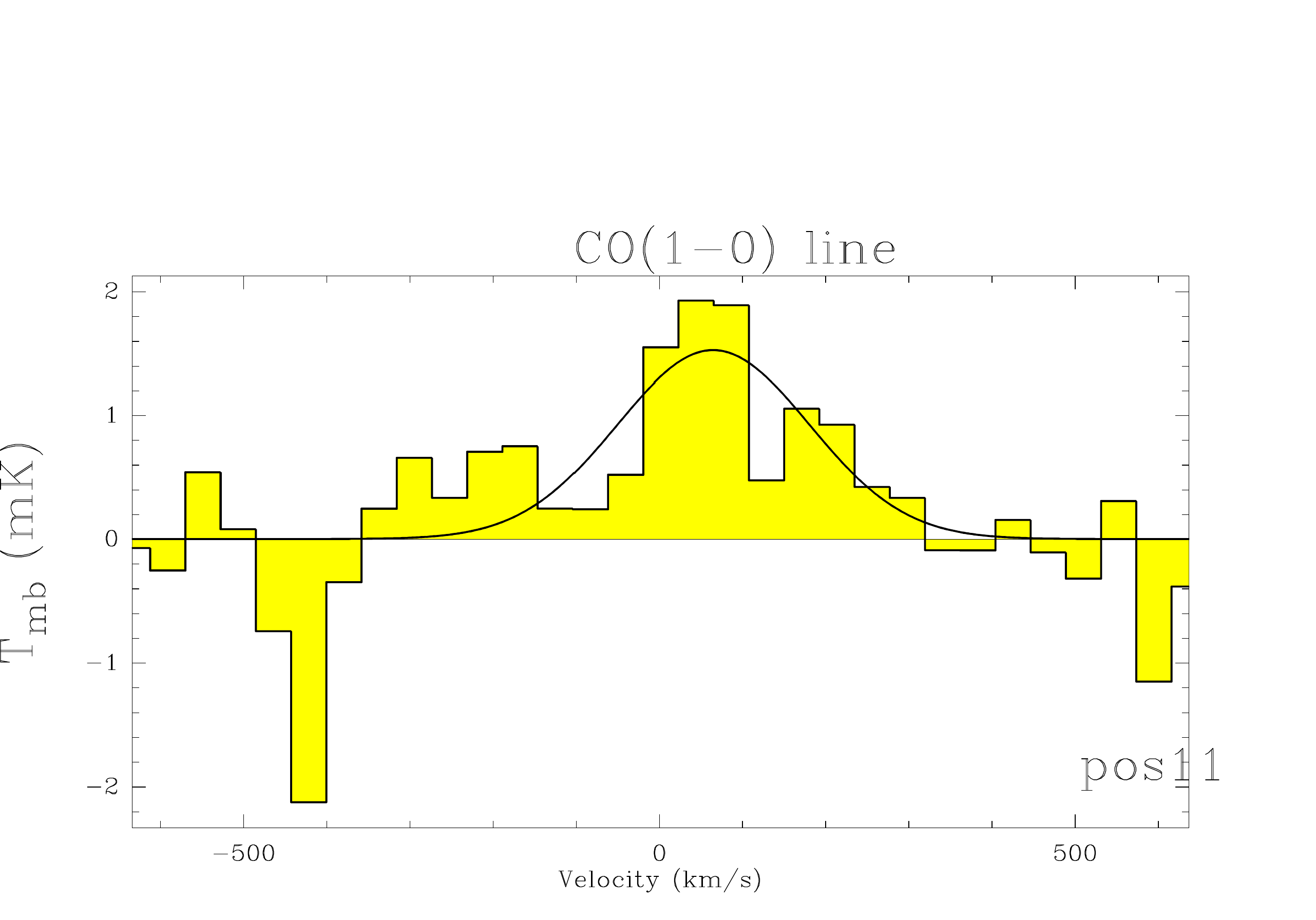} &  
\includegraphics[width=4cm,angle=-90]{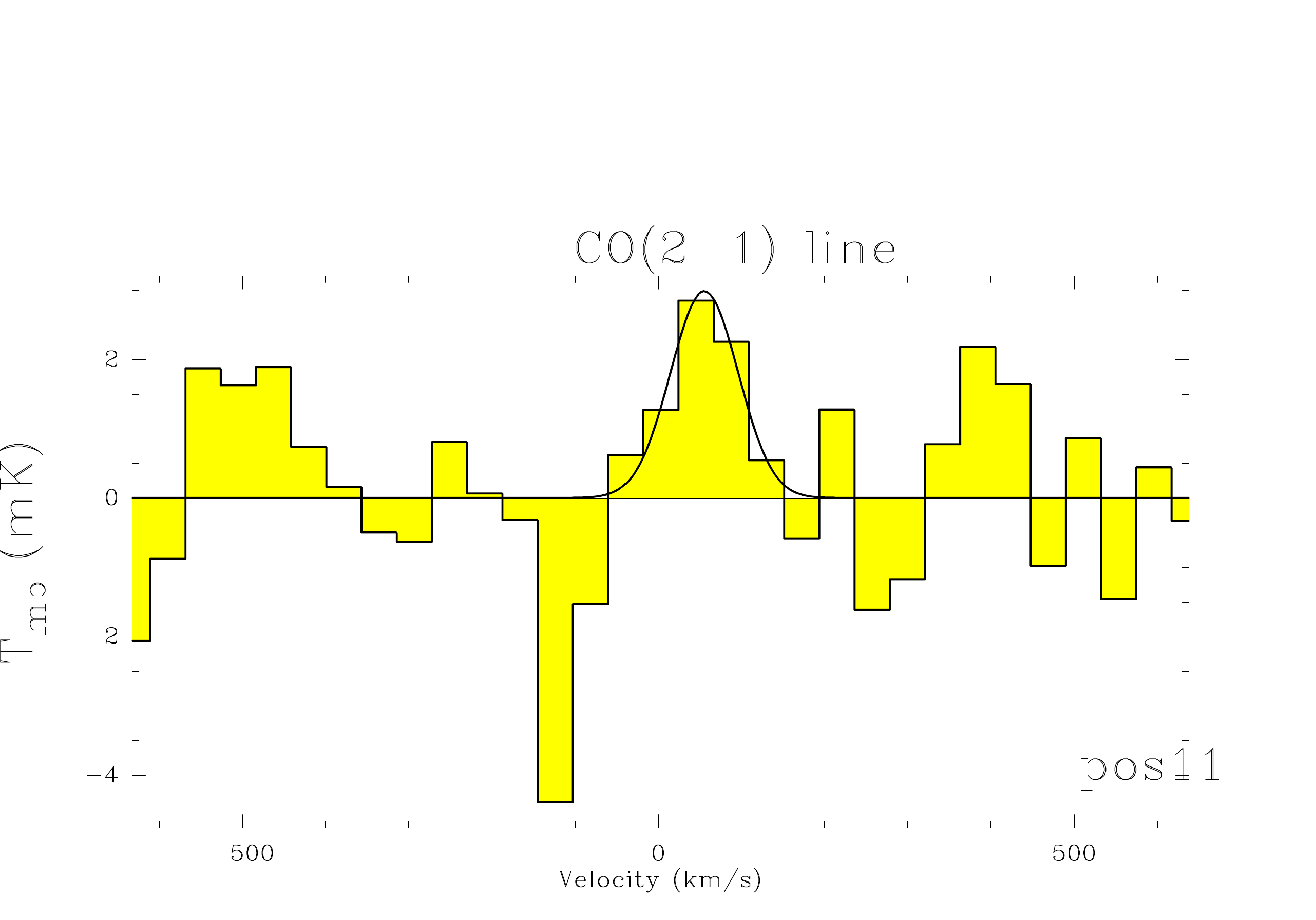} & 
\includegraphics[width=4cm,angle=-90]{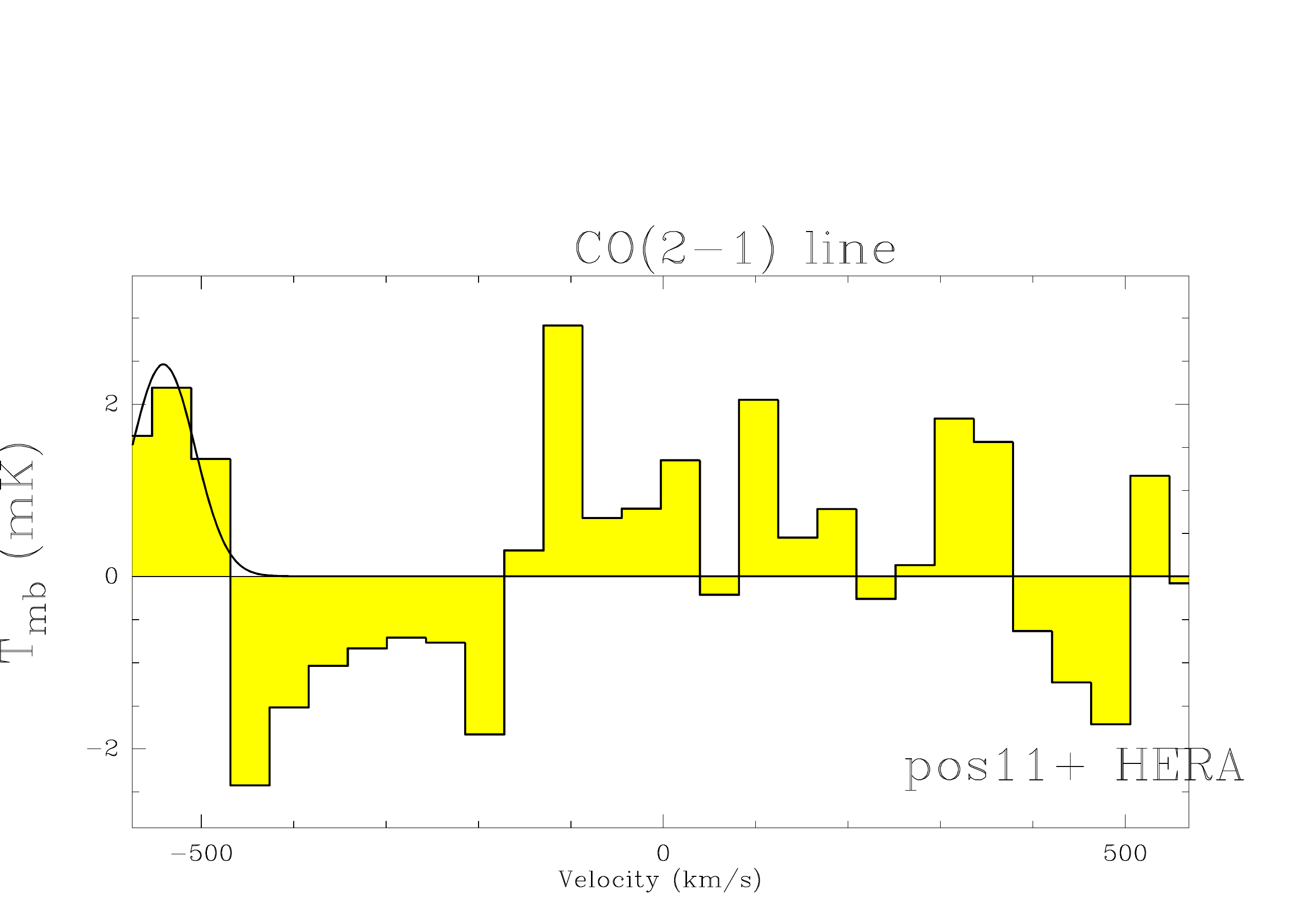} \\ 
\vspace{-1cm}
\includegraphics[width=4cm,angle=-90]{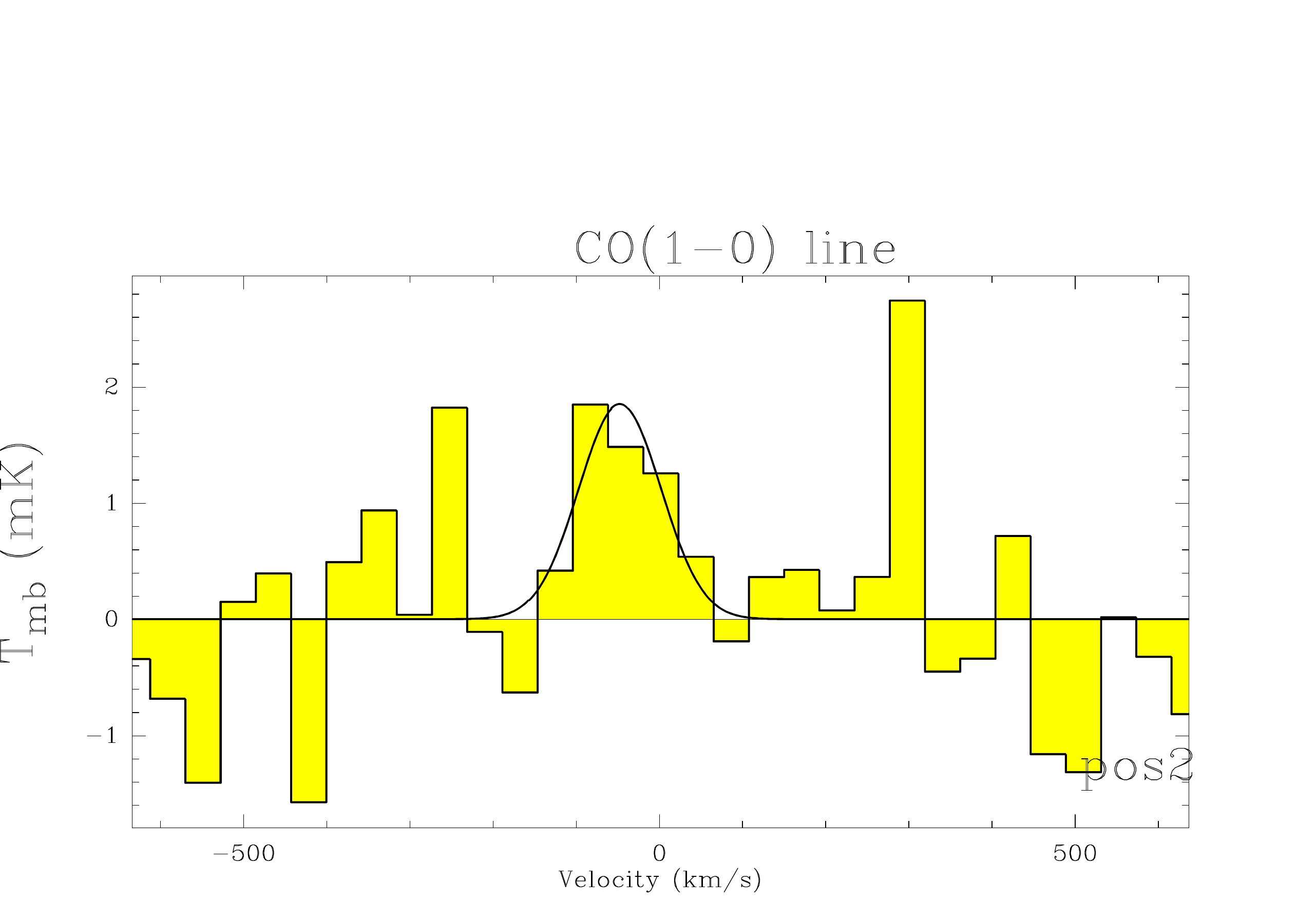} &  
\includegraphics[width=4cm,angle=-90]{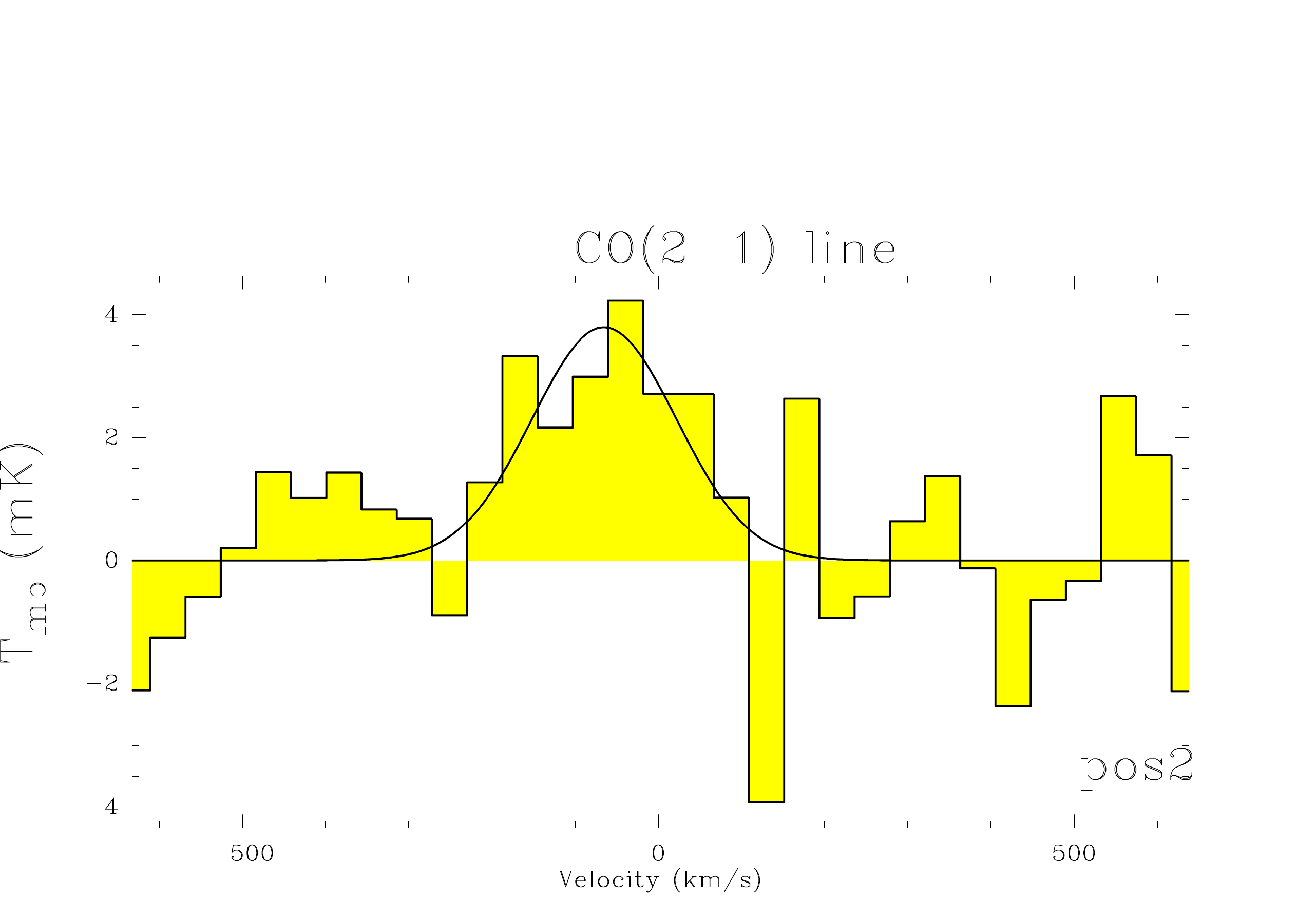} & 
\includegraphics[width=4cm,angle=-90]{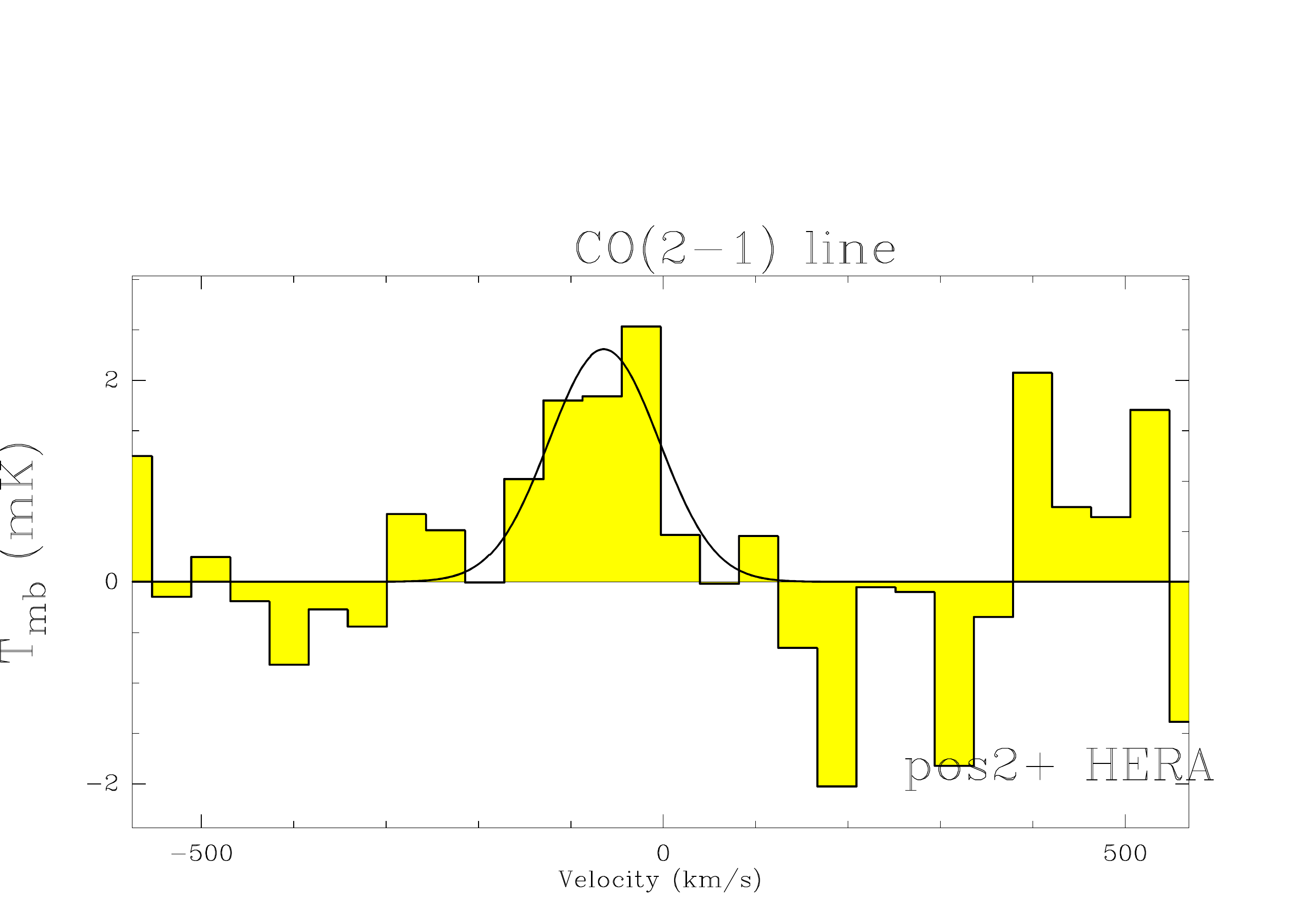} \\ 
\vspace{1.5cm} 
\end{tabular} 
\caption{CO(1--0) and CO(2--1) spectra obtained at all the positions
observed as indicated at lower right in each diagram.  The channel
width is 42 km/s. On the left hand side are the CO(1--0) lines
detected with the a100 and b100 receivers. In the middle are the
results obtained for the CO(2--1) line with the A230 and B230
receivers. On the right hand side are the CO(2--1) lines computed with
both A230 and B230 merged with previous HERA data and smoothed to the
3mm beam size.}
\label{spectra-filaments}  
\end{figure*}  
%
%
\begin{figure*} 
\centering 
\begin{tabular}{ccc} 
\includegraphics[width=4cm,angle=-90]{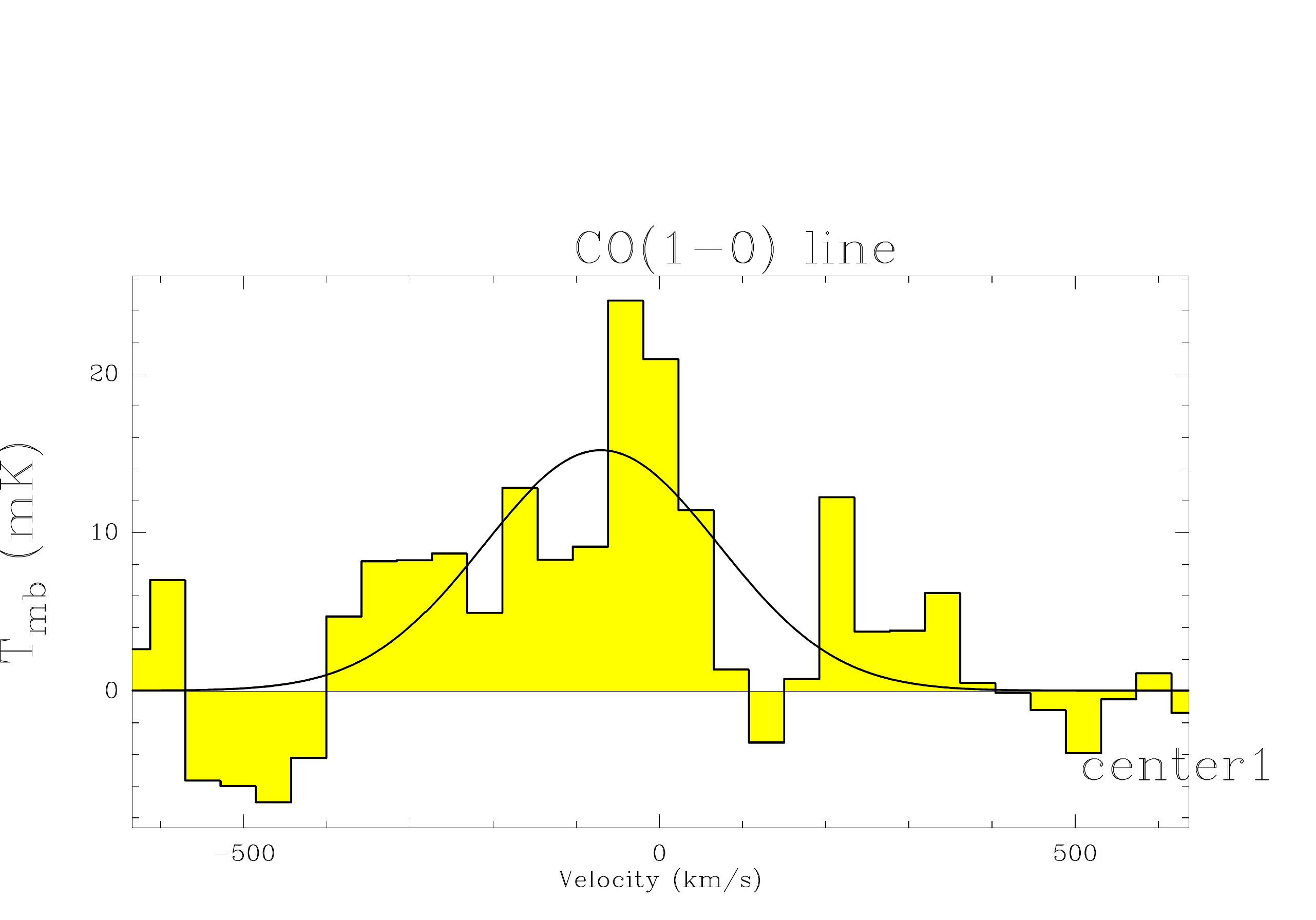} & 
\includegraphics[width=4cm,angle=-90]{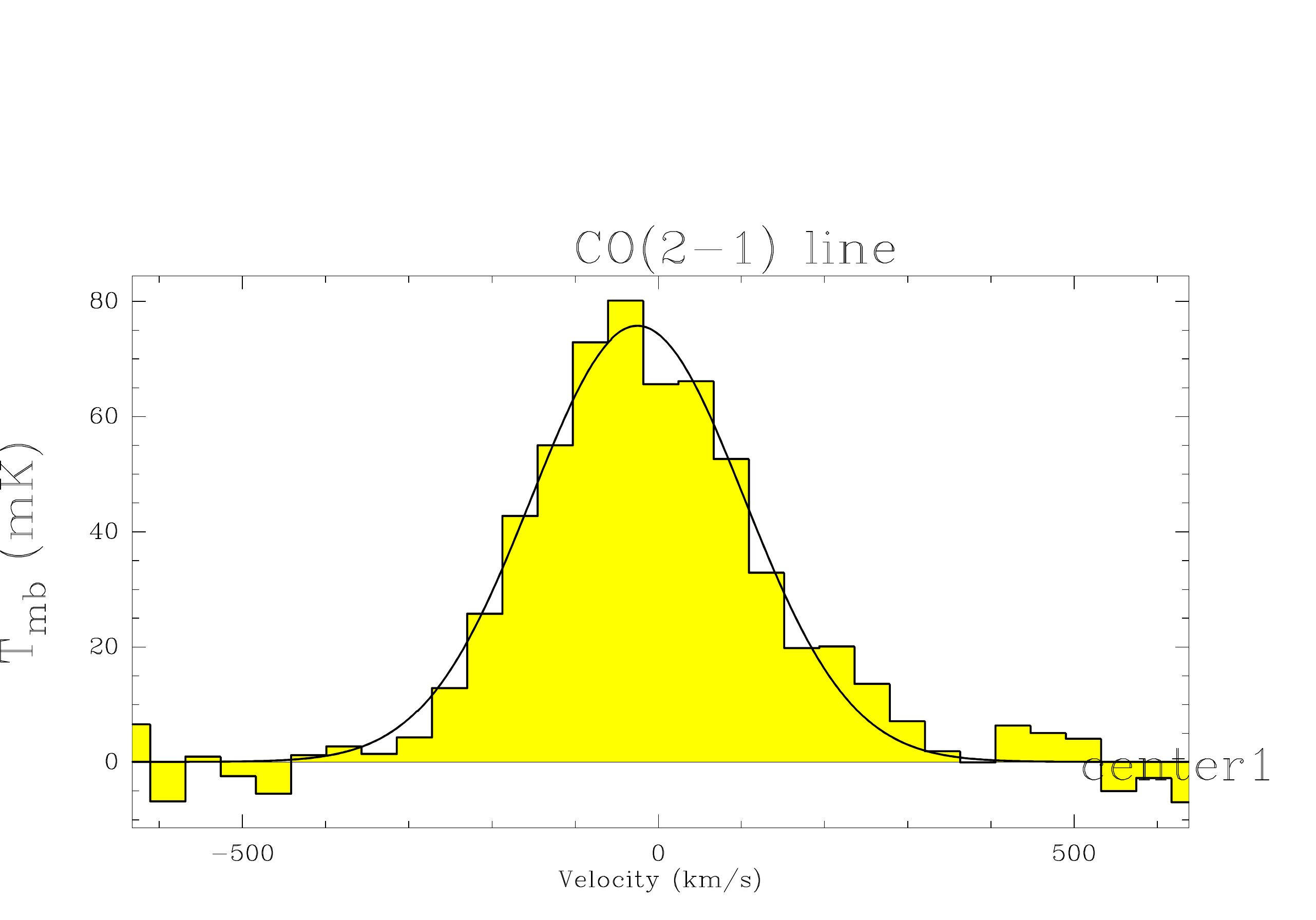} &
\includegraphics[width=4cm,angle=-90]{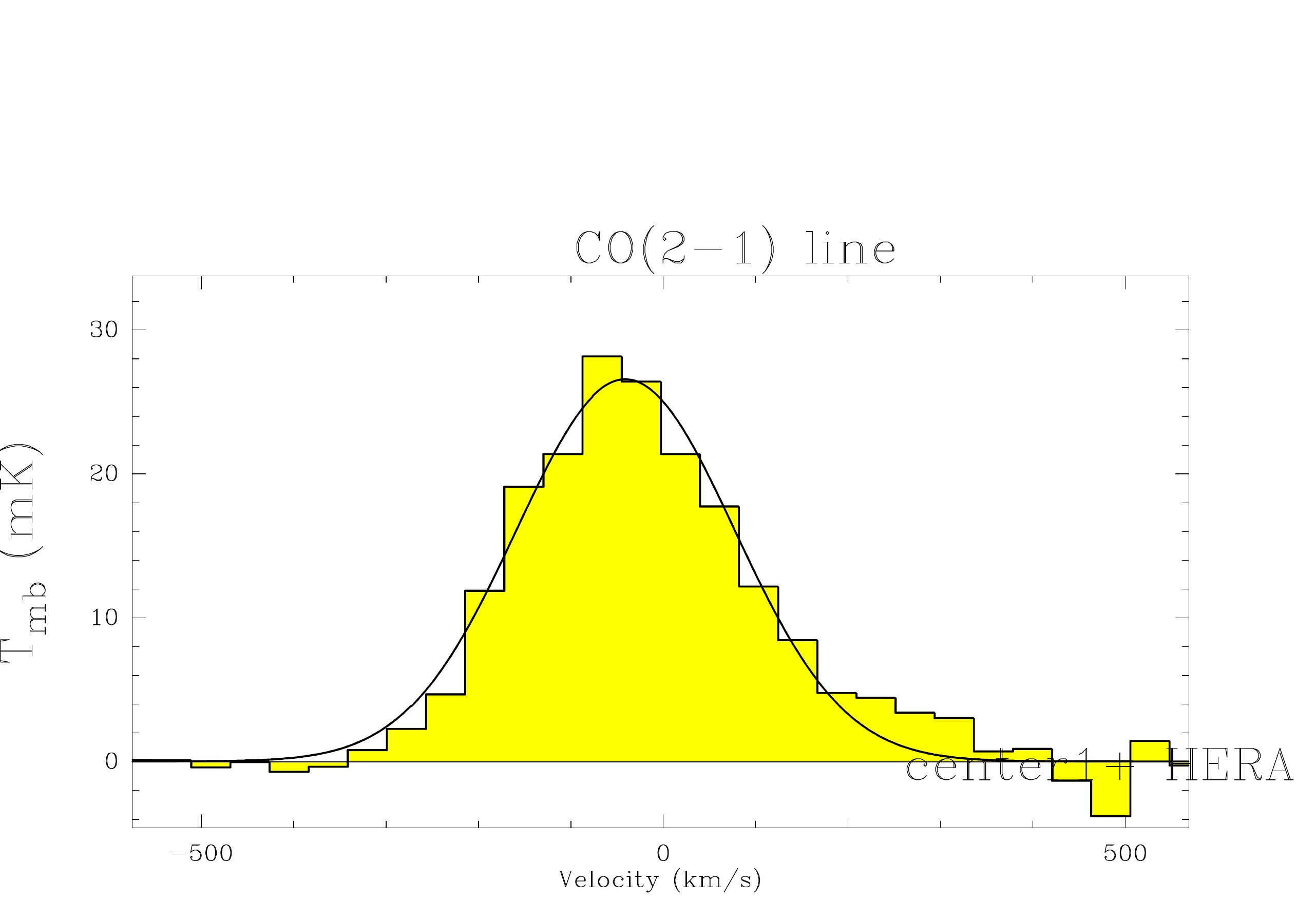} \\ 
\includegraphics[width=4cm,angle=-90]{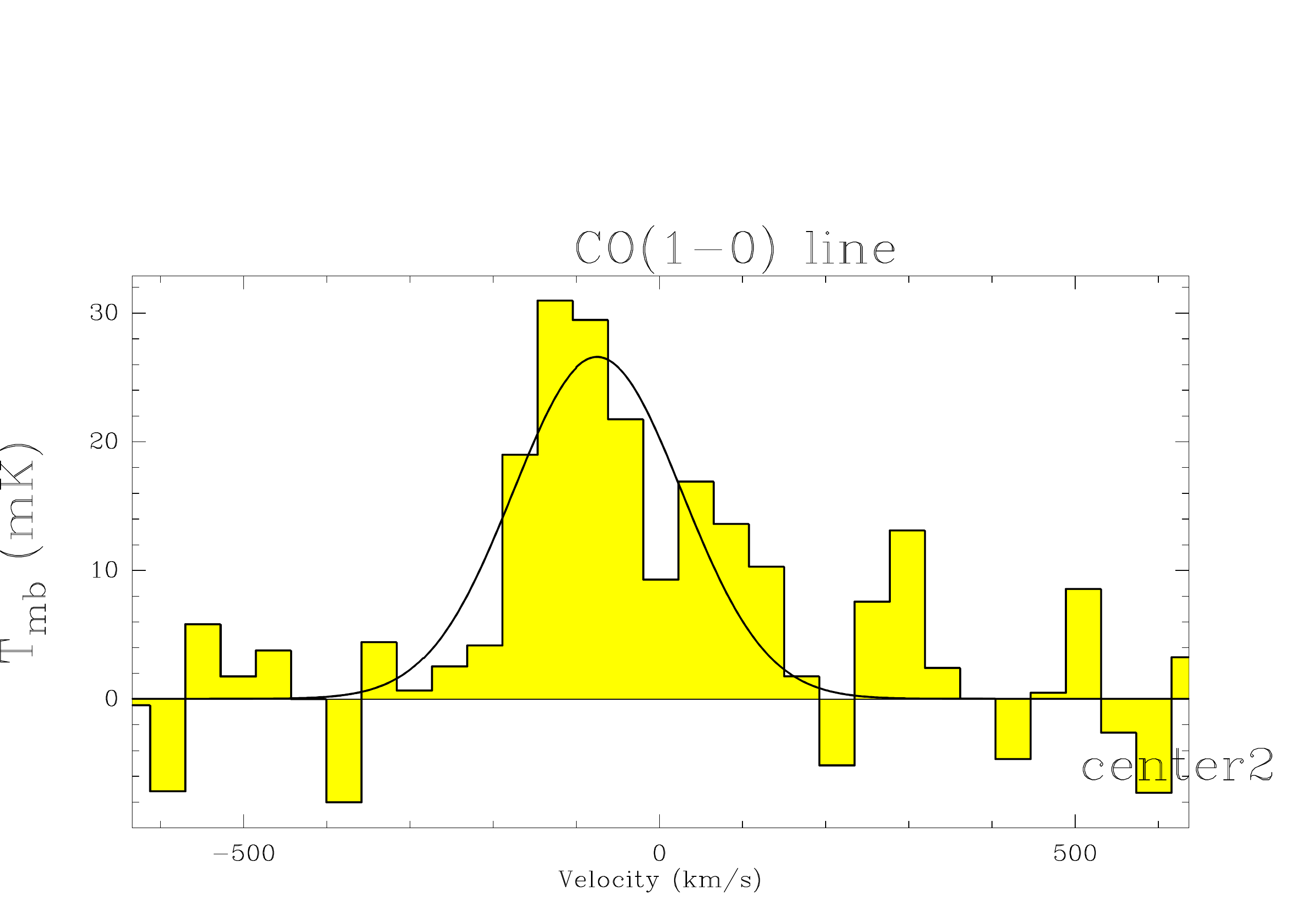} &
\includegraphics[width=4cm,angle=-90]{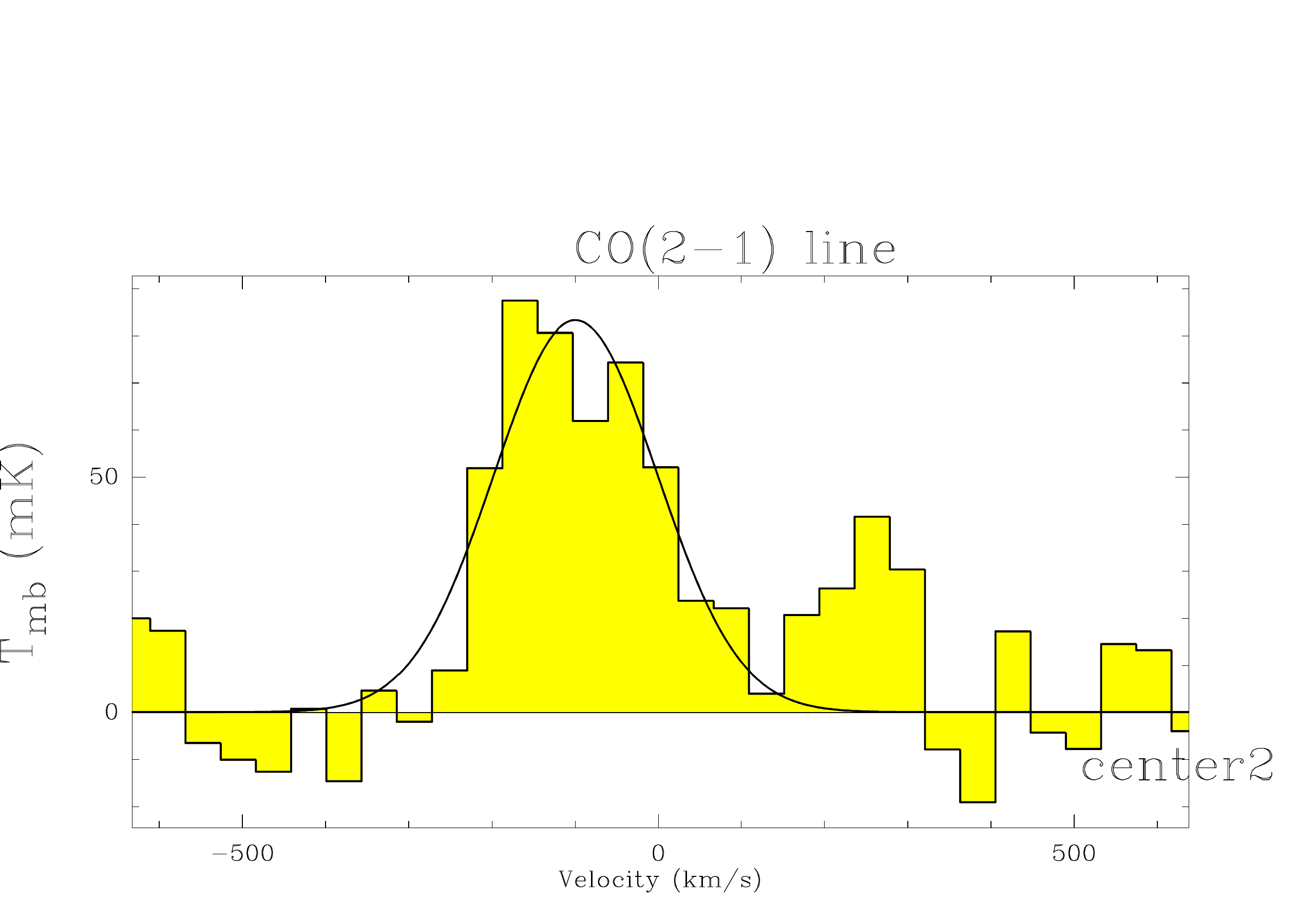} &
\includegraphics[width=4cm,angle=-90]{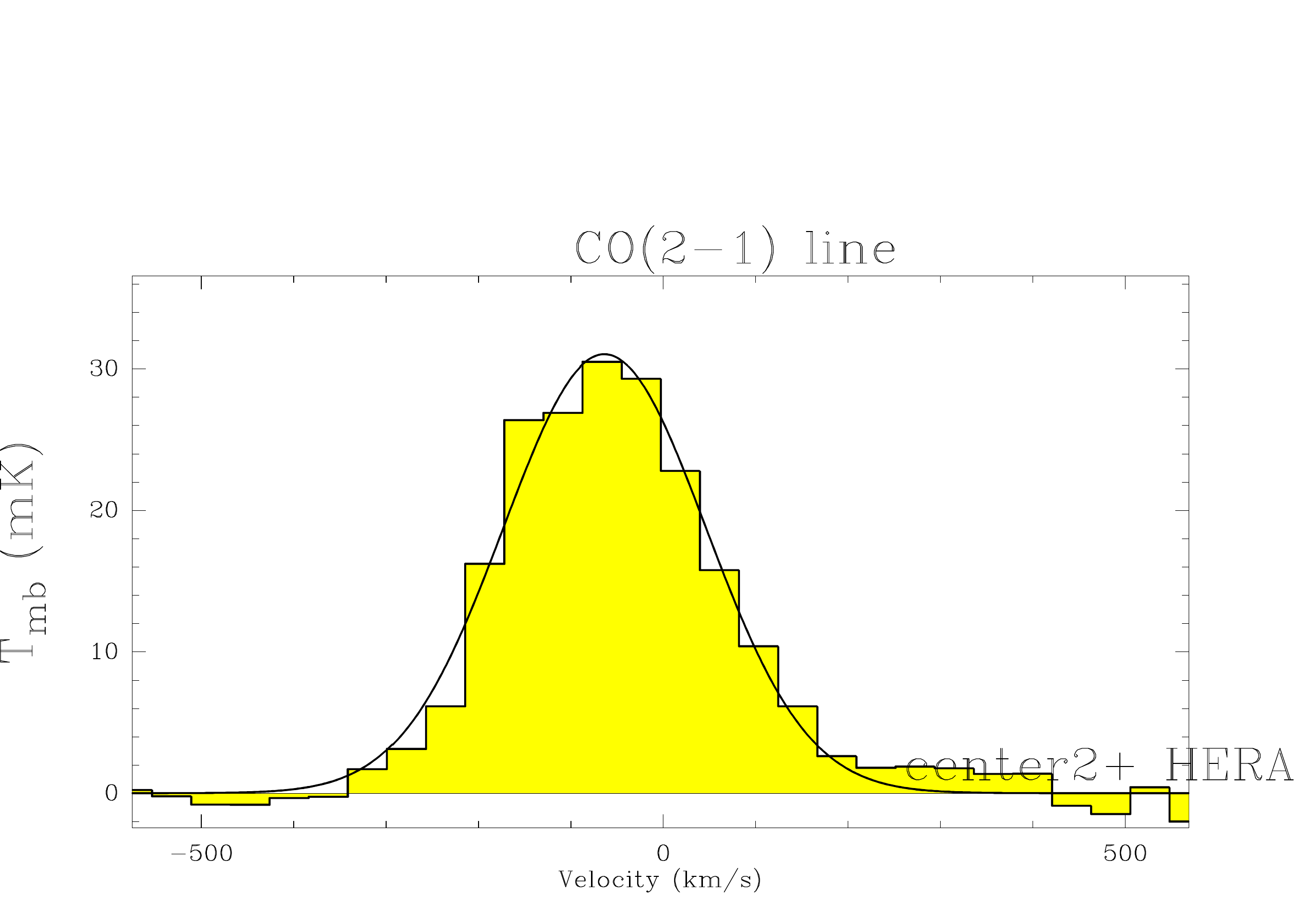} \\
\includegraphics[width=4cm,angle=-90]{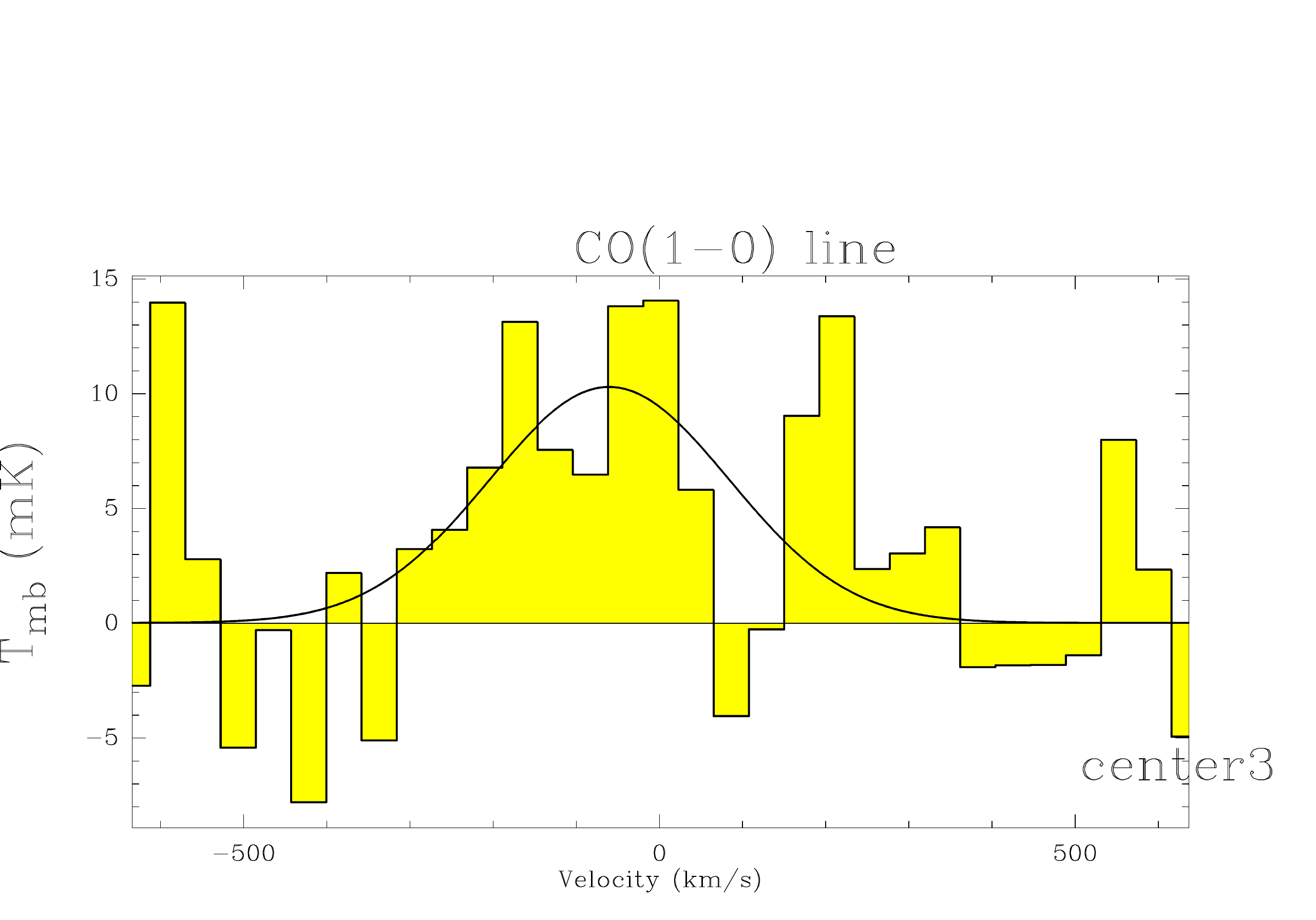} &
\includegraphics[width=4cm,angle=-90]{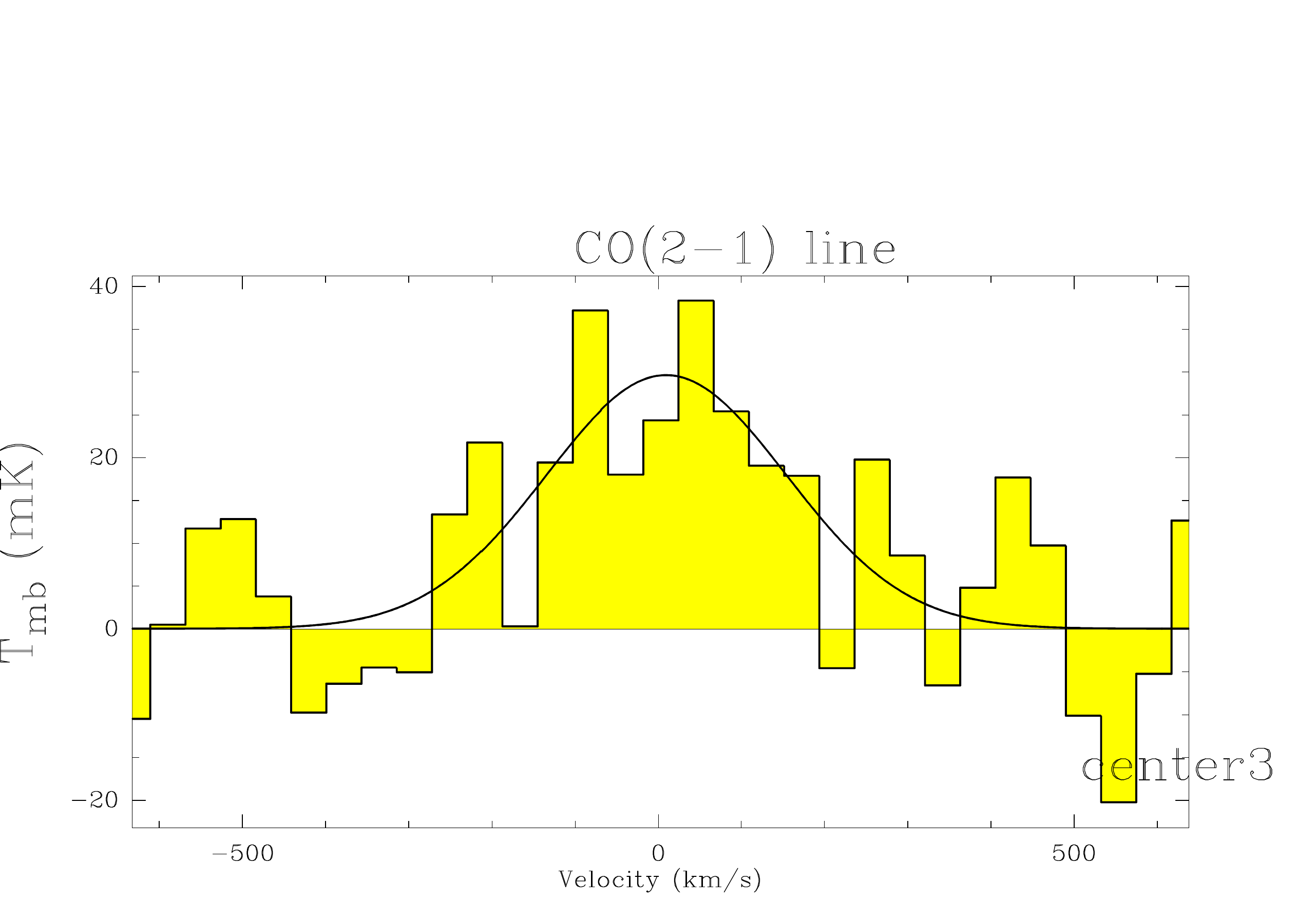} &
\includegraphics[width=4cm,angle=-90]{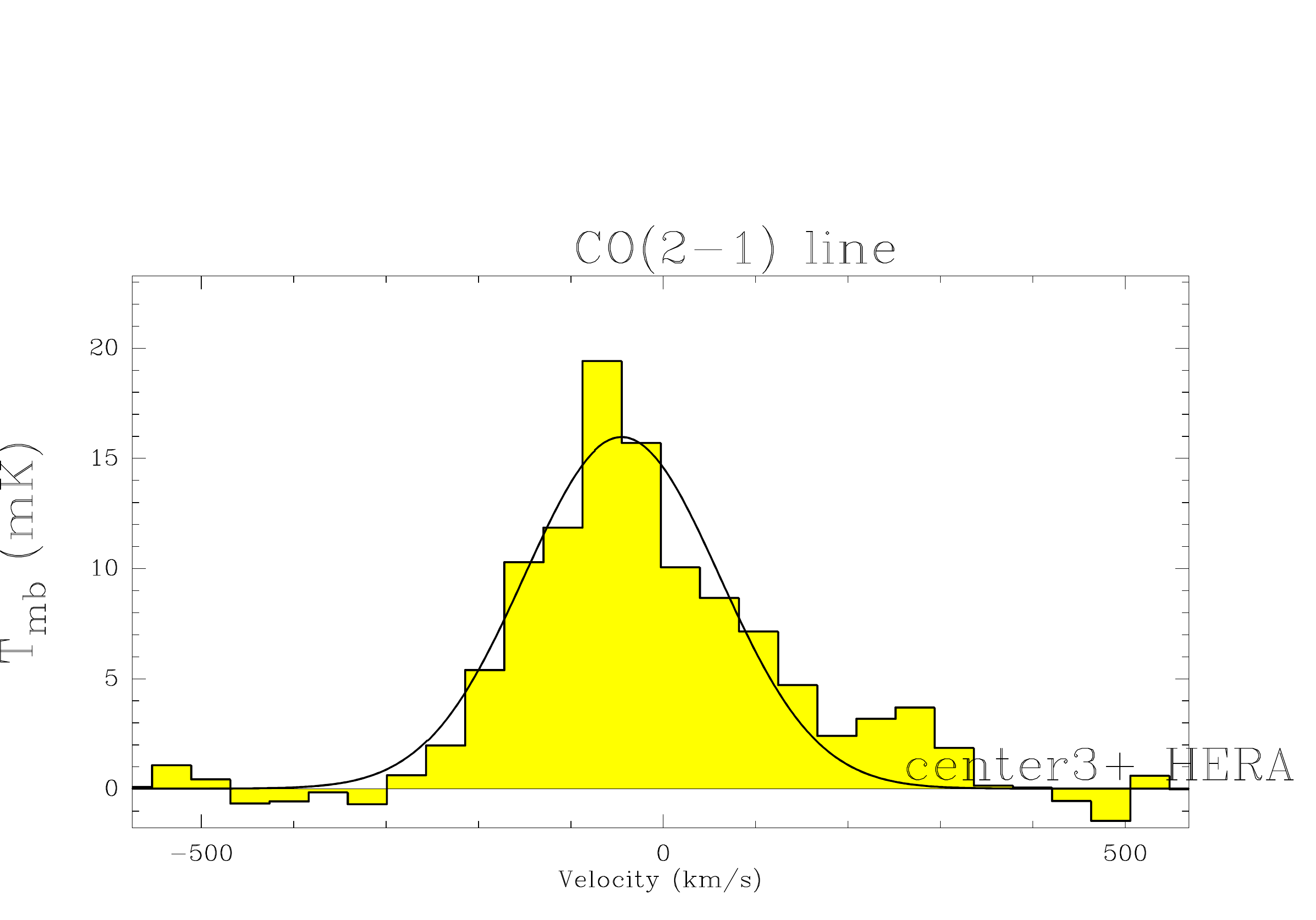} \\
\includegraphics[width=4cm,angle=-90]{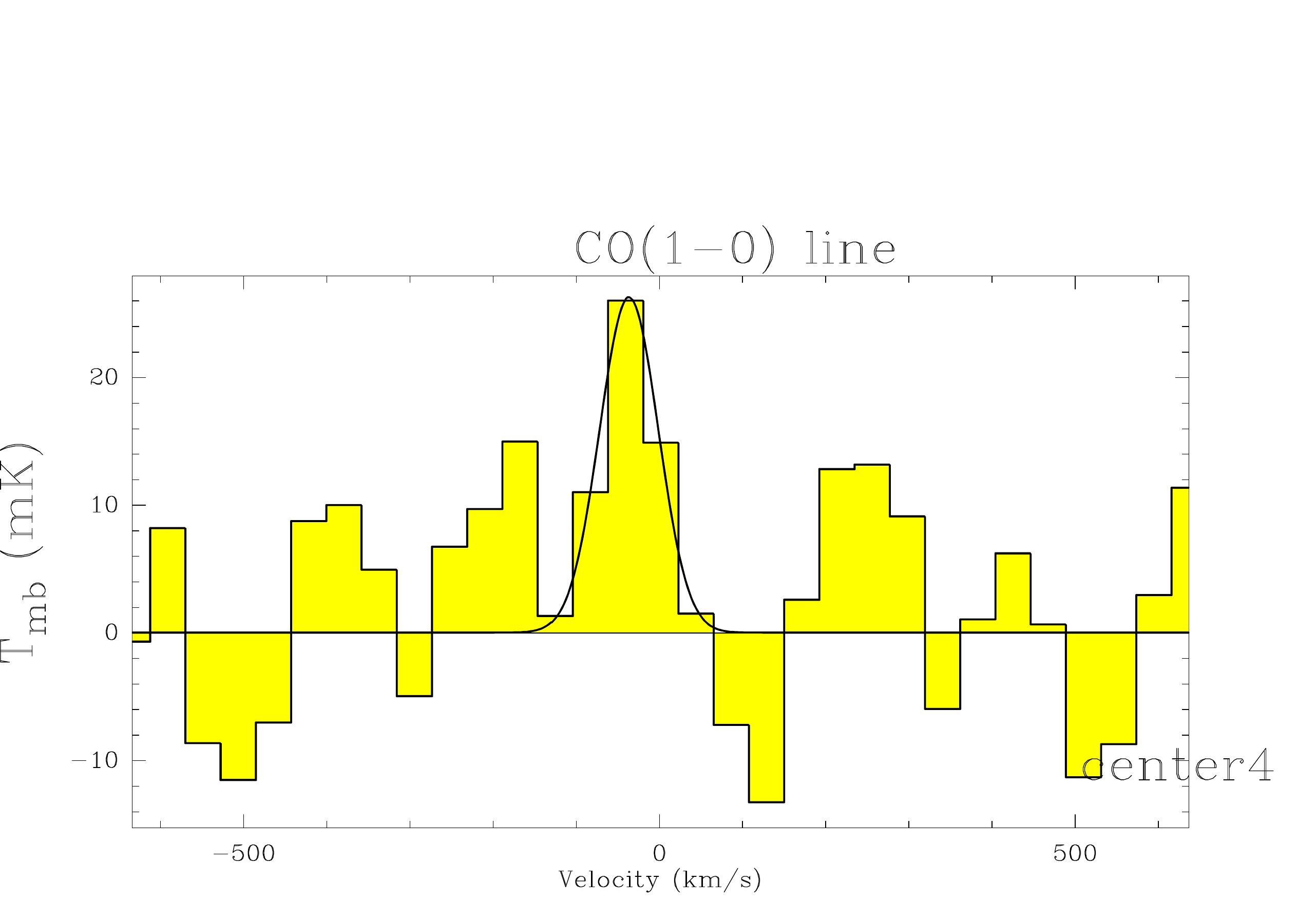} &
\includegraphics[width=4cm,angle=-90]{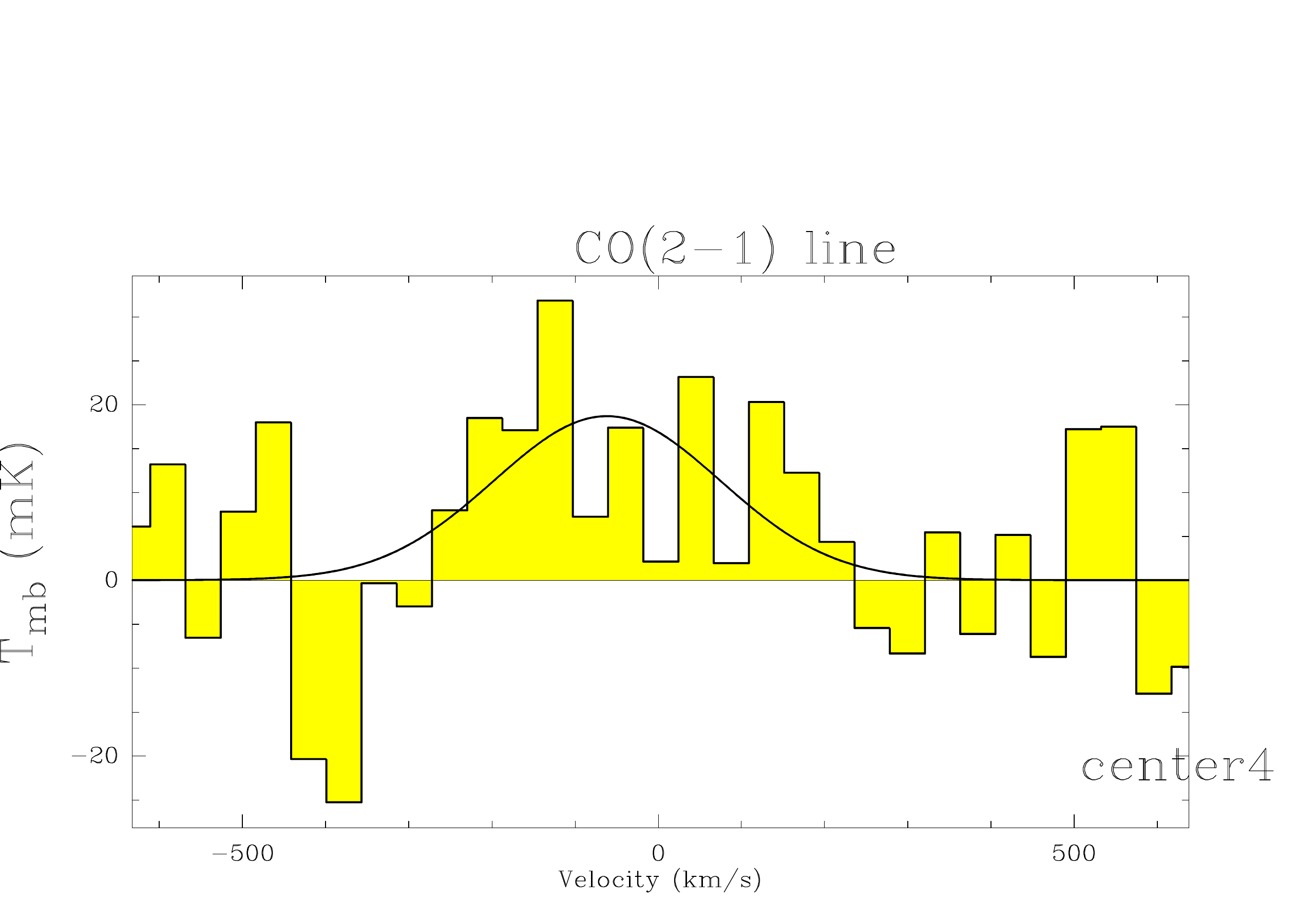} &
\includegraphics[width=4cm,angle=-90]{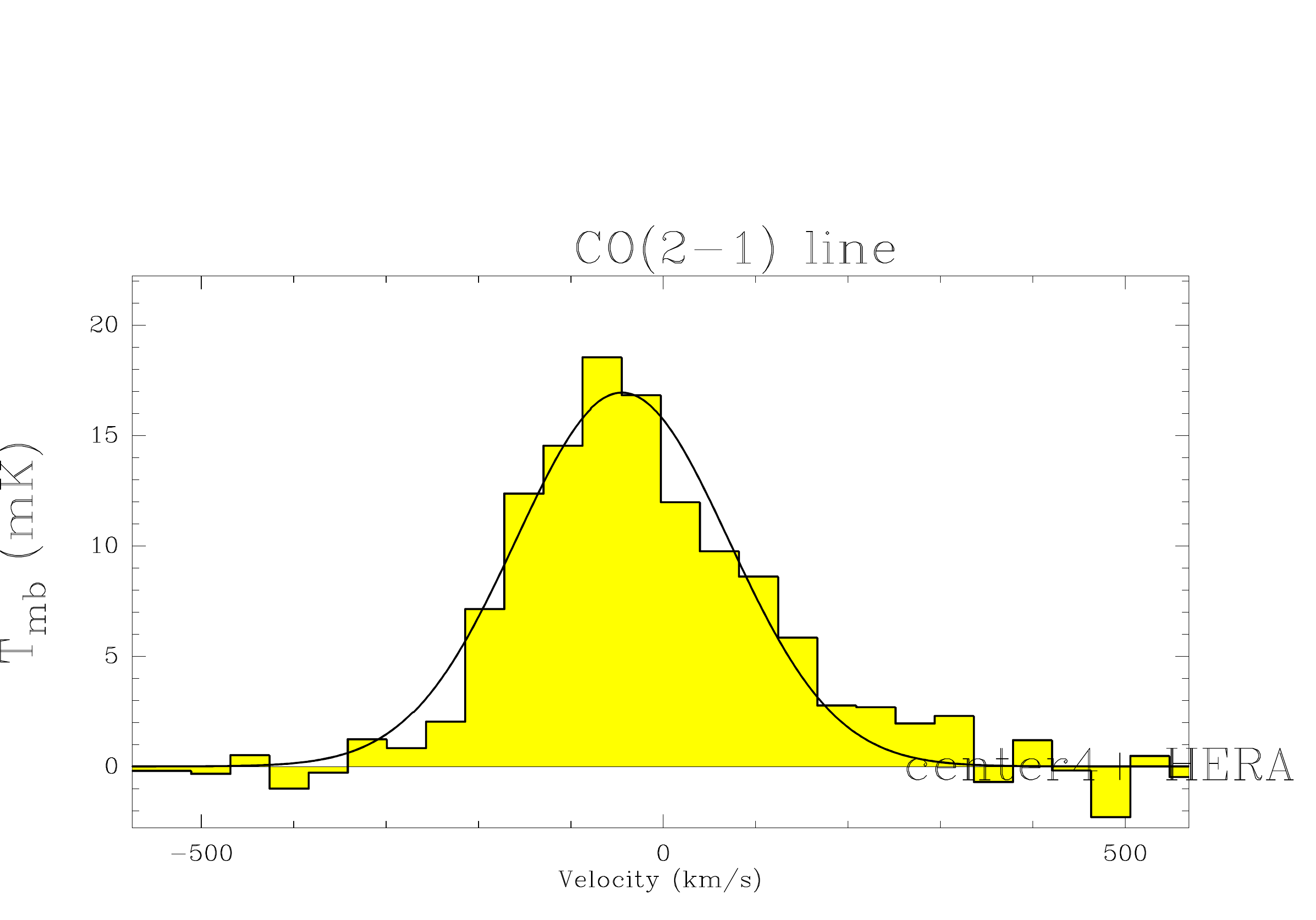} \\
\includegraphics[width=4cm,angle=-90]{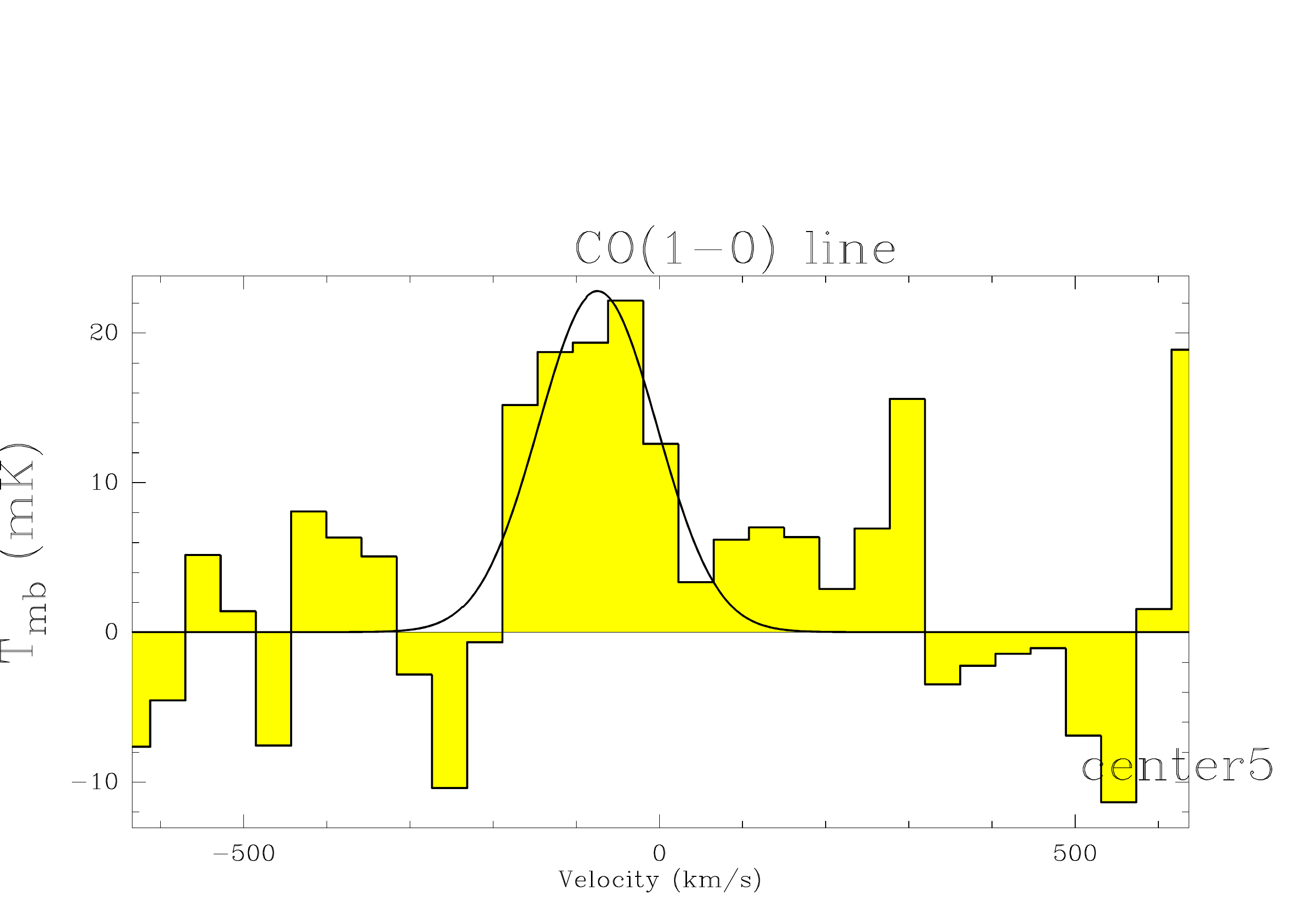} &
\includegraphics[width=4cm,angle=-90]{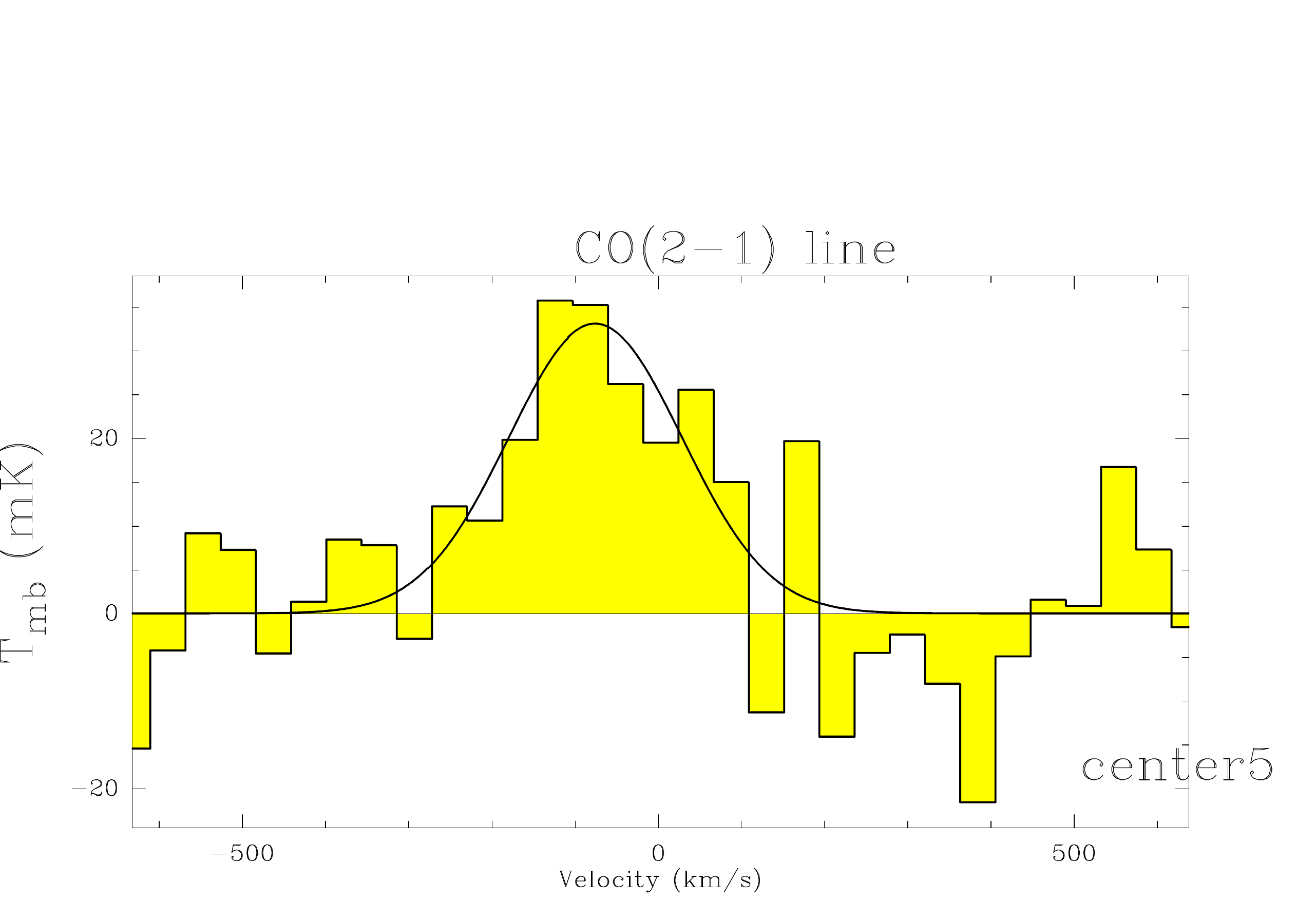} &
\includegraphics[width=4cm,angle=-90]{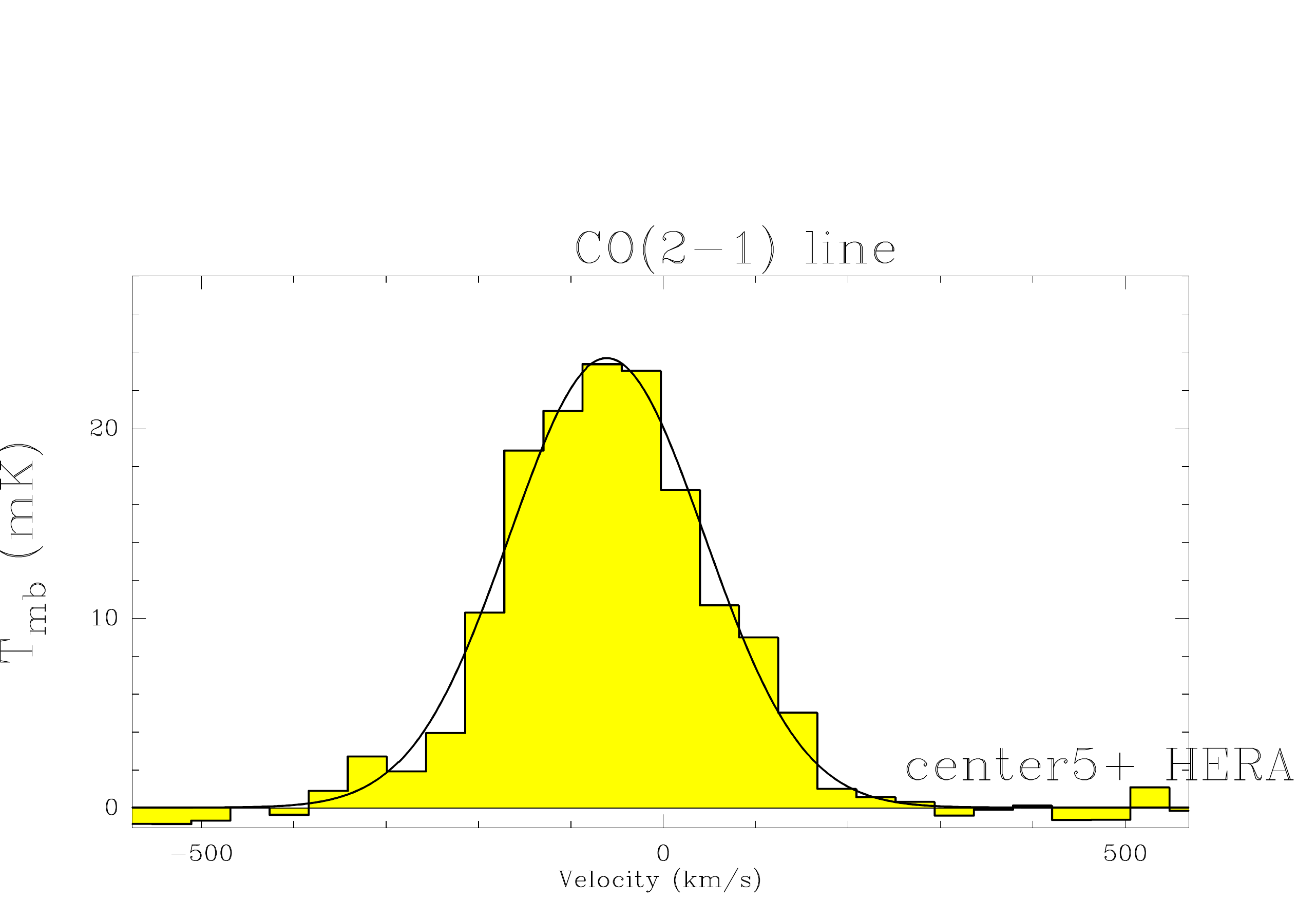} \\
\end{tabular} 
\caption{CO(1--0) and CO(2--1) spectra obtained at all the positions
near the centre of the galaxy as labeled in the lower right of each
diagram. The channel width is 42 km/s, see Table
\ref{table2-center}.}
\label{spectra-center1}  
\end{figure*}  
\begin{figure*} 
\centering 
\begin{tabular}{ccc} 
\includegraphics[width=4cm,angle=-90]{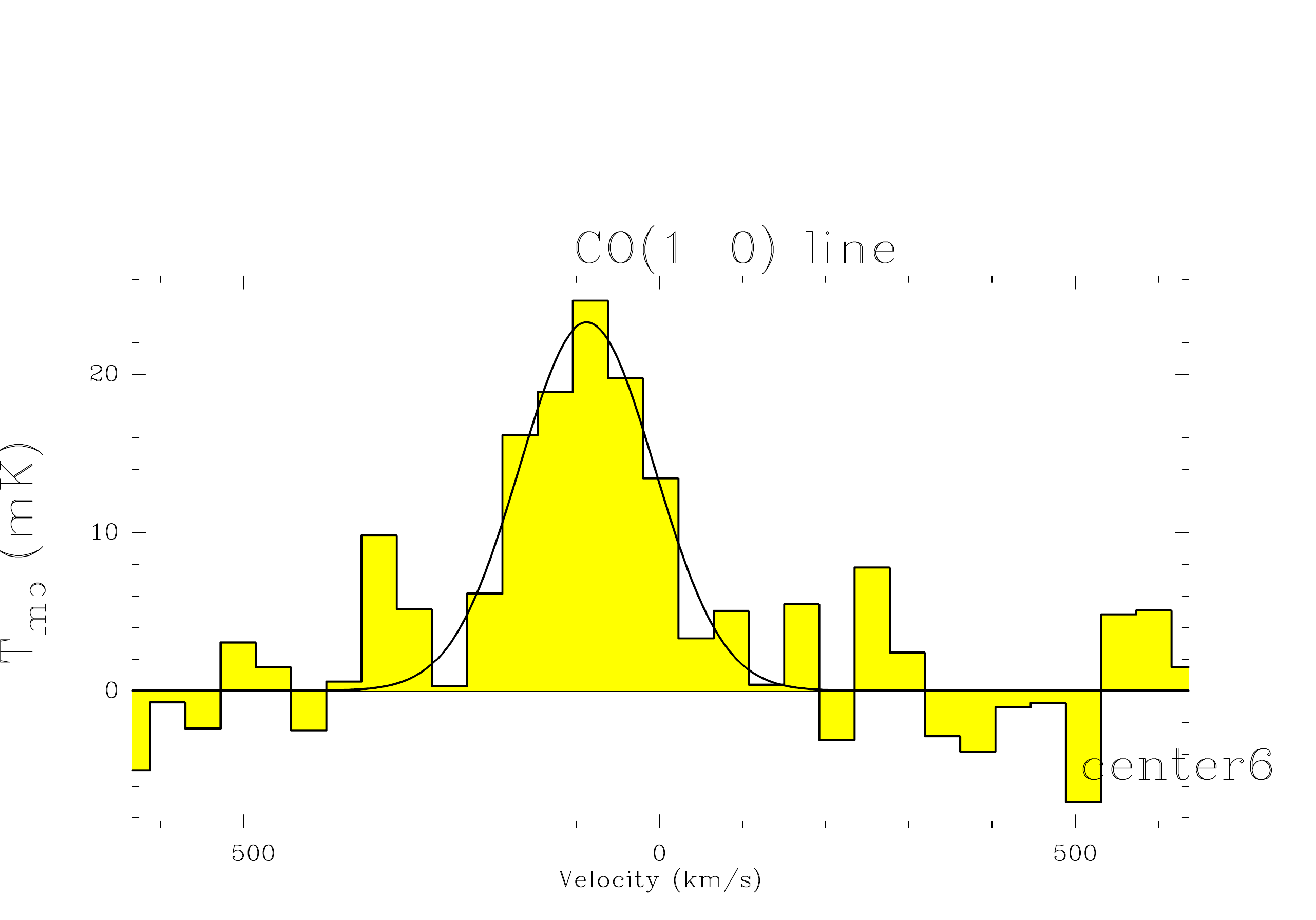} &
\includegraphics[width=4cm,angle=-90]{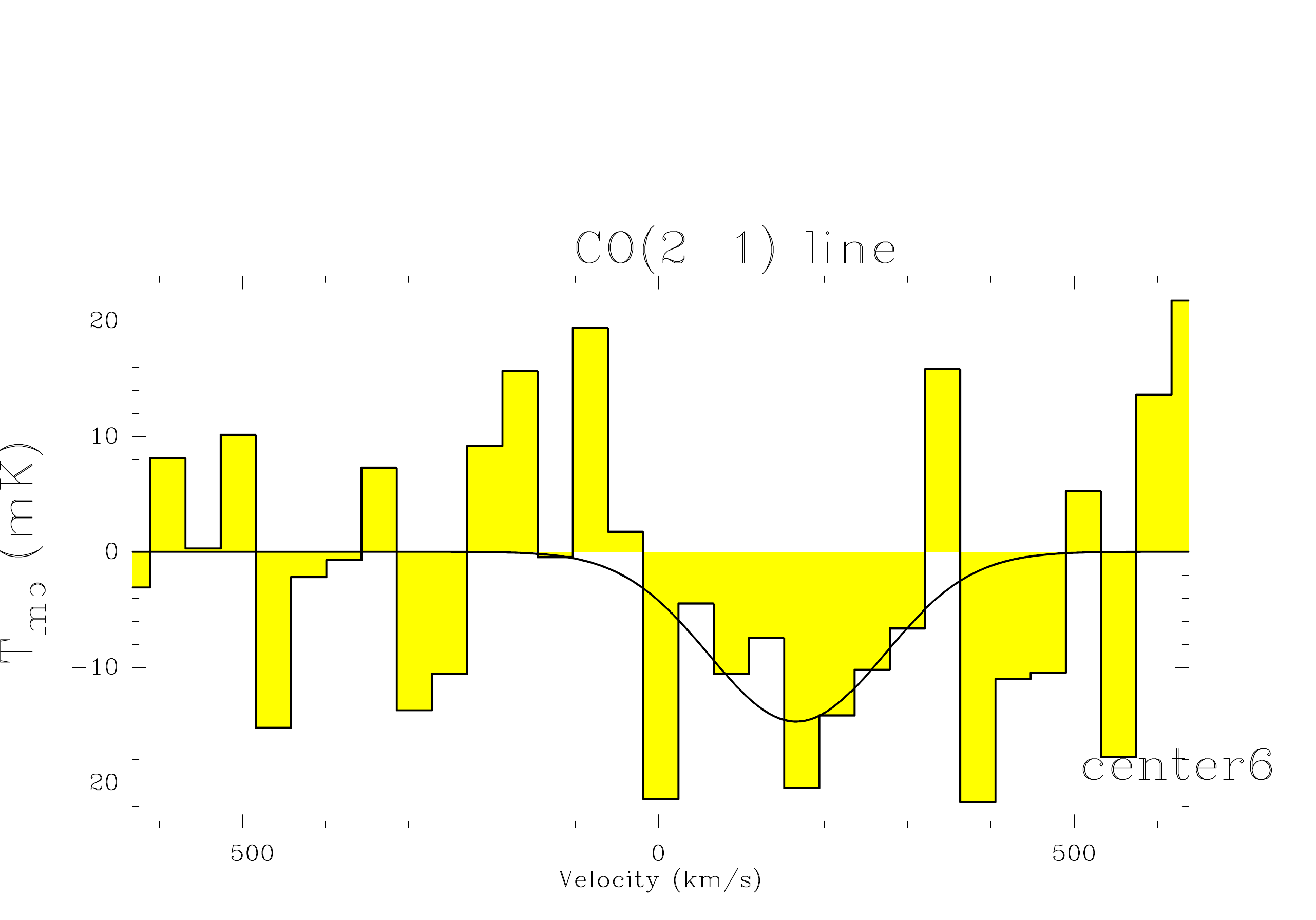} &
\includegraphics[width=4cm,angle=-90]{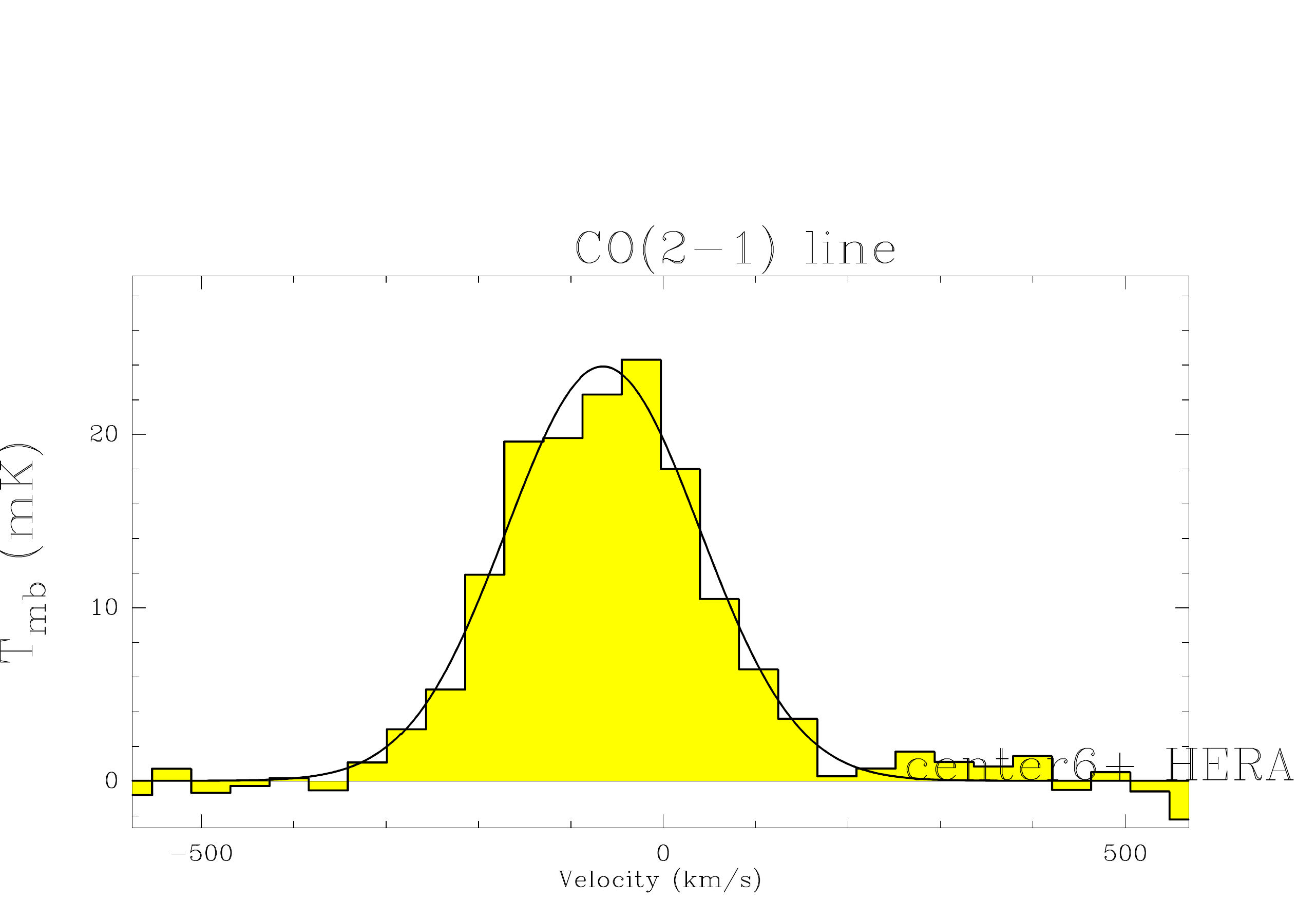} \\
\includegraphics[width=4cm,angle=-90]{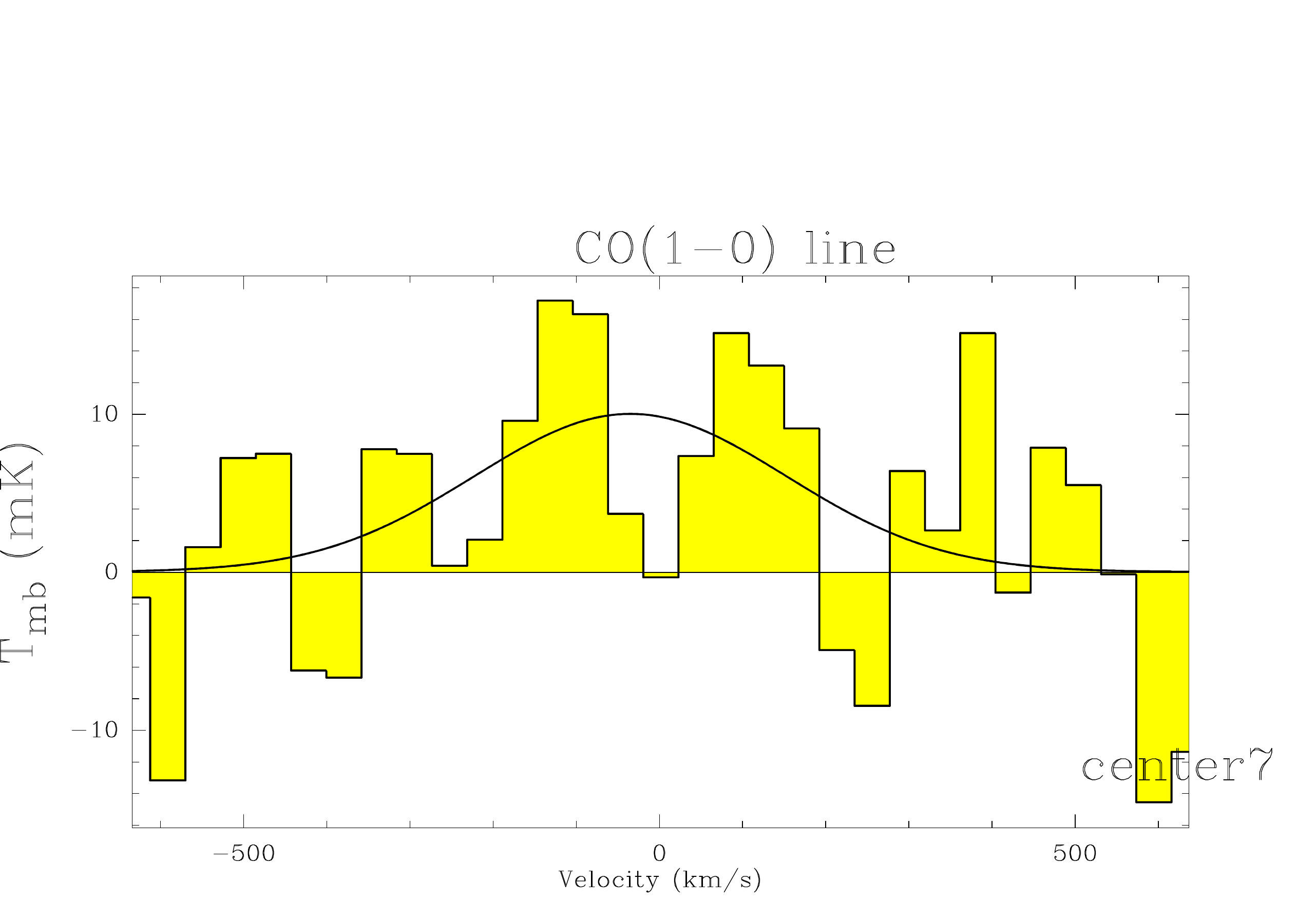} &
\includegraphics[width=4cm,angle=-90]{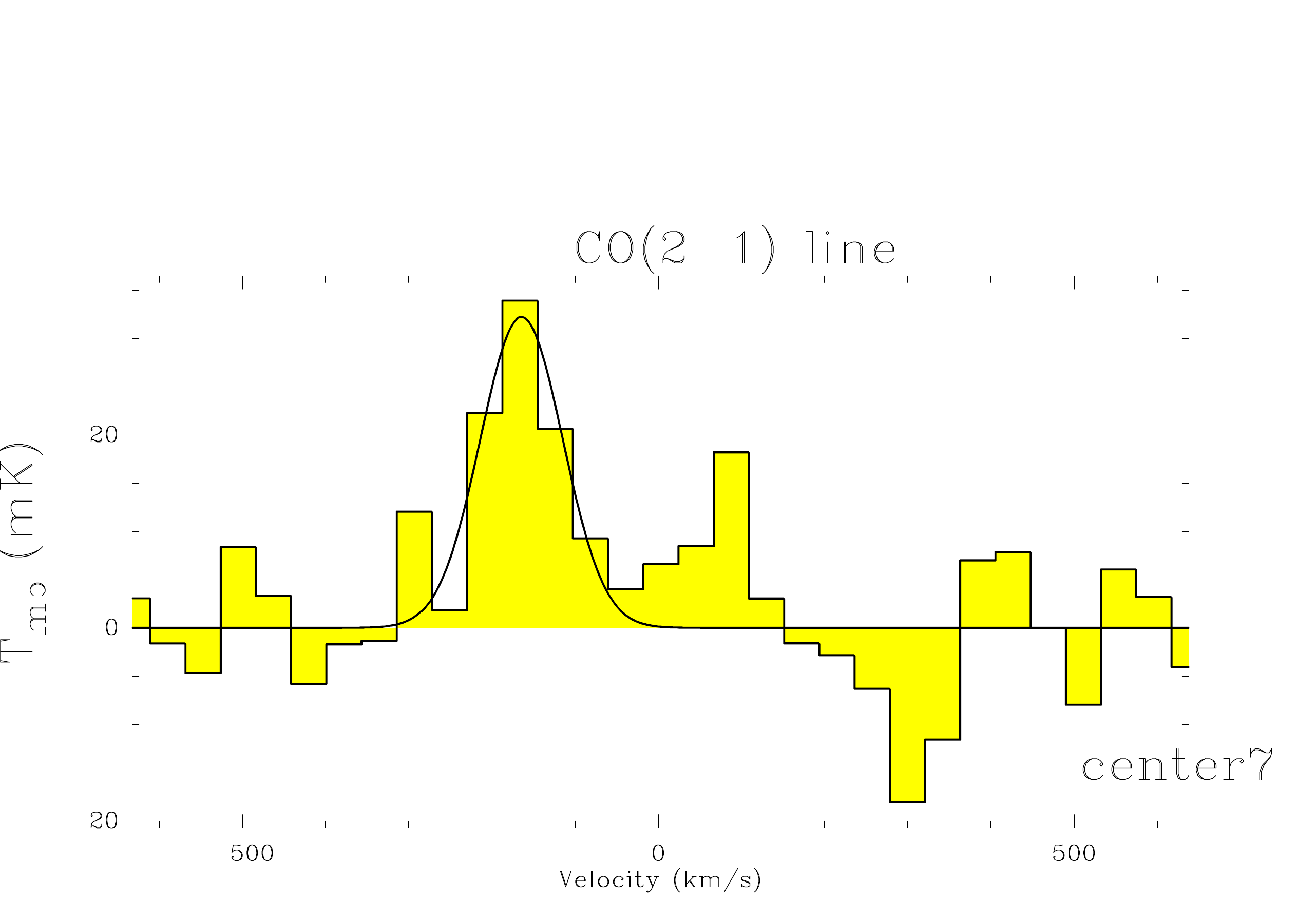} &
\includegraphics[width=4cm,angle=-90]{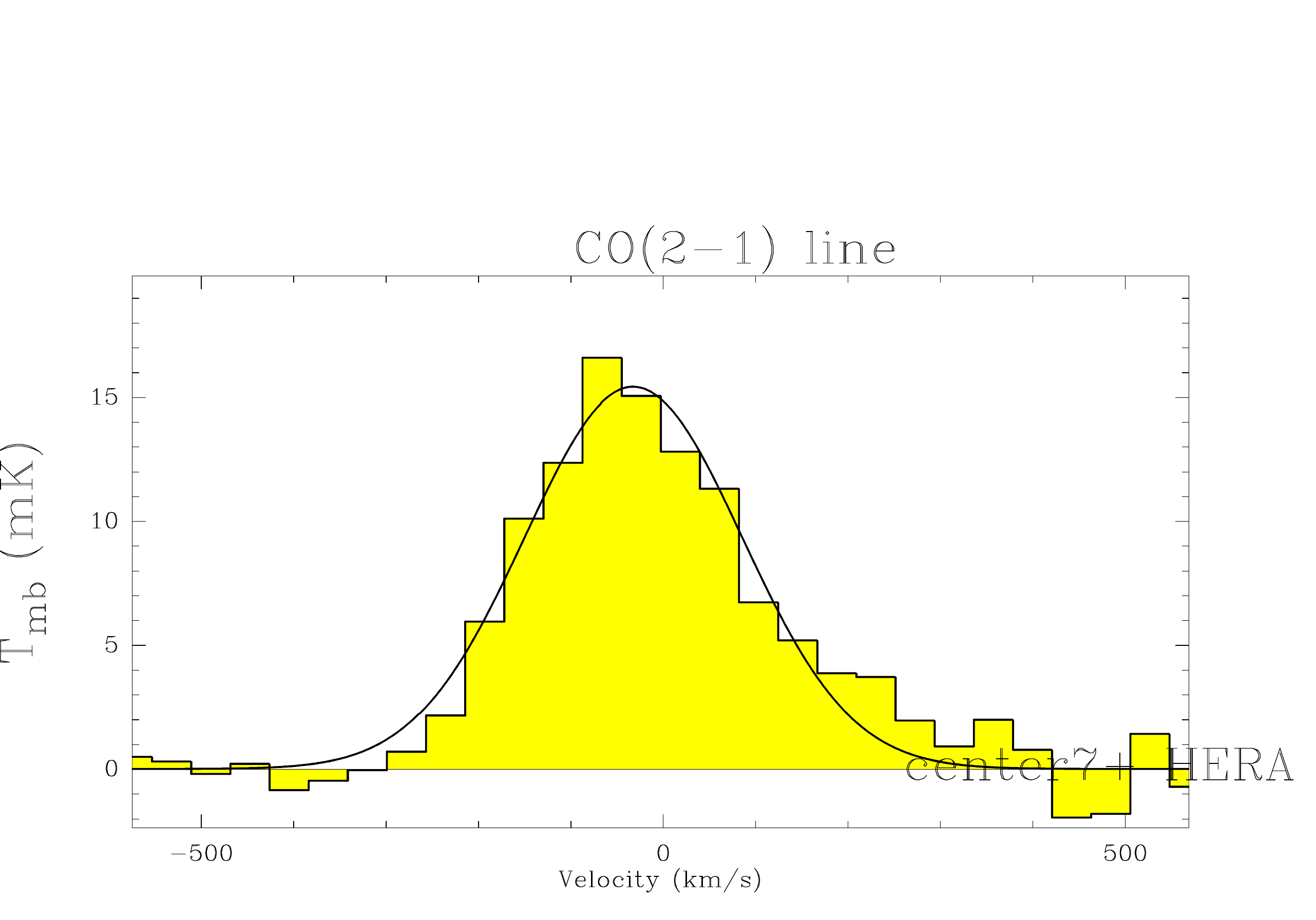} \\
\includegraphics[width=4cm,angle=-90]{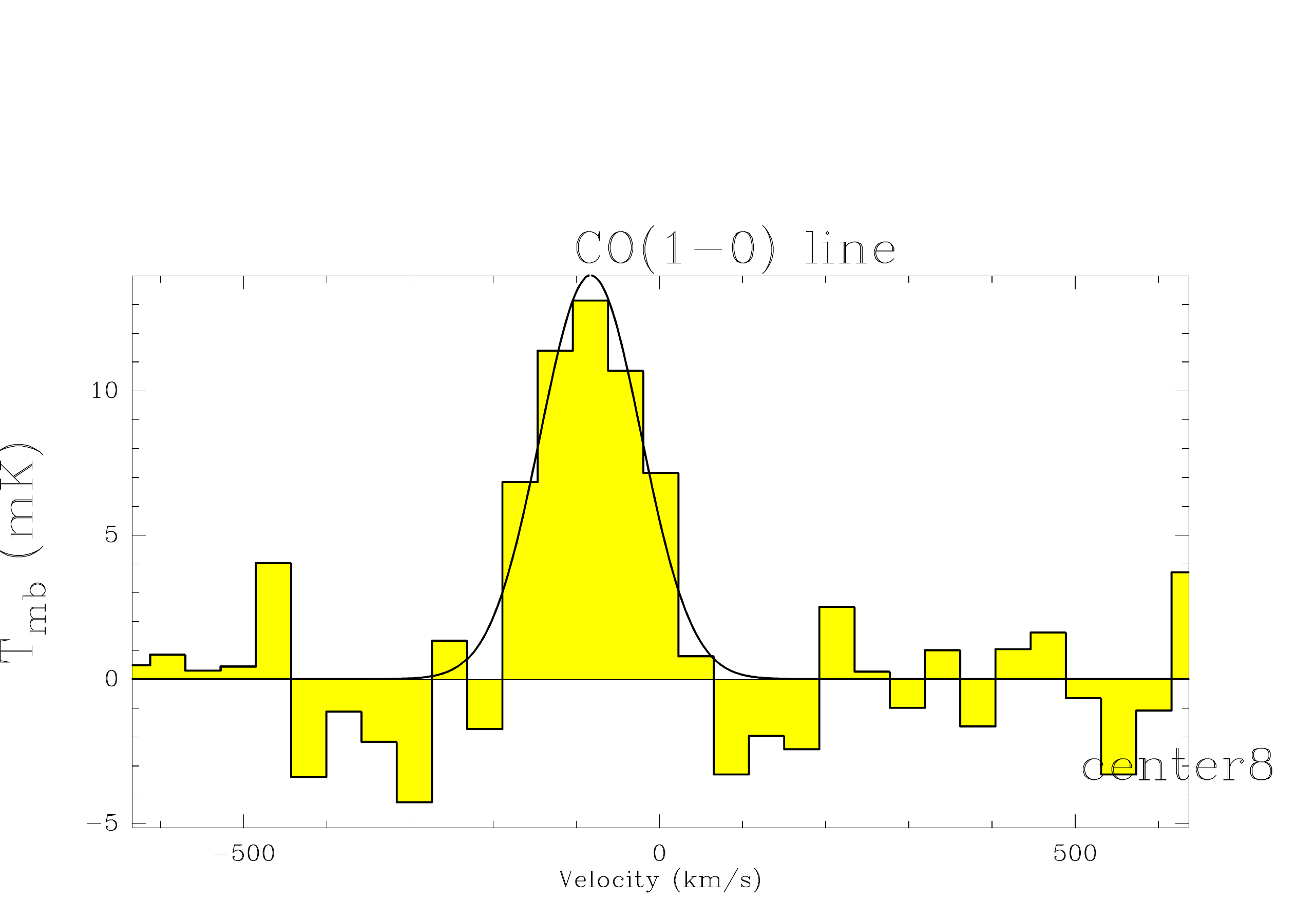} &
\includegraphics[width=4cm,angle=-90]{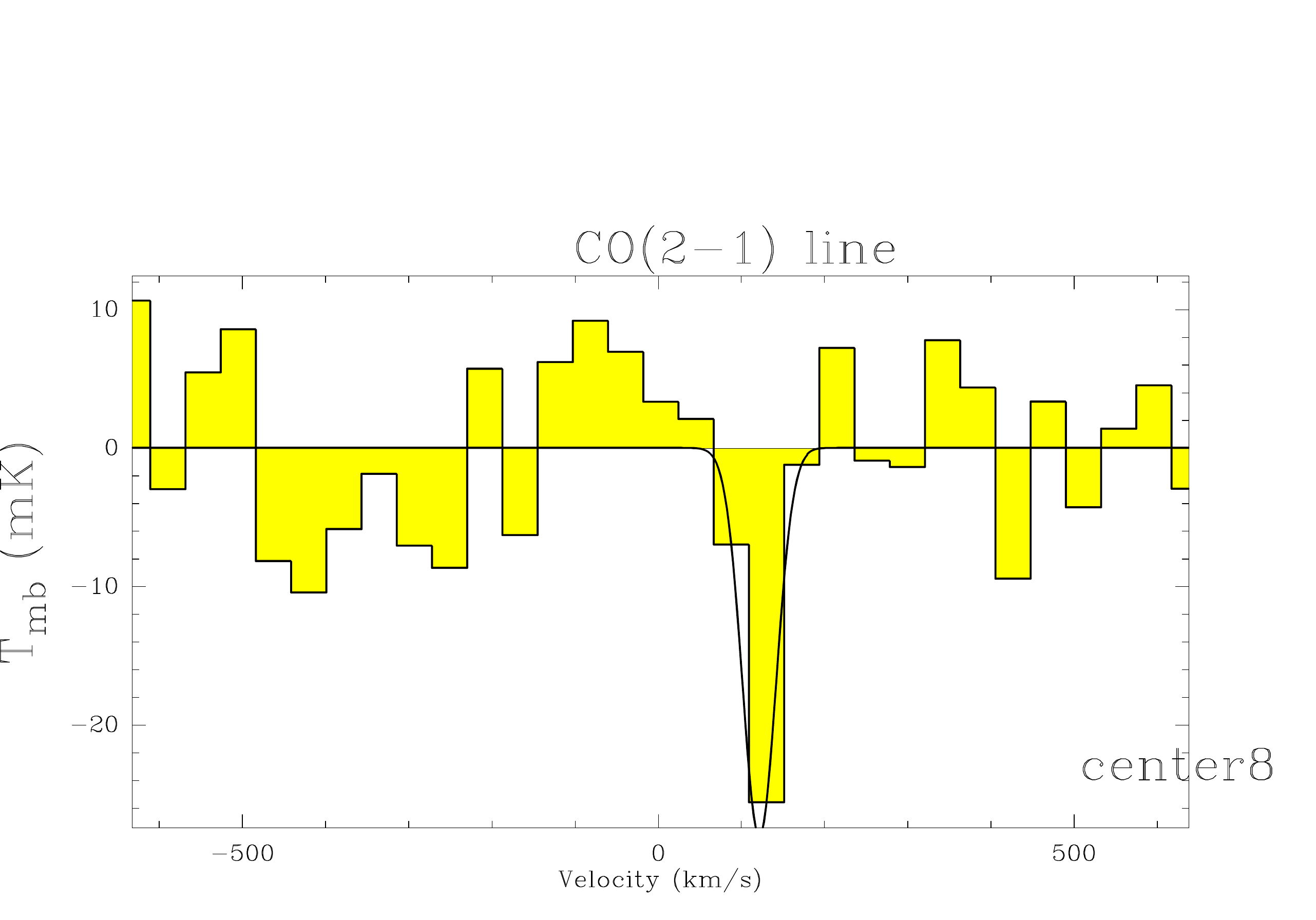} &
\includegraphics[width=4cm,angle=-90]{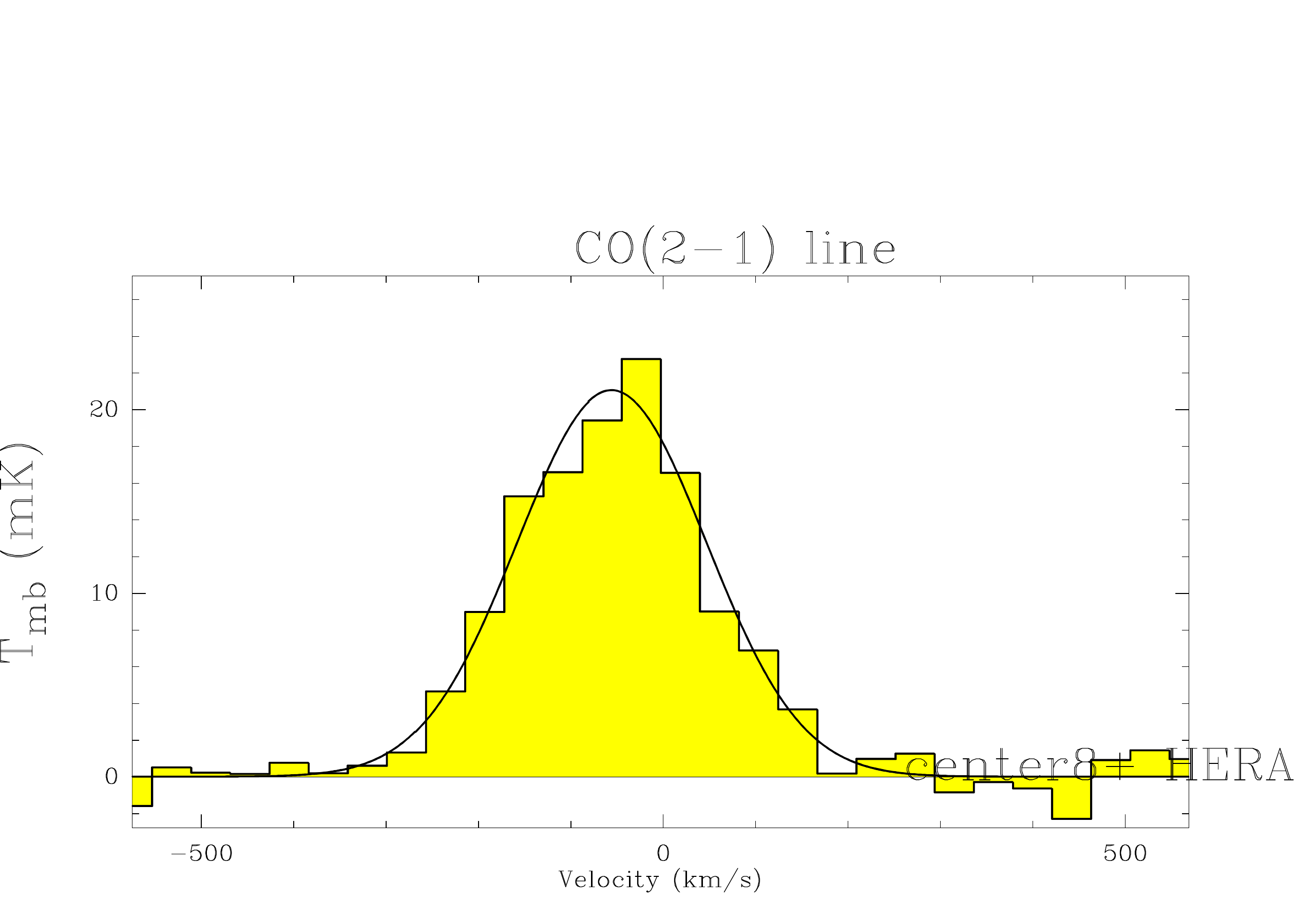} \\
\includegraphics[width=4cm,angle=-90]{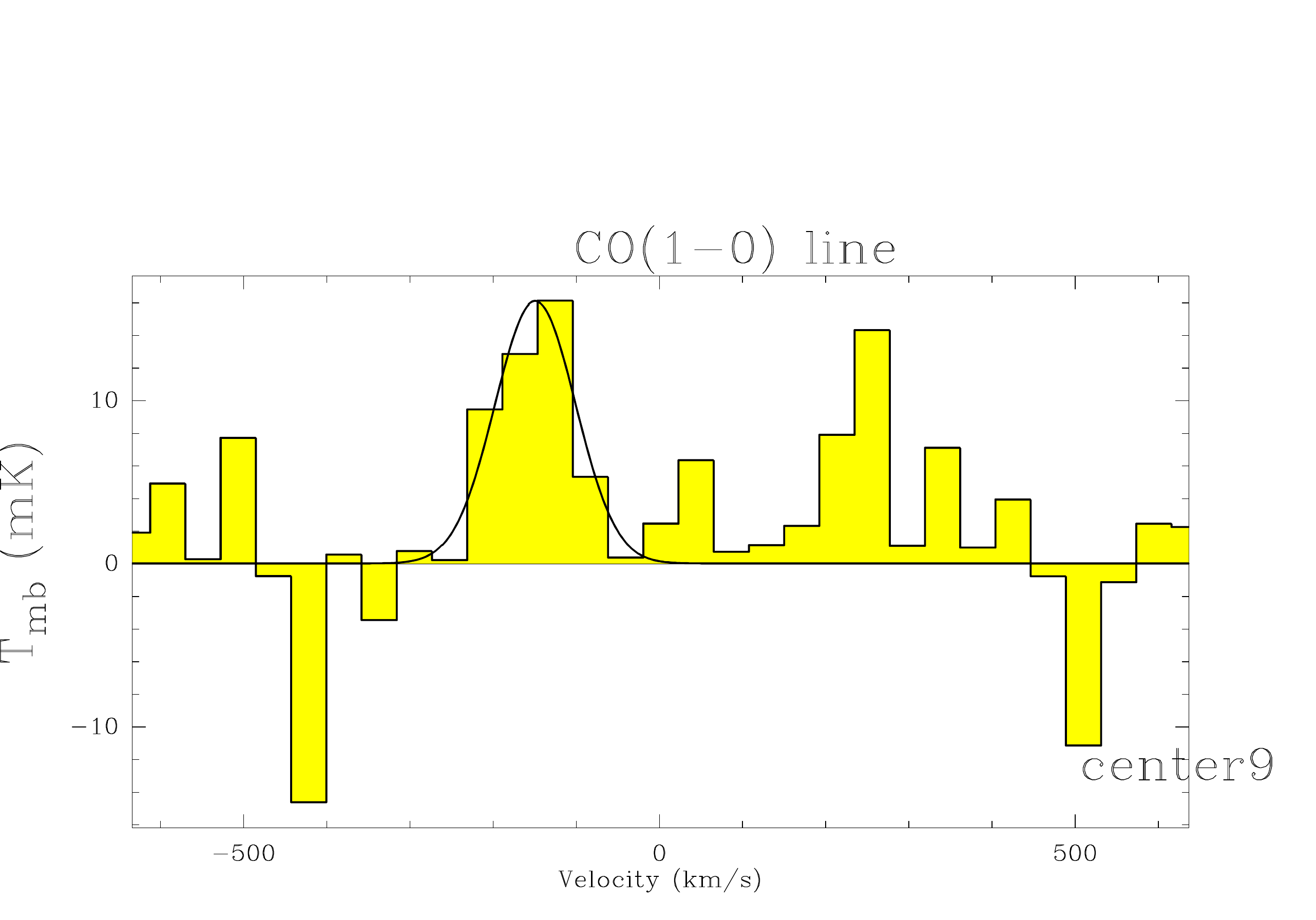} &
\includegraphics[width=4cm,angle=-90]{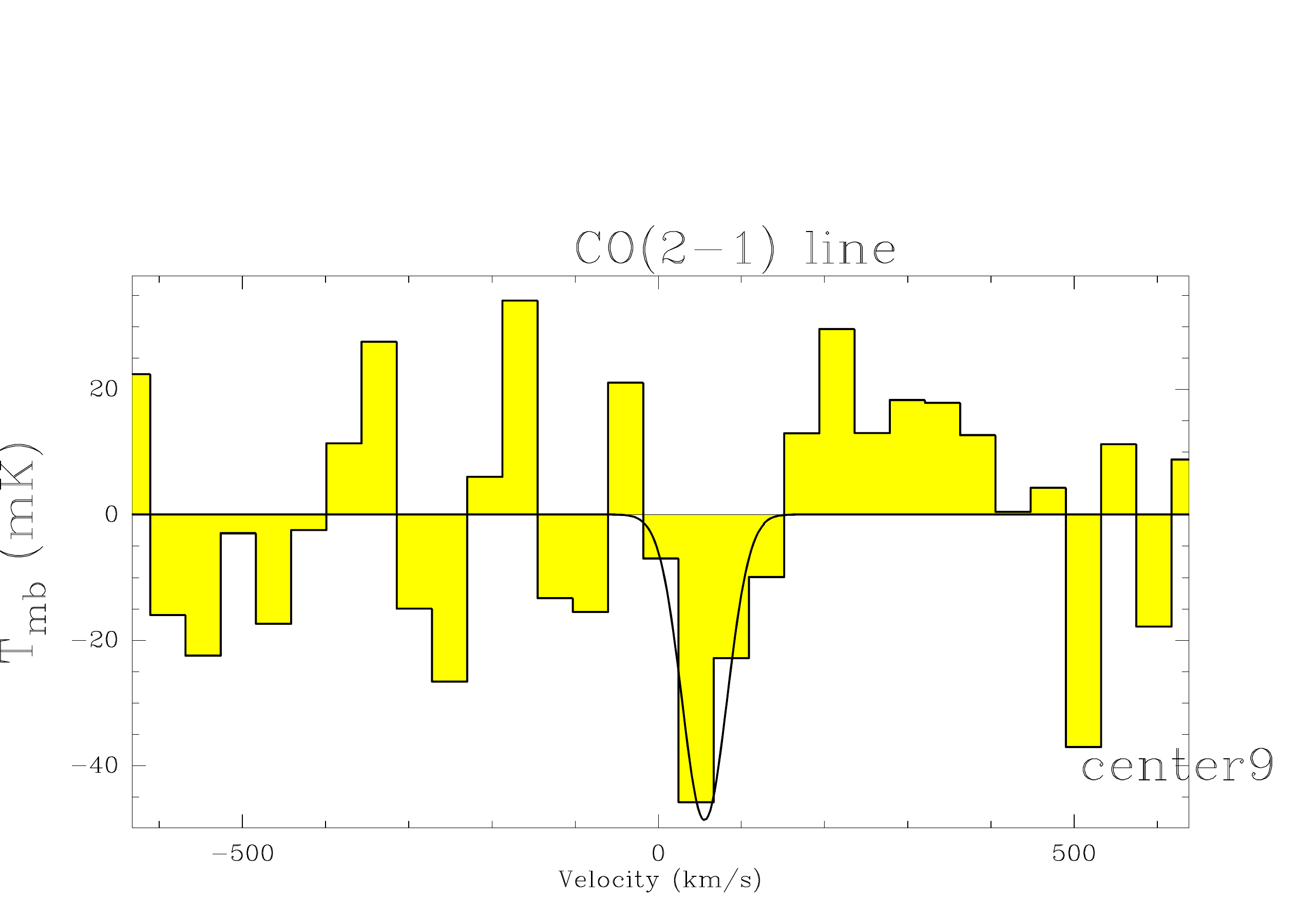} &
\includegraphics[width=4cm,angle=-90]{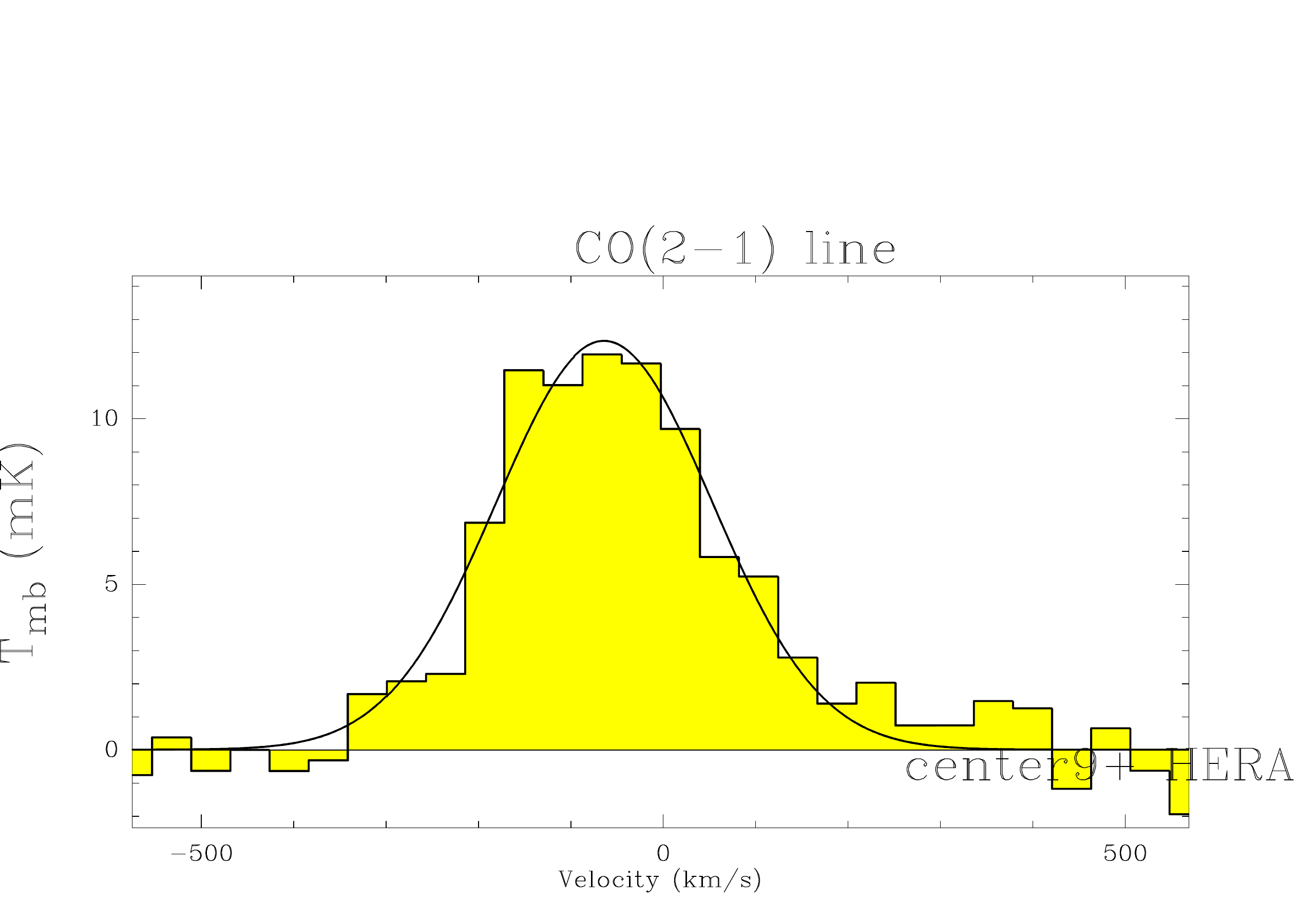} \\
\includegraphics[width=4cm,angle=-90]{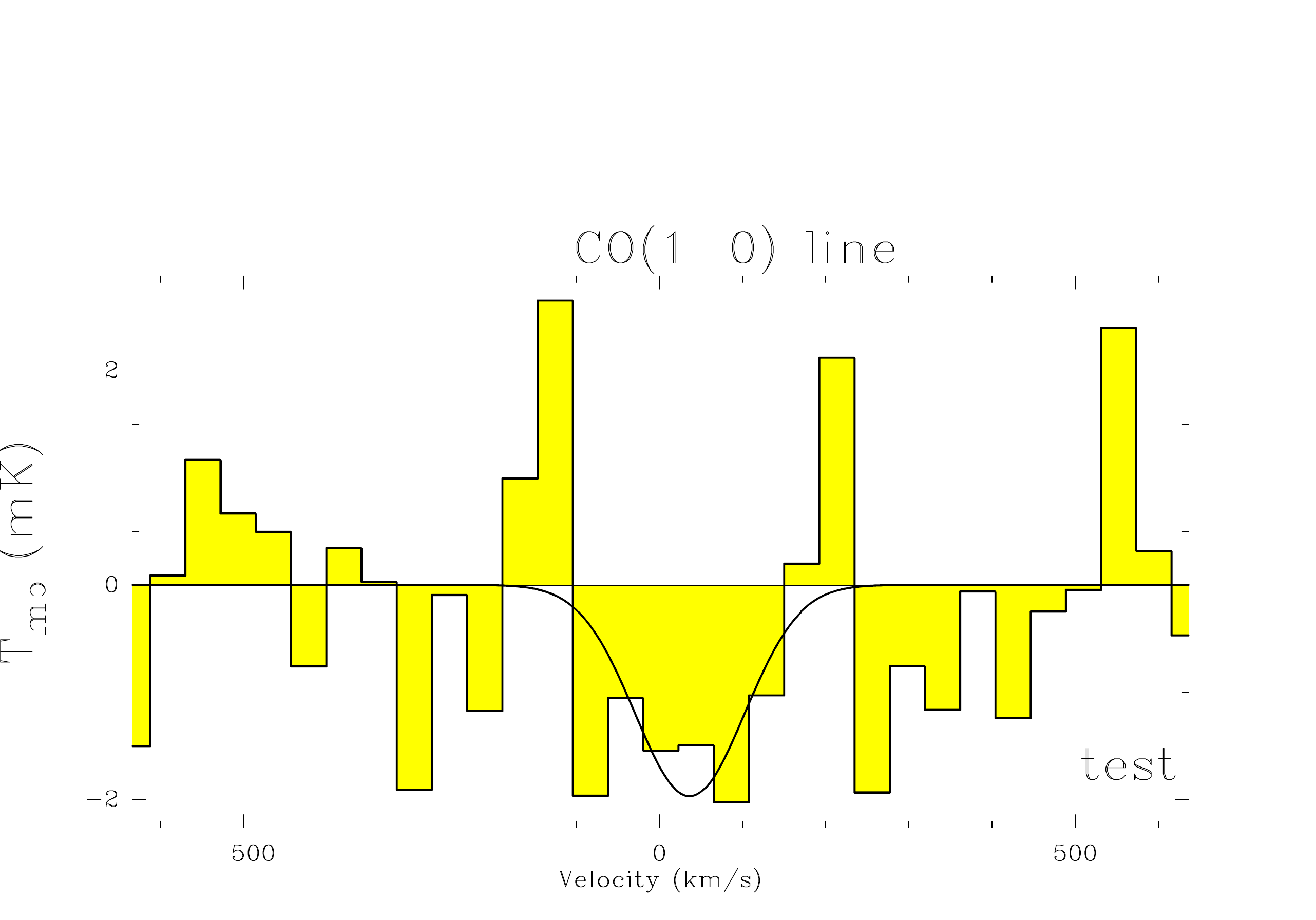} &
\includegraphics[width=4cm,angle=-90]{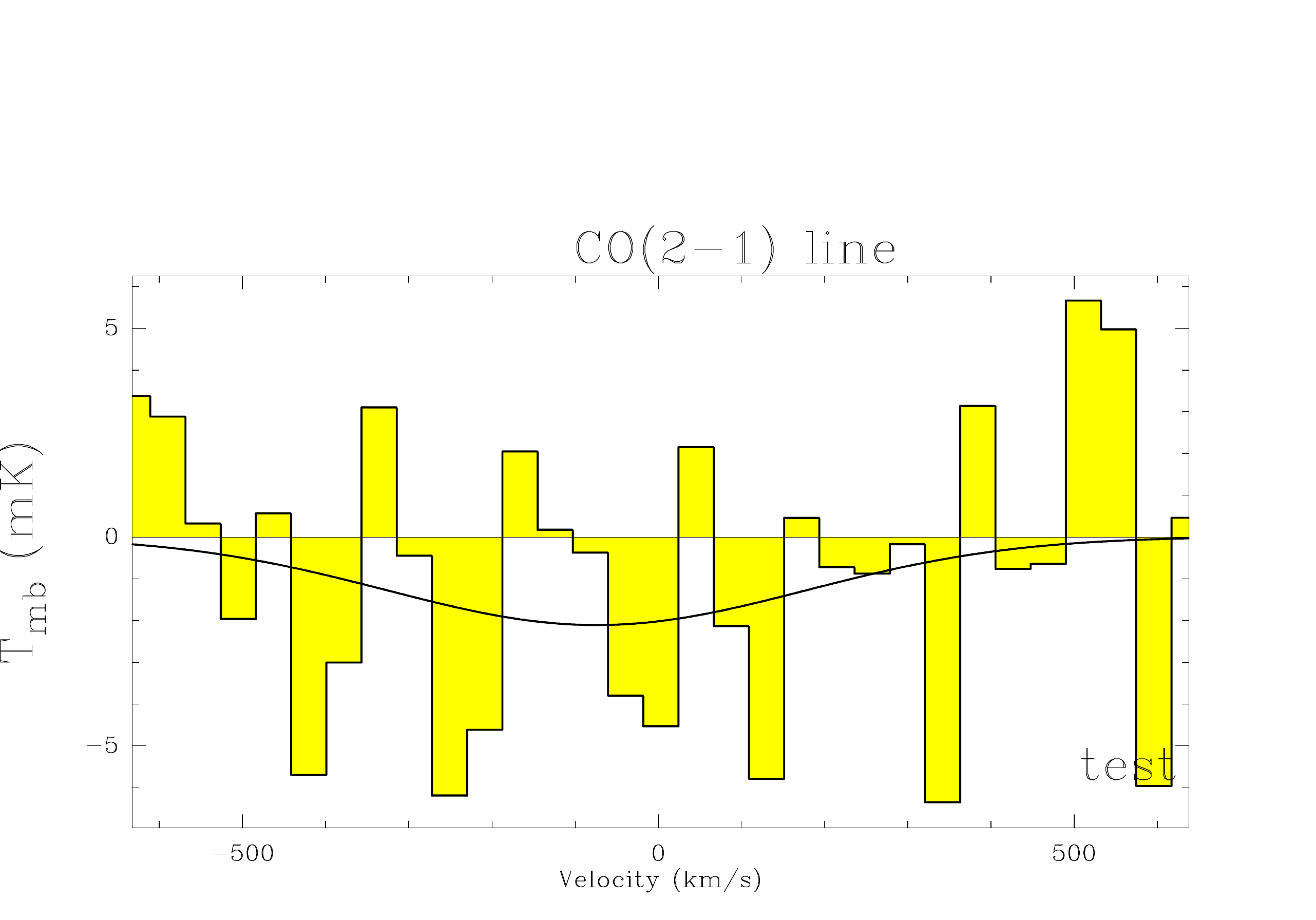} & \\
 \\
\end{tabular} 
\caption{CO(1--0) and CO(2--1) spectra obtained at all the positions
near the centre of the galaxy as labeled in the lower right of each
diagram. The channel width is 42 km/s, see Table
\ref{table2-center}.}  
\label{spectra-center2}  
\end{figure*}  
%
%
\begin{table*}
\begin{center}
\caption{Results of the observations.} 
\begin{tabular}{cccccccccc}
\hline
\hline
 Position& Offsets & Line & T$_{{\rm mb}}$ & Velocity & Width & I$_{{\rm CO}}$ & M$_{{\rm gas}}$  & T$_{21}$/T$_{10}$ \\
 & [$\prime\prime$ $\times$ $\prime\prime$] & & [mK] & [km/s] & [km/s] & [K.km/s] & [10$^8$M$_\odot$]&\\
\hline
centre1& [0, 0] & CO(1-0) & 15.2 $\pm$ 4.2 &-70.2 $\pm$ 22.8 &333 $\pm$ 69.1 &5.4 $\pm$ 0.8 & 12.6 & \\
centre1& [0, 0] & CO(2-1) & 76 $\pm$ 4.6 &-25 $\pm$ 4.8 &301 $\pm$ 11.5 &24.3 $\pm$ 0.8 & & \\
centre1& [0, 0] & CO(2-1) & 26.7 $\pm$ 1.6 &-41.2 $\pm$ 4.5 &277.9 $\pm$ 10.9 &7.9 $\pm$ 0.26 & & 1.7\\
\hline
centre2& [-11, 3] & CO(1-0) & 26.7 $\pm$ 5.6 &-74.8 $\pm$ 16.4 &238.6 $\pm$ 41.3 &6.8 $\pm$ 0.9 & 15.8 & \\
centre2& [-11, 3] & CO(2-1) & 83.7 $\pm$ 17.2 &-99.7 $\pm$ 14.8 &231.7 $\pm$ 34.6 &20.6 $\pm$ 2.6 & & \\
centre2& [-11, 3] & CO(2-1) & 31.1 $\pm$ 1.2 &-63.8 $\pm$ 2.9 &257.6 $\pm$ 6.7 &8.5 $\pm$ 0.2 & & 1.2\\
\hline
centre3& [11, -3] & CO(1-0) & 10.3 $\pm$ 5.1 &-60.4 $\pm$ 81.2 &340.3 $\pm$ 216.2 &3.7 $\pm$ 1.4 & 8.7 & \\
centre3& [11, -3] & CO(2-1) & 29.7 $\pm$ 10.2 &9.4 $\pm$ 29.1 &340.2 $\pm$ 69.2 &10.7 $\pm$ 1.9 & & \\
centre3& [11, -3] & CO(2-1) & 16 $\pm$ 1.3 &-45.1 $\pm$ 5.9 &248.4 $\pm$ 15.4 &4.2 $\pm$ 0.2 & & 1.6\\
\hline
centre4& [3, 11] & CO(1-0) & 26.4 $\pm$ 8.4 &-36.9 $\pm$ 13.4 &83.1 $\pm$ 29.6 &2.3 $\pm$ 0.8 & 5.4 & \\
centre4& [3, 11] & CO(2-1) & 18.7 $\pm$ 12.2 &-61.5 $\pm$ 60.7 &320.4 $\pm$ 97.8 &6.4 $\pm$ 2.0 & & \\
centre4& [3, 11] & CO(2-1) & 17 $\pm$ 1.1 &-44.7 $\pm$ 4.9 &271.2 $\pm$ 12.1 &4.9 $\pm$ 0.2 & & 0.6\\
\hline
centre5& [-8, 14] & CO(1-0) & 22.9 $\pm$ 7.6 &-74.5 $\pm$ 20.4 &167.5 $\pm$ 43.3 &4.1 $\pm$ 1.0 & 9.5 & \\
centre5& [-8, 14] & CO(2-1) & 33.2 $\pm$ 10.1 &-76 $\pm$ 21.6 &246 $\pm$ 50 &8.7 $\pm$ 1.5 & & \\
centre5& [-8, 14] & CO(2-1) & 23.8 $\pm$ 0.6 &-61.3 $\pm$ 1.7 &248.2 $\pm$ 4.2 &6.3 $\pm$ 0.1 & & 1.0\\
\hline
centre6& [-21, 6] & CO(1-0) & 23.4 $\pm$ 4.3 &-87.4 $\pm$ 11.5 &191.3 $\pm$ 31.9 &4.7 $\pm$ 0.6 & 11.1 & \\
centre6& [-21, 6] & CO(2-1) & -14.8 $\pm$ 12.2 &165.9 $\pm$ 68 &247.3 $\pm$ 162.8 &-3.9 $\pm$ 2.0 & & \\
centre6& [-21, 6] & CO(2-1) & 24 $\pm$ 1.1 &-65.4 $\pm$ 3.1 &246.7 $\pm$ 7.1 &6.3 $\pm$ 0.2 & & 1\\
\hline
centre7& [-3, -11] & CO(1-0) & 10 $\pm$ 8.2 &-34.9 $\pm$ 79.2 &441.4 $\pm$ 209.8 &4.7 $\pm$ 1.8 & 11 & \\
centre7& [-3, -11] & CO(2-1) & 32.4 $\pm$ 7.6 &-164.8 $\pm$ 12.8 &117 $\pm$ 39.8 &4.0 $\pm$ 1.0 & & \\
centre7& [-3, -11] & CO(2-1) & 15.5 $\pm$ 1.1 &-32.7 $\pm$ 5.6 &277.4 $\pm$ 13.8 &4.5 $\pm$ 0.2 & & 1.6\\
\hline
centre8& [-18, 17] & CO(1-0) & 14.1 $\pm$ 2.2 &-82.9 $\pm$ 8.5 &142.8 $\pm$ 16.2 &2.1 $\pm$ 0.2 & 5 & \\
centre8& [-18, 17] & CO(2-1) & -28.2 $\pm$ 6.2 &121.1 $\pm$ 14.5 &47.6 $\pm$ 28.2 &-1.4 $\pm$ 0.5 & & \\
centre8& [-18, 17] & CO(2-1) & 21.1 $\pm$ 1.1 &-56.1 $\pm$ 3.5 &241.4 $\pm$ 8.2 &5.4 $\pm$ 0.2 & & 1.5\\
\hline
centre9& [-24, -5] & CO(1-0) & 16.2 $\pm$ 5.6 &-149.5 $\pm$ 16.3 &114.3 $\pm$ 36.4 &2.0 $\pm$ 0.6 & 4.6 & \\
centre9& [-24, -5] & CO(2-1) & -48.9 $\pm$ 18.4 &55.2 $\pm$ 15.4 &63.7 $\pm$ 33.1 &-3.3 $\pm$ 1.4 & & \\
centre9& [-24, -5] & CO(2-1) & 12.4 $\pm$ 1.0 &-64.2 $\pm$ 5.9 &275.1 $\pm$ 14.3 &3.6 $\pm$ 0.2 & & 0.8\\
\hline
\end{tabular}
\label{table2-center}  
\end{center}
\end{table*}

\subsection{The filaments}
%
We describe below the regions selected from our HERA map (see
S06). These regions were chosen to be distant from the East-West
filament seen in CO(2--1) and known for their peculiar optical
morphology.  The first aim was to confirm the CO detections far away
from the central galaxy. The regions called Off 1, Off 2 and Pos 2
trace the northern and southern filaments visible in H$\alpha$ up to a
projected distance of 25 kpc north and south of the galaxy.  Pos 11 is
centred on a looped-back filament seen in optical emission lines that
may trace uplifted gas behind a rising cold gas bubble inside the hot
intracluster medium (Fabian et al., 2003; Hatch et al.,
2006). Finally, closer to the centre, East and Off 3, are two fields
in the eastern filament. The East region includes a young star cluster
(Shields and Filipenko, 1990).

%
We now compare the velocity shift of our CO detections with the velocities 
of the warm (2000 K) H$_2$ gas detected by Hatch et al., (2005) and the hot 
(10$^4$K) optical line emitting gas measured by Hatch et al. (2006).

{\bf The eastern filament: } refers to positions East and
Off 3, at about 8 kpc from the galaxy's centre. It
corresponds to regions named A, B, C and D in Hatch et al. (2005). The
CO emission lines are centred at $-$100 km/s relative to the systemic
velocity. This blueshift is the same as the velocity found for the
P$\alpha$, Br$\gamma$ and H$_2$ v=1--0 S(1), S(2)and S(3) lines
detected by the above mentioned authors.

{\bf The southern filament: }refers to position Pos 2 at 25 kpc from the galaxy's
centre. It corresponds to regions named SW1 and SW2 in Hatch et
al. 2005. Here again, the CO emission is blueshifted relative to the
nucleus, like the infrared H$_2$ tracers.

{\bf The horseshoe: } refers to the position Pos 11, at least 25 kpc
from the galaxy's centre.  The CO(1--0) line is detected at the limit
of the sensitivity reached. There is also a tentative detection of
CO(2--1) line which disappears when merged with the HERA data, so this
detection is not very strong. Nevertheless, the two emission lines
have positive velocities (36 km/s and 120 km/s).  This velocity range
agrees with that from the infrared H$_2$ tracers by Hatch et
al. (2005). It is also the same velocity shift as the hot H$\alpha$
emitting gas found by Hatch et al. (2006) in the loop of the horseshoe
filament (see their Figure 6).

{\bf The northern filament: } 
This region refers to the position Off 2, at about 25 kpc from the
galaxy centre. The H$\alpha$ emitting gas has a velocity range between 
$+$6 and $+$41 km/s relative to the systematic velocity. This is similar 
to the velocity of the CO line.

{\bf The tangential filament: }
This region refers to position Off 1, at least 12.3 kpc from the galaxy
centre. It corresponds to a region inside the tangential filament
(slit 5) of Hatch et al. 2006. The CO velocity is not well defined
here as it is not consistent between the 3mm and the 1.3mm
data. However, there is a hint of a two-component velocity structure,
at $\pm$ 100km/s. This is consistent with the H$\alpha$ velocity found
in the same region (distance between $\sim$ 15 and $\sim$ 20 kpc in
Fig. 7 of Hatch et al., 2006).
%

{\bf The test region: } A hint of a CO(2--1) detection was present in
the S06 HERA data, in a region that corresponds to no H$\alpha$
filament. We re-observed this region and found no CO(1--0) or CO(2--1)
emission (i.e. M$_{\rm gas}$ $<$10$^8$ M$_\odot$). We conclude that there is
no, or very little, cold gas outside the H$\alpha$ filaments.
%
%
\subsection{The central region}
%
We observed 9 regions to map the CO emission close to 3C84. We
present here a fully sampled CO(1--0) map of about 50$''$$\times$
50$''$ with a spatial resolution of 22$''$. The CO(2--1) map does
not fully sample the same region but has a better resolution of
11$''$. All the results are described in more detail in Table
\ref{table2-center} and Figs \ref{spectra-center1} and \ref{spectra-center2}.
In the central regions the CO(1--0) and CO(2--1) line profiles are
similar. The CO(2--1) line fluxes, when not convolved with the
CO(1--0) beam, are 4 times larger than the CO(1--0) fluxes. So within
the uncertainties, this is in agreement with an optically thick cold
gas. After convolution, all the ratio are close
to one.  Some regions have a ratio greater than one, but as mentioned
before, the CO(1--0) lines are affected by baseline ripples close to
3C84. The main beam brightness temperatures in CO(1--0) are thus very
likely under-estimated in this region.
%
%
\subsection{Detection of HCN(3--2)} 
\begin{figure}[htbp]
\centering 
\begin{tabular}{c}
\includegraphics[width=4.5cm,angle=-90]{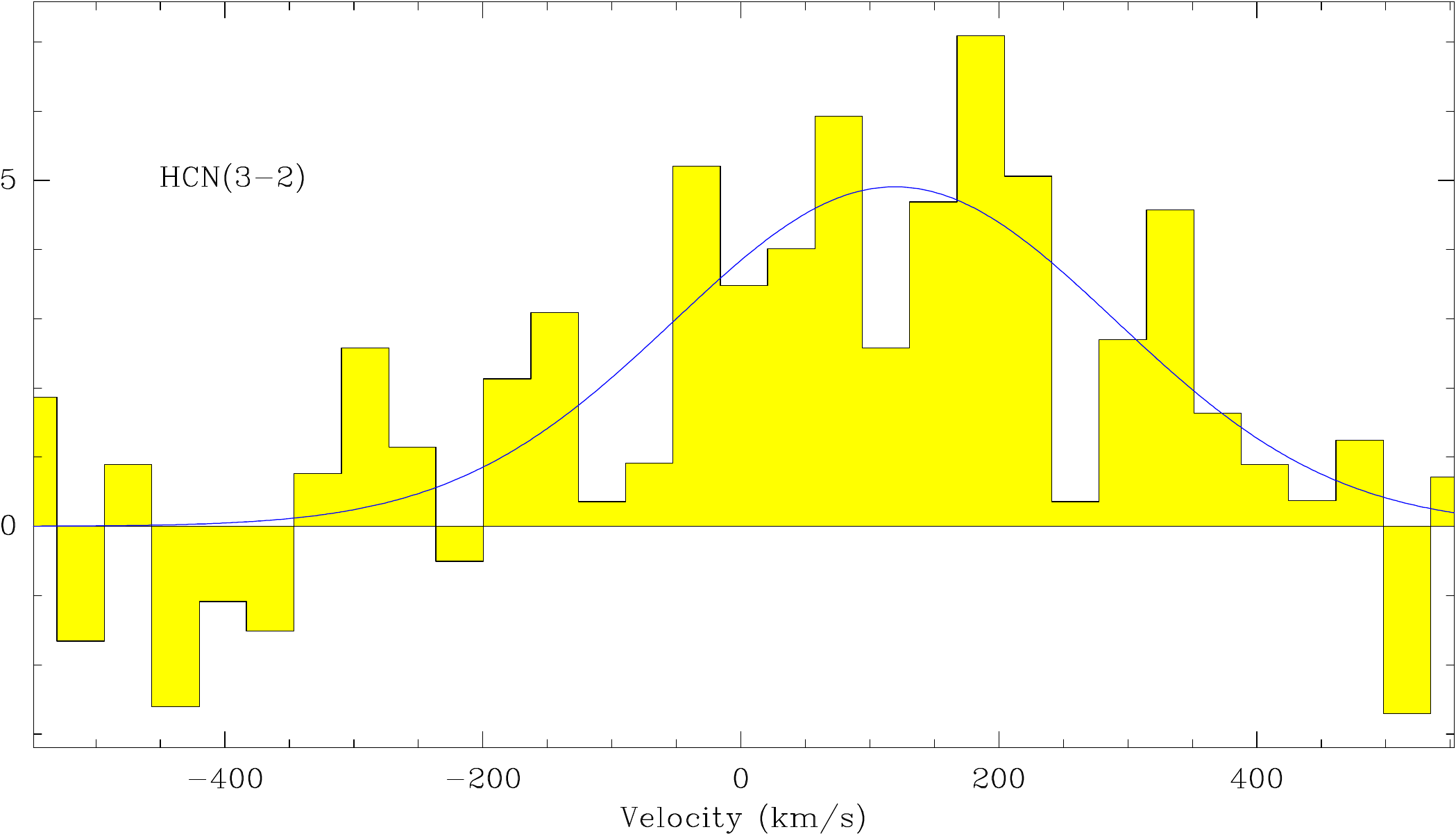} \\
\end{tabular}
\caption{HCN(3--2) emission lines in Tmb (mK), observed with the 30m
telescope (290 min). A gaussian fit gives I$_{\rm HCN}$ 2.1 K.km/s with
a peak temperature of 5 mK, a velocity of 120 km/s and a linewidth of
400 km/s. The rms is 1.6 mK with a mean T$_{\rm sys}$ of 445 K. The
channel width is 36 km/s.}
\label{hcn32}  
\end{figure}

We searched for HCN(1--0), HCN(2--1) and HCN(3--2) emission lines in
the centre of NGC 1275. The observations at low frequency were
corrupted by strong baseline ripples due to the strong continuum
source 3C84.  So we focus on the HCN(3--2) observations at 261 GHz,
observed in average weather conditions. The spectrum presented in Fig
\ref{hcn32} is the average of data taken during a first run with the
two receivers detecting two polarisations (A230, B230) and during a
second run with the two other receivers C270 and D270. The signal
appears marginally in both data sets and even better in the averaged
spectrum. So we claim a detection of HCN(3--2).  The S/N ratio on the
integrated line is 6.\\ 
The HCN molecule has a large dipole moment that requires n(H$_2) >$
10$^4$cm$^{-3}$ for significant excitation. This tracer of dense gas
regions is thus considered as an indicator of star formation activity
(Gao \& Solomon, 2004).  We computed the apparent CO and HCN
luminosities (using equation 1 from Downes et al., 1999) from the
integrated line flux measured in CO(1--0) and HCN(3--2). We assumed a
HCN(3--2)/HCN(1--0) ratio close to one. At densities higher than a few
10$^4$cm$^{-3}$ for a temperature higher that 40-50 K, radiative
transfer models predict a ratio larger than $\sim$0.25. We found
L$'_{\rm CO}$\,=\,5$\times$10$^8$ K\,km/s\,pc$^2$ and L$'_{\rm
HCN}$\,=\,5$\times$10$^7$ K\,km/s\,pc$^2$. When compared to the
L$_{\rm FIR}$-L$'_{CO}$ and L$_{\rm FIR}$-L$'_{\rm HCN}$ relations as
described by Solomon et al. (1992, 1997), our measurements fall in the
normal spiral galaxies range and independently predict an L$_{\rm FIR}$
of 2$\times$10$^{10}$ L$_\odot$. Lester et al., (1995) estimated the
thermal FIR emission to be 10$^{11}$ L$_\odot$, with the approximation
that only 20$\%$ of the total luminosity was non-thermal. This is an
order of magnitude higher than the L$_{\rm FIR}$ predicted here. So
either the contribution of the synchrotron emission is larger than
expected or there is an excess of FIR emission in NGC1275 compared to
the well-known L$_{\rm FIR}$-L$'_{\rm HCN}$ relation for star forming
galaxies.

%
\subsection{Molecular gas from the High Velocity Component}
NGC1275 is surrounded by some star formation regions, with young star
clusters, that could be formed in a merger but also in the cooling
flow (see S06).  Some are associated with the High Velocity System
(HVS) at 8200 km/s, which appears in absorption in front of the
optical emission of NGC 1275 (Gillmon et al., 2004).  No stellar
component is detected, even in the near-infrared, corresponding to the
HVS, which could be a gas-rich disrupted dwarf, or tidal debris, as
proposed by Hu et al (1983).

\begin{figure} 
\centering
\includegraphics[width=4cm, angle=-90]{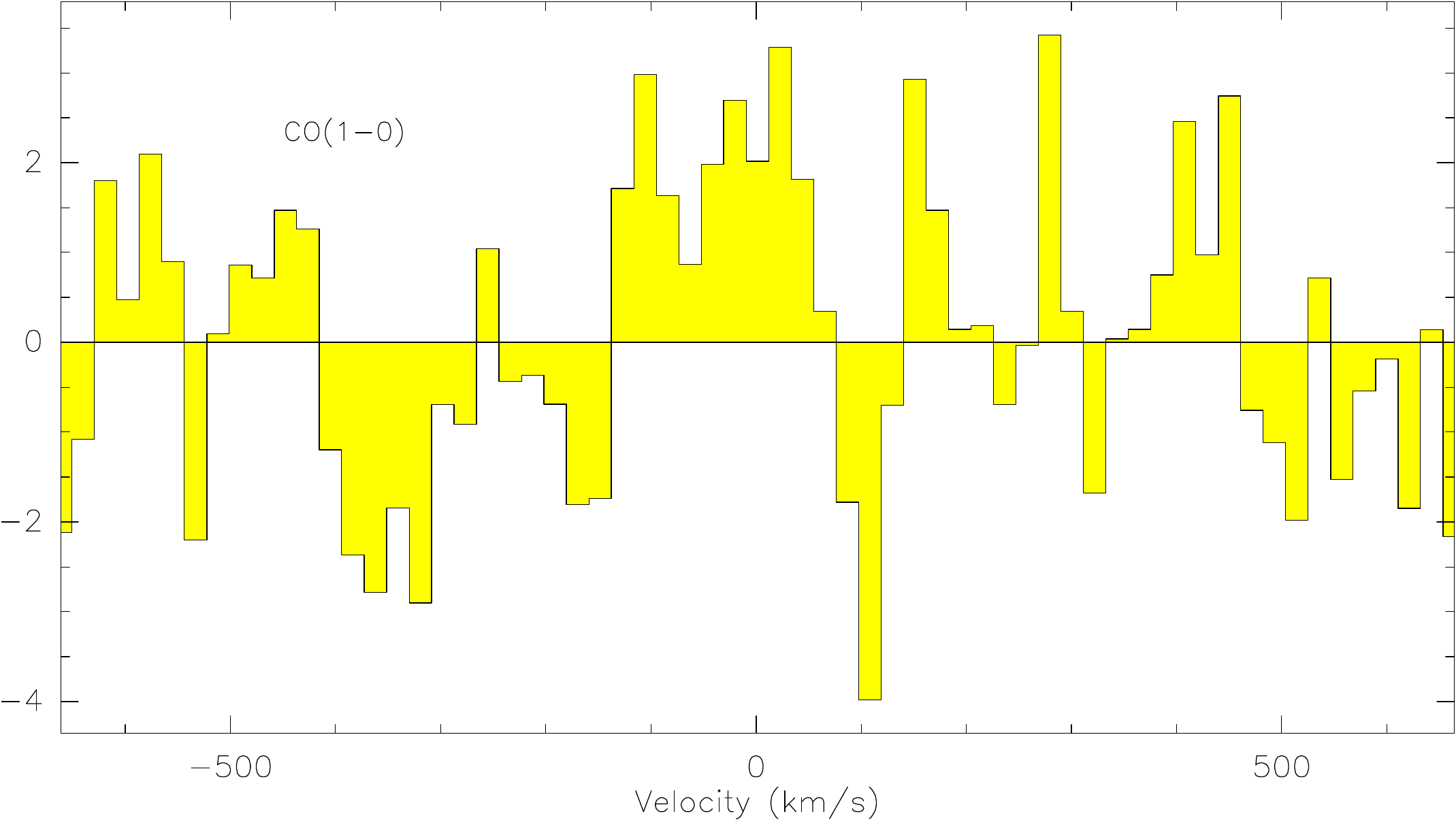} \\
\includegraphics[width=4cm, angle=-90]{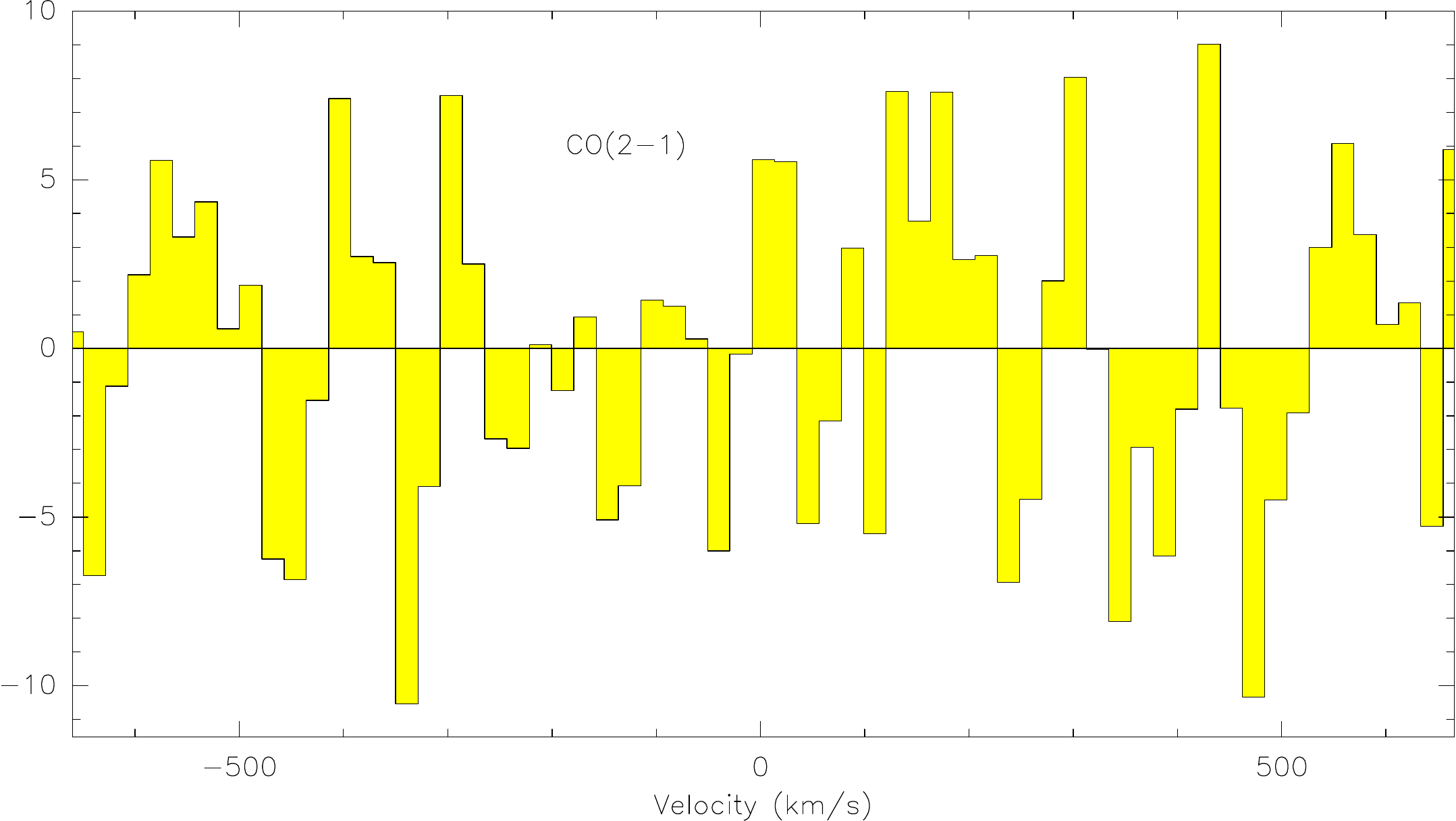} \\
\caption{Observations at 112.175 GHz (top) and 224.346 GHz (bottom),
the CO(1--0) and CO(2--1) frequencies expected in the HVS restframe at
8280 km$s^{-1}$. No CO(2--1) is detected. There is a hint of detection
in CO(1--0), with I$_{CO}$ = 0.4$\pm$0.1 K.km/s.  This would give a
mass of about 1$\times$10$^8$M$_\odot$. The channel width is 21 km/s.}
\label{hvs} 
\end{figure}

We searched for CO emission at 8280 km$s^{-1}$ on one position in the
HVS (centred on 3C84's position). The results are shown in Fig
\ref{hvs}. There is no clear detection but a hint of CO(1--0)
emission. The line intensity we derive is up to a maximum of the order of 
0.4 K.km/s which corresponds to $\le$ 10$^8$M$_\odot$ of cold gas.
So even if interacting with the ambient medium
closely surrounding NGC~1275, the HVS contains a very small amount of
cold gas. This mass is comparable or less than what is found in the
cold filaments around the cD galaxy. Therefore, if the HVS merges with 
the cD galaxy, it will not deposit a significant amount of cool gas.
%
%
\section{Discussion}
%
\subsection{CO(2--1)/CO(1--0) line ratios}
We have computed the CO(2--1)/CO(1--0) line ratios. The data at 1.3mm
were combined with the HERA data described in S06. We then convolved the
CO(2--1) data by the CO(1--0) beam size in order to make a meaningful
comparison. 
We list, in Table \ref{table-filaments}, the CO(2--1)/CO(1--0)
line ratios of the different regions observed.
If we compare the main beam temperature in the CO(2--1) spectra with
and without the 3mm beam shape convolution (central and right hand
side column of Fig. \ref{spectra-filaments}), we can see that we have
lost some signal by diluting the 1.3mm emission with a larger beam
size. This means that the molecular gas emission is very likely to come
from a region that is smaller than the 1.3mm beam size
(11$^{\prime\prime}$).

At Pos 11, the tentative detection of CO(2--1) disappears when
convolved to the CO(1--0) beam.  Regions Off 2, Pos 2, Pos 11 and East
show lower main beam temperatures at 1.3mm when we applied the 3mm
beam convolution. Therefore, the emission from these positions
probably comes from regions with an angular size less than
22$^{\prime\prime}$ (the beam size at 3mm). Regions Off 1 and Off 3 do
not show lower main beam temperatures when convolved with the 3mm beam
and are therefore probably extended by at least 22$^{\prime\prime}$.
We have recently confirmed this last result by mapping the
eastern filament with the
Plateau de Bure interferometer. This shows that the molecular gas lies
in thin ($\le$ 2$''$) and elongated structures, exactly coincident
with the H$\alpha$ filaments (Salom\'e et al., 2008).

After convolution to the same beam size,
we find the CO(2--1)/CO(1--0) line brightness
temperature ratio is close to one in regions East, Off3, Pos2 and
close to 3C84. Such a ratio indicates optically thick CO(1-0) and
CO(2-1) emission lines (Eckart et al., 1988). For regions farther out
from the galaxy centre like Off2, the ratio is slightly larger than
one. However, we sometimes find the line widths smaller in CO(1--0)
than in CO(2--1). This means that the apparently high ratio could
actually be closer to one if the line widths were constrained to be the
same. Note that the Off1 emission-line ratio is about 2.5 while the
line shapes are the same at 1.3mm and 3mm.  So we see hints of higher
CO(2--1)/CO(1--0) ratios at larger radius from the galaxy centre. This
particular high ratio (larger than one) indicates optically thin
emission (Goldsmith et al., 1999).

Closer to the centre, the strong 3C84 source
produced baseline ripples that affected the 3mm line shapes. However,  
we estimated CO(2--1)/CO(1--0) line 
ratios close to one, in agreement with an optically thick 
cold gas (see Table \ref{table2-center}).
%
%
%
\subsection{Molecular gas reservoir and star formation}
The scenario for the origin of the molecular gas we detect is that the
cold gas accumulating towards the centre of the cluster gets dragged
out with the bubbles and compressed in curved shocks, or flows back
down around the borders of cavities. Feedback from the AGN not only
heats the gas (negative feedback), but also compresses it in thin
layers, that favors cooling and formation of molecular filaments
(positive feedback) at a distance greater than 10 kpc (Revaz et al.,
2008).
 
We expect no significant amounts of dust in the filaments, due to high
sputtering rates.  The question is then whether, in these conditions,
there is still sufficient dust (which acts as a catalyst) to form
molecules quickly. It may be that some dust from the ISM of NGC~1275
has been captured inside the uplifted cooler gas and this is
sufficient to catalyse the formation of molecules. Dust could also
form in interstellar shocks, if present. The presence of dust grains
at large radii would then help to form molecules in the outer
filaments. A small quantity of dust grains is sufficient.  We have
proposed further observational investigations in order to constrain
the amount of dust inside the filaments of NGC~1275.

For the assumed standard CO/H$_2$ conversion factor, the amount of
molecular gas detected in the different regions observed varies from
7$\times$10$^7$ M$_\odot$ to 8$\times$10$^8$ M$_\odot$, which when 
added together gives a total mass of 1.3$\times$10$^9$ M$_\odot$.
These filaments therefore contain large quantities of cold molecular gas.
The regions are also conspicuous in the emission lines of H$\alpha$ and [NII]
but the optical line ratios are incompatible with young stars as the source of
excitation as might be expected if stars were forming from the molecular gas.
The excitation of these lines is not coming from the AGN either, since
it does not decrease with distance from the nucleus of NGC~1275
(Hatch et al 2006). The filaments
are emitting extensively in UV and optical lines, much more than in X-rays, and
a possible explanation for this is through heating by shocks at the interface
between the relativistic jet plasma and the entrained gas at the exterior of
the bubbles.

One of the most CO-rich regions is the eastern extension, which
contains an unresolved ($<$\,1\,kpc) young star cluster (Shields et al.
1990). The molecular gas mass of this region is greatly in excess of
the mass of the star cluster and confirms a low star formation
efficiency within the filaments (see also Salom\'e et al, 2008 for
more details).
%
%
\section{Conclusions}
%
New CO(1--0) and CO(2--1) observations towards selected optical
filaments around NGC 1275 confirm the presence of CO
associated with the H$\alpha$ filaments. The gas
kinematics are the same in CO and as in H$\alpha$. We detect for the first
time, a large amount of cold gas inside the outer filaments (a few
10$^8$M$_\odot$).

The excitation of the gas, derived from the comparison of
CO(2--1) and CO(1--0) convolved to the same spatial resolution,
indicates a range of densities are present. Optically thick emission 
is present close to the galaxy centre, whilst further out the emission 
could be optically thin.
  
We report a detection of HCN(3--2) emission in the centre,
indicating high density gas.
The association of the molecular gas with the H$\alpha$
filaments is consistent with the scenario in which cooling of the gas is
taking place at large radii. This could happen in compressed
hot gas at the border of the X-ray cavities. A detailed
interpretation awaits specific simulations taking into account the
cold, warm and hot gas phases.
%
%
\begin{acknowledgements}
IRAM is supported by INSU/CNRS (France), MPG (Germany) and IGN
(Spain). RMJ acknowledges support by STFC and the Royal Society. 
We thank D. Downes for helpful comments and discussions.
\end{acknowledgements}

%
%
%
%
\end{document}